\documentclass[11pt,letterpaper]{article}
\usepackage[utf8]{inputenc}
\usepackage[T1]{fontenc}
\usepackage{amsmath,amssymb,amsfonts,mathtools,bm}
\usepackage{newpxtext}
\usepackage{newpxmath}
\usepackage{graphicx}
\usepackage{booktabs,threeparttable,tabularx,array,multirow,makecell}
\usepackage{enumitem}
\usepackage{subcaption}
\usepackage[table]{xcolor}
\usepackage[hyphens]{url}
\usepackage{hyperref}
\hypersetup{colorlinks=true,linkcolor=black,citecolor=black,urlcolor=black,
            breaklinks=true}
\usepackage{natbib}
\usepackage{geometry}
\usepackage{setspace}
\usepackage{float}
\title{Equilibrium Transition from Loss-Leader Competition: 
       How Advertising Restrictions Facilitate Price Coordination 
       in Chilean Pharmaceutical Retail}
\author{Yu (Jasmine) Hao\thanks{I thank the Chilean Competition Authority (FNE) for
data access, and the editor and anonymous referees for invaluable comments.
Financial support from the 2020 Gambling Award Fellowship is gratefully
acknowledged. All errors are my own. An open-access version of this paper is
available at \url{https://arxiv.org/abs/2512.22917}.}
\\[2pt]
\normalsize Faculty of Business and Economics\\ University of Hong Kong \\
\normalsize \texttt{haoyu@hku.hk}}
\date{\today}

\begin{document}\maketitle

\begin{abstract}\noindent
Between December 2007 and April 2008 Chile's three retail pharmacy chains coordinated price increases on 220 medicines, weeks after advertising restrictions ended a comparative-price war that had driven prices below cost.  I study the transition with a demand-grounded structural model.  Two forces sustained the war: comparative-price ads broadcast who was cheapest, so undercutting paid; and a coordinated increase holds only if rivals expect it matched.  The ban moves both.  By collapsing price sensitivity ($\hat\alpha$ from $0.103$ to $0.029$) it makes undercutting unprofitable for the inelastic majority; as a public event it shifts beliefs, releasing the wave.  A dynamic model estimated by simulated method of moments reproduces the war, the failed attempts, the cartel wave, and the post-cartel rent.  The harm is a transfer to supra-competitive rents, with small deadweight loss because post-ban demand is inelastic.


\end{abstract}

% =====================================================================
%  Section 1: Introduction (Revision 2)
% =====================================================================

\section{Introduction}\label{sec:intro}

Between December 2007 and April 2008 the three retail pharmacy
chains that together account for $92\%$ of pharmacy sales in
Chile (Cruz Verde, Farmacias Ahumada (FASA) and Salcobrand)
coordinated price increases on more than two hundred
prescription and over-the-counter medicines.  The episode was
documented in detail by Chile's National Economic Prosecutor's Office
(\emph{Fiscal\'ia Nacional Econ\'omica}, FNE) and the Competition Tribunal
(\emph{Tribunal de Defensa de la Libre Competencia}, TDLC) in two contemporaneous
court records (the FNE \emph{Requerimiento} of $9$~December~$2008$
and the TDLC \emph{Sentencia} N$^\circ$~$119$/$2012$%
\footnote{FNE, ``Requerimiento contra Farmacias Cruz Verde,
Farmacias Ahumada y Salcobrand,'' $9$ December~$2008$, available at
\url{https://www.fne.gob.cl/wp-content/uploads/2011/03/requ_0009_2008.pdf};
TDLC, ``Sentencia N$^\circ$ $119$/$2012$,'' $31$ January~$2012$,
available at
\url{https://www.fne.gob.cl/wp-content/uploads/2012/01/Sentencia_119_2012.pdf}.})
and is the setting I study here, following earlier work on collusion
among retail pharmacy chains \citep{AleChilet2016,AleChilet2018}.

Through 2006 the chains discounted these medicines only occasionally; a persistent below-cost
price war set in from January~$2007$.  In August~$2007$ Cruz~Verde launched a comparative-price
advertising campaign (\emph{Desaf\'\i o}) that intensified it; in September~$2007$
the advertising self-regulation council (CONAR) ruled the campaign out, and on $6$~November~$2007$ a
Civil Court precautionary order upheld the restriction.  The first coordinated price increase that
held followed within weeks, on $3$~December~$2007$, and the coordinated period ran to the FNE's
investigation notice of $31$~March~$2008$.

Two facts about the episode come from different sources.  The first is documented: the court record shows that the chains
coordinated through the drug manufacturers.  A lab's \emph{visitadores} delivered a single
coordinated price proposal (a batch call) to the three chains in the same week, so the
communication channel between competitors is on the record, not inferred.  The second is what
the event panel adds.\footnote{\S\ref{sec:background_data} defines a coordinated increase as all three chains raising a given drug by at least $15\%$ within a week, with the level held.}  I date each
attempt and separate the ones that failed from the ones that held.  Among the latter, I distinguish a
first increase that restores a $20$--$25\%$ margin from a second increase to a supra-competitive
level, which I call the rent.  Three patterns follow from this coding (\S\ref{sec:descriptive}): the
frequency of failed attempts is higher before and after the cartel than during it; many
drugs receive two increases in sequence; and the increases arrive in batches. On a
given day several products move together, and they are typically the portfolio of a single
laboratory's supplier.

The transition is visible in the first batch calls (\S\ref{sec:descriptive}).  Days after the
$6$~November court order Salcobrand raised a manufacturer's batch that its rivals did not follow,
reverting within about ten days; the first batch that held came four weeks later, on
$3$~December, matched by both rivals.  Because the channel is explicit, that gap is the time the
manufacturer-and-chains negotiation (\S\ref{sec:mechanisms}) took to organize a batch all three
would follow, not a learning-by-doing or tacit-coordination delay.

This article studies the transition from the loss-leader competition that preceded the
coordinated period to the cartel that followed.  The object is the move from one
equilibrium to another, not the post-cartel price levels.  The convictions establish that the chains colluded, so I do not test for collusion.  The explicit lab channel is how they selected the coordinated price.  What the model explains is why that price became self-sustaining once the ban removed the incentive to undercut.  How collusion is initiated, rather
than how it is sustained \citep{GreenPorter1984}, has received less empirical attention.  Two
features set this episode apart from the closest work.  First, the coordination is explicit: the
lab-mediated channel is in the court record, so I do not infer tacit coordination from prices.
Second, the build-up is fast, not gradual.
\citet{ByrneDeRoos2019}, the closest empirical predecessor, documents a $2.5$-year gradual rollout
in retail gasoline; this episode is a $4$-month analogue.  It was triggered by a single
regulatory event, the $6$~November~$2007$ Civil Court order that banned Cruz~Verde's comparative-price
advertising.  The onset is sharp rather than a slow build, a near-discrete flip in the static
incentive rather than years of learning.  The wholesale-cost record from the legal proceedings lets
me identify a per-period spillover in CLP units and compute welfare across the transition.

Two forces sustained the war, and the episode itself shows both are real.  A chain prices below cost only because the shopper drawn in by a cheap headline drug fills the rest of her basket at a profit; that store-traffic value is what repays the lost margin, so the war's persistence is itself evidence that undercutting paid.  The second force explains why the war did not end on its own.  A coordinated increase holds only if rivals are expected to match it.  Through the war none did, so no chain tried to lead one out: belief was low and self-fulfilling.  The advertising ban moves both at once.  By collapsing price sensitivity ($\hat\alpha$ from $0.103$ to $0.029$) it makes a cut steal too little traffic to pay, so for the inelastic majority of drugs no chain wants to undercut the coordinated price.  As a public, dated event it also lifts the shared belief that a raise will be matched, releasing the wave.  Neither change alone reproduces the path (\S\ref{ssec:mech_robust}).

This article makes three contributions to the empirical study of collusion.

The first is an estimable model of how a cartel begins.  Most empirical work studies collusion once it is already sustained, so the transition itself is rarely modelled.  I model the move from a below-cost price war to coordinated supra-competitive pricing with one set of primitives that rationalises both regimes, estimating the store-traffic value $\mu$ inside the dynamic model rather than backing it out of coordinated prices.  Rolled forward from the January-$2007$ onset, the model reproduces the intensifying war, the failed attempts, and the cartel wave, and the emergence and Cruz~Verde leadership of the post-cartel rent.  One feature of that last stage lies beyond it: a synchronized cluster of rent increases in late $2008$, after the investigation window closes, which the model's smooth-arrival dynamics cannot generate.  It also matches the documented order in which larger laboratories coordinate first.

The second is that a public event can coordinate beliefs rather than only change payoffs.  The advertising ban acts as a focal, dated signal that shifts the shared expectation a raise will be matched, so a regulatory event can select an equilibrium like a commitment device.  The ban does not create the coordinated equilibrium, which is dynamically
sustainable in both regimes given a high enough belief.  It selects it.  I embed this in a two-layer
laboratory-to-drug dynamic game estimated by simulated method of moments, and the knockouts show that
coordination needs both the undercut incentive to vanish and the belief to shift.  Eleven parameters are
estimated; the chain-leadership primitives, the firms' loss cushions and within-market shares, are
calibrated from the case record and held fixed.  Who leads is not identified from prices; I take the ordering the documents make most plausible and treat the leadership it implies as a prediction to check, not a fitted result.

The third is that one advertising policy facilitated collusion.  The comparative-price ad ban collapsed the very price sensitivity that made loss-leading pay, so undercutting no longer repaid its margin for the inelastic majority and the loss-leader war gave way to coordination.  A consumer-protection rule aimed at advertising thus had an unintended pro-collusion side effect.  The resulting welfare cost is distributional, a transfer of about
CLP$\,3.4$bn over the cartel window, with small deadweight loss because post-ban demand is inelastic.

\textbf{Related literature.}
Most empirical work on collusion studies how it is sustained
\citep{GreenPorter1984,Harrington2018} or recovers collusive conduct from prices and
quantities \citep{igami2017innovator,clark2014effect,miller2017understanding,
miller2021oligopolistic}; how a cartel begins has drawn less attention.  The
mechanics of reaching coordination run through communication and observable price
moves: Sugar Institute communication harmonising practices \citep{genesove2001rules},
asymmetric gasoline adjustments acting as cartel-holding transfers
\citep{clark2013collusion}, and distributor-run hub-and-spoke coordination
\citep{chaves2025inner}.  The closest study of initiation, \citet{ByrneDeRoos2019}, documents a two-and-a-half-year experimental
rollout in retail gasoline; my episode is its short, regulatory-triggered
counterpart, where one advertising restriction flips a static undercut
incentive so coordination forms in four months without multi-year learning.  Two
studies by Al\'e Chilet examine this same Chilean episode: \citet{AleChilet2016} on
the within-cartel ordering and \citet{AleChilet2018} on Salcobrand's leadership as
Bayesian signalling.

The model's primitive is the cross-category traffic a multi-product retailer wins by
being cheapest: the loss-leader logic in which below-cost ``leaders'' draw shoppers
whose baskets repay the lost margin \citep{lal1994retail,thomassen2017multicategory,
chen2012loss,degraba2003volume,florez2020multiproduct,rao2001equilibrium}.  There
store traffic sustains below-cost pricing; I run it in reverse.  That
restricting price advertising raises prices is an old finding, for eyeglasses
\citep{benham1972advertising} and retail pharmacy
\citep{cady1976estimate,MilyoWaldfogel1999,sinkinson2019ask}, as is the broader result that an
advertising ban can soften competition \citep{Eckard1991}; I supply the channel:
the ban leaves the store-traffic value untouched and acts through consumers' price
sensitivity, whose estimated collapse shrinks the traffic a cut wins so loss-leading
stops paying.

The episode also belongs to a literature on regulation that reshapes
conduct in unintended ways \citep{carranza2015price,ryan2012costs,cicala2015does}.  Here the
consumer-protection advertising rule worked on two margins.  It changed payoffs (collapsing
the price sensitivity that made loss-leading pay, which made the coordinated price self-enforcing), and
it shifted beliefs: a public, dated event that let the chains expect a coordinated increase to
be matched.  In this second role the rule acts like a focal point or commitment device, much as a
price ceiling can become a focal point for tacit coordination \citep{knittel2003price,lewis2015odd};
the difference here is that the coordination is explicit and the rule released it rather than
capping it.  Closest is \citet{AleChiletAtal2020}: a trade association lets many physicians
coordinate the same low-to-high move; here the device is instead a public regulatory event, working
through demand (a belief shift and the elasticity collapse) among only three chains.

Finally, firms often reach a new equilibrium by adaptive learning rather than jumping
to it: fictitious play over three years in UK electricity \citep{doraszelski2018just},
learning-to-price after liquor privatisation \citep{huang2022learning}, strategic
ability in deregulation \citep{goldfarb2011thinks}, biased beliefs in market power
\citep{aguirregabiria2020firms}, menu-cost rigidity
\citep{aguirregabiria1999dynamics,kano2013menu}.  My transition is the opposite
limiting case: a dated event flips the undercut incentive and the belief I estimate
is a fast transient, so the wave is mechanical rather than learned.  For the
multi-product cartel structure I draw on \citet{igami2022measuring} and for
elasticity benchmarks on \citet{grabowski1992brand}; methodologically I follow the
simulated-method-of-moments tradition for dynamic games
\citep{mcfadden1989smm,pakespollard1989simulation,bajari2007estimating} and identify
demand as in \citet{berry1994estimating}.

\textbf{Outline.}
Section~\ref{sec:background_data} describes the data and the regulatory record;
Section~\ref{sec:descriptive} the descriptive patterns the model rationalises
(below-cost pricing, the price-tier transition, and Salcobrand-led timing);
Section~\ref{sec:demand_estimation} estimates nested-logit demand and the
ban-induced collapse in price sensitivity, the pre-ban level on the competitive
window and the post-ban level on the cartel-excluded weeks \citep{berry1994estimating}; Section~\ref{sec:mechanisms} sets out the structural
model, shows it reproduces the war-to-cartel transition, and quantifies the welfare consequences;
and Section~\ref{sec:conclusion} concludes.

% =====================================================================
% End of intro
% =====================================================================

% =====================================================================
%  Section 2: Background and Data (Revision 2)
%  Every cited number carries an explicit source with title + URL.
% =====================================================================

\section{Institutional Setting and Data}\label{sec:background_data}

\subsection{Industry and regulatory record}\label{ssec:background_industry}

The Chilean retail pharmacy market in 2007 was concentrated.  Cruz
Verde, FASA, and Salcobrand together held approximately $92\%$ of
nationwide pharmacy sales (FNE Requerimiento $2008$,
chap.~II)\footnote{The $92\%$ figure is from the FNE complaint, p.~$7$;
the $95\%$ figure from \citet{DiazGaletovic2015} is a later
academic estimate covering both prescription and OTC pharmacy
revenue.}, and were estimated to account for as much as $95\%$ of
pharmacy revenue more broadly \citep{DiazGaletovic2015}.  Each chain set
its list price for each medicine centrally at the corporate level,
valid across all of that chain's stores nationwide; the
``coordination'' I study in this article therefore refers to
coordinated movements of the three chains' centrally set
national list prices, not to local store-level pricing.

Demand for these medicines is price-inelastic and physician-driven: the
prescribing doctor, not the consumer, chooses the molecule, and a patient cannot
substitute across branded products without a new prescription.  The Competition
Tribunal describes consumers as captive through the prescription
(\emph{cautivos a trav\'es de la receta m\'edica}) \citep{tdlc2012sentencia}.
Generic competition is limited and largely outside the chains: only $22.5\%$ of
the molecules in my sample face a generic substitute, and those are sold mainly
through independent pharmacies rather than the three chains, which stock
predominantly branded products bought through common wholesale networks.  This
institutional inelasticity is the basis for the small quantity response, and
hence the transfer-not-efficiency-loss welfare finding, of \S\ref{sec:welfare}.
This was not the first such episode: in $1993$--$94$ the Comisi\'on Resolutiva
found the same three chains to have raised the prices of $80$ ethical medicines
in concert, a sanction the Corte~Suprema affirmed \citep{ciper1995r432}.

\textbf{Marginal-cost symmetry across the three chains.}  The three
chains face nearly identical wholesale-cost schedules: manufacturers publish a
single suggested retail price (\emph{precio de venta al p\'ublico sugerido},
PVPS) per molecule, incorporating a $20$--$25\%$ gross-margin band and sent to
each chain alike, so wholesale-acquisition cost is common across chains for each
drug.  The one substantive asymmetry is Cruz~Verde's vertical integration with the
distributor Socofar, which affects margin retained at the wholesale stage, not the
marginal cost of acquiring the drug.  I therefore treat the wholesale cost $c_{jt}$ as common across
chains, using Salcobrand's disclosed wholesale-cost series (\S\ref{ssec:background_data})
as that common $c_{jt}$.  Chain-level
demand, by contrast, is not forced symmetric: the demand stage
(\S\ref{ssec:demand_est}) estimates a chain-by-drug intercept $\phi_{ij}$ absorbing persistent
differences in how a drug sells across the three chains, whereas the dynamic model adds a chain-specific loss-cushion asymmetry $\psi_i$ that carries the leadership: the non-pharmacy business that absorbs part of the war loss, so the least-cushioned chain is most exposed and leads the escape (calibrated from the business mix, \S\ref{ssec:mech_estim}).

\textbf{Origin of the price war.}  The loss-leader price war
that this article analyses was triggered by an advertising campaign
that Cruz~Verde launched in August~$2007$ under the slogan
``Cruz~Verde Challenge: Low Prices Without Competition''
(\emph{Desaf\'\i o Cruz~Verde, Precios Bajos sin Competencia}).
The campaign published side-by-side price comparisons for selected
products against FASA, claiming systematically lower
prices.\footnote{FNE, ``Requerimiento contra Farmacias Cruz Verde,
Farmacias Ahumada y Salcobrand,'' $9$~December~$2008$,
chap.~II~$\S~3$, identifies the price war (\emph{guerra de precios}) as launched
by Cruz~Verde in August~$2007$, with FASA and Salcobrand reacting
by cutting their own prices on the same products to defend share.
The press coverage in El Mercurio, Econom\'\i a y
Negocios (late September~/ early October~$2007$, article
ID~$34637$) and the \emph{Diario Financiero} of $12$~September~$2007$
documents the campaign and FASA's reaction.}  All three chains
then cut prices well below cost on the products that consumers
were most likely to be familiar with, to defend pharmacy traffic
share.  This is the loss-leader equilibrium whose collapse drives
the cartel-formation episode in this article.  Two primitives keep
it below cost.  Store traffic makes loss-leading pay, because the
basket profit on the traffic of being cheapest repays the lost drug
margin.  Belief keeps the chains trapped: each expects that a
coordinated raise will be undercut, so the war is a self-enforcing
low-belief equilibrium.  Escaping it needs both the undercut
incentive to vanish and the belief to shift, and
\S\ref{sec:mechanisms} shows that neither change alone reproduces
the path.  I measure the
intensity of the war by the below-cost prevalence
(Fig.~\ref{fig:loss_intensity}, monthly revenue shortfall from
below-cost selling), which rises tenfold from June to
November~$2007$.

\textbf{The regulatory record.}  Two regulatory events bracket the
period of interest.  The chronology below documents them; \S\ref{sec:mechanisms} takes up the roles.  The first event is a sequence of three
mutually reinforcing instruments that, between September and
November~$2007$, terminated Cruz~Verde's comparative-advertising
campaign and the loss-leader price war it had driven.  On
\textbf{$7$~September~$2007$}, the board of the Advertising Self-Regulation
Council (\emph{Consejo de Autorregulaci\'on y \'Etica Publicitaria}, CONAR), acting on a
complaint filed by FASA, issued a ruling in Case ROL~$704$/$07$
finding that Cruz~Verde's campaign infringed Articles~$4$, $6$,
$10$ and~$22$ of the Chilean Code of Advertising Ethics (\emph{C\'odigo Chileno de \'Etica Publicitaria}) and
ordering Cruz~Verde to cease the
campaign.\footnote{CONAR, Caso ROL~$704$/$07$ (``Farmacias Ahumada
S.A.\ con Farmacias Cruz Verde S.A.''), Directorio resolution of
$7$~September~$2007$; reconsideraci\'on of $5$ October~$2007$; and
Tribunal de \'Etica decision later that month.  The compiled
jurisprudence is at
\url{https://www.conar.cl/wp-content/uploads/2011/04/Jurisprudencia_Conar-2007.doc}.}
CONAR, founded in $1987$, is the self-regulatory organ of the Chilean
National Advertisers Association (Asociaci\'on Nacional de
Avisadores, ANDA); it adjudicates advertising-ethics complaints
under that private code and its rulings are not judicially
enforceable.\footnote{The Sep--Oct CONAR rulings were morally
binding but not coercive, which is why the price-war prevalence kept
rising through October as Cruz~Verde appealed; the binding
instrument was the $6$~Nov Civil Court injunction documented below.}
The board confirmed the ruling on
reconsideration on $5$~October~$2007$ and the CONAR ethics tribunal
(\emph{Tribunal de \'Etica}) confirmed it on appeal later in October.  On
\textbf{$31$~October~$2007$} the National Consumer Service (Servicio Nacional del
Consumidor, SERNAC) added an administrative judicial complaint under the
consumer-protection law.\footnote{SERNAC press communication,
``Sernac denuncia a Cruz Verde por su publicidad,''
$31$~October~$2007$, available at
\url{https://www.sernac.cl/portal/604/w3-article-918.html}.}
The first coercive judicial instrument came on
\textbf{$6$~November~$2007$}, when the
$17$\textsuperscript{o}~Juzgado Civil de Santiago granted a
\emph{medida precautoria} requested by FASA in its parallel civil
suit under Ley~$20{,}169$ (competencia desleal) and ordered
Cruz~Verde to discontinue the campaign, with the suit subsequently
producing a definitive ruling on $2$~July~$2010$ and a confirming
sentence from the Court of Appeals of Santiago on
$23$~July~$2012$.\footnote{$17$\textsuperscript{o} Juzgado Civil de
Santiago, ``Medida precautoria,'' $6$~November~$2007$ (reported by
CIPER Chile, ``El dossier del caso farmacias,'' $9$~April~$2009$,
available at
\url{https://www.ciperchile.cl/2009/04/09/el-dossier-del-caso-farmacias-asi-se-subieron-los-precios-segun-fasa/});
and definitive ruling of $2$~July~$2010$, confirmed by the Court
of Appeals of Santiago on $23$~July~$2012$ (Rol $4155$-$2010$).}
The first coordinated price increase in the event panel occurs
$27$~days later, on \textbf{$3$~December~$2007$}.

The roughly thirteen-week gap from the CONAR ruling to the first
coordinated increase is consistent with this dating. The non-binding CONAR
ruling did not halt the war: below-cost selling intensified rather than
abated through September--October~$2007$
(Figure~\ref{fig:loss_intensity}). The binding $6$~November Civil Court
\emph{medida precautoria} was the first instrument with the legal force to
compel withdrawal.  The four-week interval from it to the first coordinated
increase on $3$~December is short relative to the multi-week lab-mediated
coordination cycle (single-manufacturer batches take weeks to organise across
the three chains).

The cartel was brought to an end by the FNE's investigation that
led to the formal complaint (\emph{Requerimiento}) of
$9$~December~$2008$.  The TDLC ultimately sanctioned all three
chains, finding (i)~that the chains had collusively coordinated
price increases on $222$ medicines, (ii)~that the suggested retail
margin on these medicines was $20$--$25\%$, and (iii)~that the
cartel's profits represented $2.8\%/4.0\%/3.1\%$ of $2007$ sales
for FASA, Cruz~Verde and Salcobrand respectively, with the $222$
coordinated medicines accounting for roughly $16\%$ of national
pharmaceutical sales (the Farma segment, $2006$--$2009$ average).\footnote{Figures
from the FNE \emph{Requerimiento} ($9$~December~$2008$) and the TDLC
\emph{Sentencia} $119$/$2012$ (both cited above with their URLs).  The profit
split in~(iii) is the FNE's quantification, reproduced in the \emph{Sentencia};
the $16\%$ figure in~(iv) is the Tribunal's own finding (Considerando~$41$),
measured as the coordinated medicines' share of national pharmaceutical sales.}  I treat these events as exogenous in
the structural model, each with its own role: the September CONAR ruling ($t_{\mathrm{CONAR}}$) is the
non-binding precursor that ends the comparative campaign; the binding November ban ($t_{\mathrm{ban}}$)
both collapses demand (the price-sensitivity drop, \S\ref{ssec:demand_est}) and, as a public
signal, shifts the common belief (the focal jump, \S\ref{ssec:mech_id}); and the March FNE notice
($t_{\mathrm{FNE}}$, Oficio~$419$) raises enforcement on the rent tier (the post-Oficio scrutiny,
the skepticism $m$ the chains work off as they survive), not the coordination belief.  The belief jump is dated to the binding November event, not the
September ruling (\S\ref{ssec:mech_robust}).

\textbf{The batch-call coordination mechanism.}
The coordination in this episode is explicit rather than
tacit: pharmaceutical manufacturers in Chile distribute updated price
recommendations through dedicated sales representatives
(\emph{visitadores}), who call on each chain's central purchasing
office on a periodic basis.  Manufacturers publish a suggested retail
price (\emph{precio de venta al p\'ublico sugerido}, PVPS) per drug (a list price that
incorporates the standard $20$--$25\%$ gross margin
(\S\ref{ssec:background_industry})) and visitadores deliver revised
PVPS schedules to each chain simultaneously.  When a chain accepted
a manufacturer's new PVPS, it raised its prices on that
manufacturer's entire current portfolio within a single weekly
window: this is a lab-batch call.  The TDLC
\emph{Sentencia} $119/2012$ documents that Salcobrand and rivals
used the visitadores channel as a bilateral coordination conduit:
Salcobrand would first commit to the manufacturer's new PVPS, and
the visitador would carry that commitment to Cruz~Verde and FASA as
the signal that triggered their follow-on raises.\footnote{TDLC
\emph{Sentencia} $119/2012$ (available at
\url{https://www.fne.gob.cl/wp-content/uploads/2012/01/Sentencia_119_2012.pdf}),
conduct findings; FNE \emph{Requerimiento} $9$~December~$2008$,
chap.~II, identifying the $222$~coordinated medicines across $36$
therapeutic categories (IMS classification).  The FNE's post-cartel market study
\citet{fneEM2018} confirmed the visitadores protocol as the
coordination mechanism.}
Because the PVPS covers every presentation of a molecule from a
single manufacturer, each visitador call naturally generates a
batch across that lab's full portfolio.  Every batch call in November~$2007$ failed: the initiating chain raised prices and
rivals declined to follow. The first sustained three-chain
coordinations emerged in the week of $3$--$9$~December (Online
Table~\ref{tab:batch_calls} lists the principal batches).

\subsection{Data}\label{ssec:background_data}

\textbf{Data.}  Three sources, all from the public case record, underlie the analysis.
\textbf{(i)~Prices and quantities.}  A transaction-level panel of daily list prices and unit sales for each of the $222$ drugs named in Annex~I of the $2008$ FNE \emph{Requerimiento}, at the three chains (Cruz~Verde, FASA, Salcobrand), obtained from the FNE under Chile's transparency law (Ley No.\ $20.285$), $2006$--$2008$.  Wholesale costs are Salcobrand's, disclosed to the TDLC.  I estimate demand on this daily panel; Online Appendix~\ref{app:data} gives the construction.
\textbf{(ii)~Event panel.}  From the price series I build one event panel, combining a $15\%$ coordination threshold with a more permissive $12\%$ daily detector (a $12\%$ move over the prior $42$-day median, within a $\pm7$-day rival-follow window).\footnote{Online Appendix~\ref{app:event_coding} validates this panel against the FNE official cartel indicator: the two largely agree, with a median timing difference of $-2$ days and $79\%$ of drugs within a week.  I use the event panel as the primary outcome because it isolates the coordination event from the trailing reaction periods that contaminate the official panel.}  It records price moves in both directions.  A coordinated increase is all three chains raising a drug by at least $15\%$ within a $7$-day window with the level held, labeled a success, a failed attempt, or a unilateral hold by whether rivals follow.  Where the automated detector is ambiguous, the initiating chain, the success-versus-failed-attempt label, and a small number of coordination successes are hand-coded from the case record (Online Appendix~\ref{app:event_coding} gives the coding rules), so these structural moments are partly hand-curated rather than purely mechanical.  A price decrease, or war deviation, is a chain cutting its price at least $15\%$ below its own $2006$ baseline for at least two weeks during the pre-cartel war; the panel records $814$ such war deviations in total (the structural moment in \S\ref{sec:mechanisms} is a narrower windowed count, not this raw total).  After coordination, a rarer defection is a chain cutting below both the cheaper rival and its own cartel plateau ($51$ events, about $16$ of them deep and lasting past the March~$2008$ FNE notice).  The depth of these below-cost decreases and the timing of the failed attempts are moments the structural estimation weights heavily.  This panel is the source for the event counts in \S\ref{sec:descriptive}.  Of the $222$ drugs named in the TDLC \emph{Sentencia}, $220$ ($99.1\%$) show at least one successful three-chain coordination (Online Appendix~\ref{app:data} details the mapping).
\textbf{(iii)~Documentary record.}  The TDLC \emph{Sentencia} No.\ $119$/$2012$ (conviction on $206$ drugs, fines near US\$$38$M, upheld by the Supreme Court), the FNE complaint (Requerimiento, Rol~C No.\ $184$-$08$), the CONAR self-regulatory ruling ($704$/$07$), the $6$~November~$2007$ civil-court injunction, and FASA's $2007$ Humphreys credit-rating report \citep{FasaHumphreys2007}.  All are publicly available with their URLs (Online Appendix~\ref{app:timeline}).

% =====================================================================
% End of background
% =====================================================================

% =====================================================================
%  Section 3: Descriptive Evidence (Revision 2)
%  Descriptive evidence section. Comes before the demand model
%  (§4), the structural mechanisms (§5), and the survival analysis
%  (§6). All numbers in this section are pre-regression, descriptive
%  cuts of the data; structural estimates are introduced later.
% =====================================================================

\section{Descriptive Evidence}\label{sec:descriptive}

\textbf{Market structure.}  Three national chains (Cruz~Verde, FASA, and
Salcobrand) together hold close to $90\%$ of formal pharmacy sales.  They compete
head-to-head for the same branded molecules, priced off a common centralized
national list and bought through shared wholesale networks
(\S\ref{ssec:background_industry}).  Two features of this structure shape every
fact below.  Demand is largely captive: the prescribing doctor picks the
molecule, so the chains compete for which store a shopper visits, not
whether she buys.  This is what makes a below-cost loss leader a rational bid for
store traffic.  The chains' own accounts show the basket is large enough to matter: in the pre-cartel years non-pharmaceutical lines (cosmetics, personal care, baby products, beverages) made up about a quarter of Cruz~Verde's profit ($24\%$) and revenue ($23\%$), so a customer a cheap headline drug draws in fills a substantial basket.  The store-traffic value $\mu$ the model estimates is the profit on the marginal traffic a deviation wins (the deal-seeking switchers), so it sits above this average per-visit margin. It is read from the war floor (\S\ref{sec:mechanisms}), not measured from these accounts.  And coordination, when it comes, runs through the manufacturers'
batch calls rather than chain-to-chain, so the unit that moves is the
laboratory's portfolio, not the single drug.

The regulatory timeline is set out in
\S\ref{ssec:background_industry}: the war that Cruz~Verde's August-$2007$ campaign
drove, the CONAR ruling and $6$~November Civil Court order that ended it, and the
FNE investigation that closed the cartel.  Read against that timeline, the facts
describe a loss-leader equilibrium whose intensity was accelerating through $2007$
toward a level that could not be sustained.  They then describe a coordinated wave
of price increases that restored the chains' weighted pharmacy margin to roughly
the historical $20$--$25\%$ band.  I establish three facts and organise the section
around them: \textbf{(i)}~pre-ban below-cost pricing whose intensity was
accelerating, not stationary (\S\ref{ssec:desc_loss_intensity}); \textbf{(ii)}~a
sharp price-tier transition that restored the historical margin and was not a cost
shock (\S\ref{ssec:desc_prices}); and \textbf{(iii)}~Salcobrand-dominant leadership
and a four-month wave organised laboratory-by-laboratory, the largest laboratories
first (\S\ref{ssec:desc_leadership_timing}).  Sections~\ref{sec:demand_estimation}
to~\ref{sec:welfare} build the structural model that rationalises the magnitudes;
this section establishes that they are there to be explained.

\subsection{Fact 1: Below-cost pricing accelerated before the ban}\label{ssec:desc_loss_intensity}

The standard formal argument for a loss-leader equilibrium is that
firms expect the per-period loss on the loss-leader to be recovered
from contemporaneous spillovers on other categories
(\S\ref{ssec:mech_spillover_id}).
A natural diagnostic of whether such an equilibrium is sustainable
is whether the loss is approximately stationary or whether it grows
over time.  The below-cost gap is visible directly in
Figure~\ref{fig:wholesale}: across 2007 each chain's quantity-weighted price
(Panel~B) runs below the wholesale cost (Panel~A), and the gap widens through the
year.  Panel~C aggregates, for each
chain, the monthly revenue shortfall from below-cost sales: that
is, $\sum_{j,t} \max(c_{j}-p_{ijt},\,0)\cdot q_{ijt}$ summed across
drugs and days within each month.\footnote{The cost
benchmark $c_{j}$ is the drug's wholesale cost from the Salcobrand cost series, which the data record
from November~$2007$; for earlier months I hold $c_{j}$ at its first observed level, because the
wholesale price of these off-patent molecules is stable over the window.  Because the pre-November
benchmark is backward-extrapolated, the level of the early shortfall is measured with more error
than the post-November level; the acceleration into late~$2007$ is robust to this, but its exact onset
month should be read as approximate.}

Through 2006 and the first half of 2007, the monthly
shortfall is roughly steady at $\sim$CLP\,$50$--$100$ million per
chain.  From mid-2007 onwards, every chain's shortfall begins to
grow, and the growth accelerates into November--December 2007, with
Cruz Verde reaching a peak of roughly $\sim$CLP\,$900$ million per
month.  The chains were bleeding roughly an order of magnitude more
in late 2007 than they were a year earlier.  Within weeks of the
first coordinated increase, the shortfall is back to baseline.

This pattern is hard to reconcile with the loss-leader equilibrium
having been ``stable'' through 2007.  It motivates the structural
measurement: the firms moved as the net payoff to staying put in the war deteriorated.

\subsection{Fact 2: The price-tier transition}\label{ssec:desc_prices}

\begin{figure}[ht]
\centering
\caption{Wholesale cost, prices, and below-cost loss}
\label{fig:shares_time}\label{fig:loss_intensity}\label{fig:wholesale}
\includegraphics[width=\linewidth]{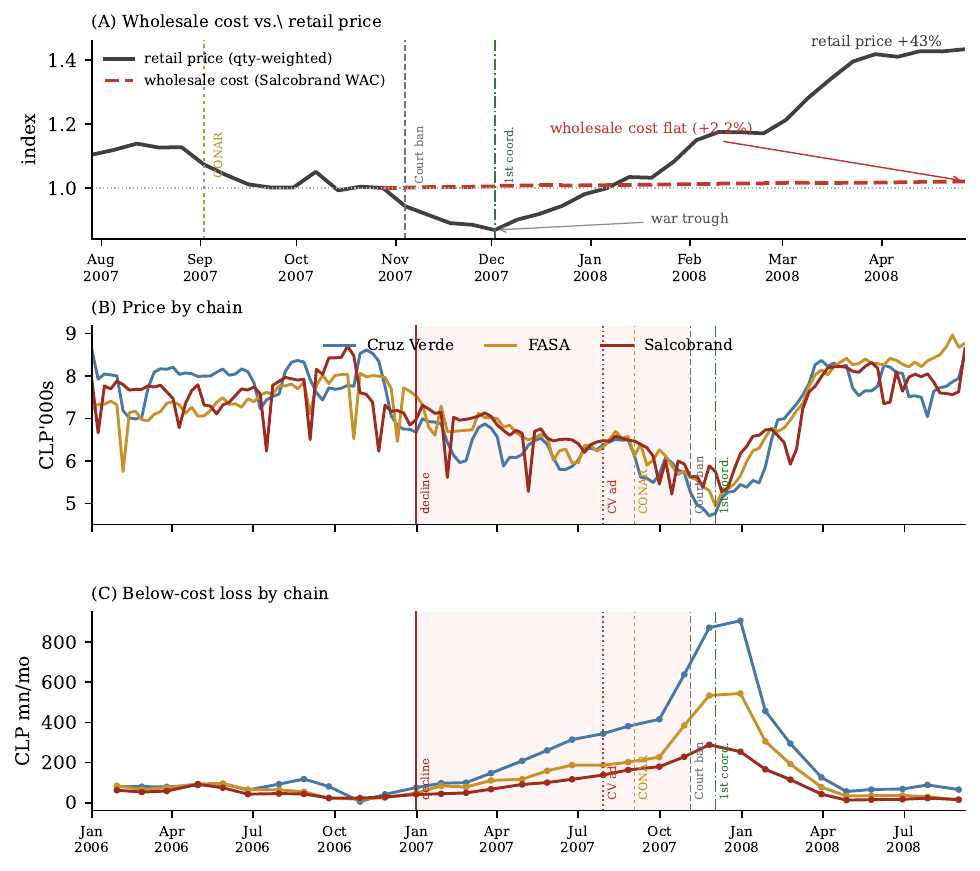}
\par\smallskip\flushleft
{\footnotesize \textbf{Note}: \textbf{(A)} Salcobrand wholesale-cost index (dashed) and
quantity-weighted retail-price index (solid), both indexed to the cost-series start,
November~$2007$--May~$2008$.  \textbf{(B)} Weekly quantity-weighted average price by chain
across the $222$ coded drugs, $2006$--$2008$.  \textbf{(C)} Monthly below-cost revenue
shortfall by chain, $\sum_j \max(c_j-p_{ijt},0)\,q_{ijt}$.  The CONAR self-regulatory ruling
($7$~September~$2007$) and the binding Civil Court order ($6$~November) are marked in
Panels~B--C.}
\end{figure}

Figure~\ref{fig:shares_time} (Panel~B) shows the weekly quantity-weighted
average price for each chain over the 2006--2008 window.  Three
features stand out.  First, prices co-move throughout: across the
$156$ weeks of data, the three chains are within roughly
$\pm$CLP\,$500$ of one another at the weekly frequency, consistent
with the centralised national list-pricing
described in \S\ref{ssec:background_industry}.  Second, prices
drift downward from mid-2006 through late 2007 (from
$\sim$CLP\,$7{,}000$ to $\sim$CLP\,$5{,}000$), with the dip steepening
between the CONAR ruling and the first coordinated increase.  Third,
in a single sharp step in December 2007--February 2008, the
chain-level price returns to and exceeds the 2006 level.
This step-up was not a cost shock.  Over the same coordination window the chains'
wholesale costs rose only a few percent, about $3.3\%$ on Salcobrand's WAC series
disclosed to the Competition Tribunal (December 2007--May 2008; $+2.2\%$ over the
longer November--May window of Figure~\ref{fig:wholesale}).  Retail prices over the
same window rose $28$--$60\%$ across therapeutic categories \citep{tdlc2012sentencia}.
A cost shock cannot account for a price path that fell during the war and then rose
far more than costs during the recovery: the reversal reflects coordinated conduct,
not input costs.

The coordinated step restores a sustainable, nonnegative margin.  Quantity-weighted
markups move from substantially negative ($-12$ to $-16\%$) at the war floor to small
positives at the first ($\ell\!=\!0\to1$) step, still below the $20$--$25\%$
suggested-retail band the TDLC documents (\emph{Sentencia} $119/2012$, Cap.~XI).  A
second push ($\ell\!=\!1\to2$) overshoots to $+34$ to $+41\%$, into the region the
Tribunal flags as supra-competitive.  This asymmetry maps onto the structural model
directly: the first step is the regime flip (the ban turns the one-shot undercut gain
$G_j$ negative, restoring a sustainable margin), the second a selective
rent-extraction step taken by only the most inelastic drugs (\S\ref{sec:mechanisms}).
Table~\ref{tab:desc_tier_markups} decomposes the markup by chain and tier.

\begin{table}[ht]
\centering\small
\caption{Markups by tier and chain}
\label{tab:desc_tier_markups}
\begin{tabularx}{0.9\linewidth}{@{}>{\raggedright\arraybackslash}X c c c@{}}
\toprule
                    & Tier~0   & Tier~1   & Tier~2  \\
                    & (pre-coord.) & (1st coord.) & (2nd coord.) \\
\midrule
\multicolumn{4}{l}{\textbf{Panel A. Unweighted mean markup}} \\
Cruz Verde          & $0.025$  & $0.302$  & $0.596$ \\
FASA                & $0.041$  & $0.304$  & $0.592$ \\
Salcobrand          & $0.047$  & $0.310$  & $0.610$ \\
\midrule[\heavyrulewidth]
\multicolumn{4}{l}{\textbf{Panel B. Quantity-weighted mean markup}} \\
Cruz Verde          & $-0.158$ & $0.105$  & $0.371$ \\
FASA                & $-0.154$ & $0.089$  & $0.342$ \\
Salcobrand          & $-0.122$ & $0.127$  & $0.405$ \\
\midrule[\heavyrulewidth]
\multicolumn{4}{l}{\textbf{Panel C. Count of products with negative markup}} \\
Cruz Verde          & $133$    & $60$     & $12$    \\
FASA                & $131$    & $59$     & $14$    \\
Salcobrand          & $127$    & $53$     & $11$    \\
\bottomrule
\end{tabularx}
\par\smallskip\flushleft
{\footnotesize \textbf{Note}: Panel B: quantity-weighted mean
markup.  Panel C: count of chain-drug-weeks with negative markup.
Markup $m_{ijt}=(p_{ijt}-c_{jt})/p_{ijt}$.  Tiers correspond to
the coded coordination events (\S\ref{sec:background_data}).
$N=222$ drugs.}
\end{table}

\subsection{Fact 3: Leadership and timing}\label{ssec:desc_leadership_timing}

Salcobrand led $247$ of the $363$ coordination events ($68\%$) and dominated the
cartel wave ($211$ of $284$ coordinations).  This holds even though Cruz~Verde bore the
heaviest pre-ban below-cost exposure ($31.3\%$ of drug-weeks below cost against
Salcobrand's $25.6\%$, FASA between).  The structural model (\S\ref{sec:mechanisms}) resolves this through the loss cushion.  Salcobrand is almost purely a pharmacy, so it has the least non-drug business to cushion the war's below-cost losses.  The war is therefore costliest for it, which gives it the strongest incentive to lead the escape even though the larger Cruz~Verde cut deeper.  The case record confirms the order: the FASA minute notes that Salcobrand ``very habitually'' raised first (\emph{muy habitualmente Salcobrand}; \citealp{tdlc2012sentencia}).  In the post-investigation rent the below-cost loss is gone and market share governs instead, so the largest chain Cruz~Verde overtakes Salcobrand ($36$ initiations to $34$), the $\mathrm{SB}\!\to\!\mathrm{CV}$ handoff the model reproduces.  The loss-cushion ordering is read from the chains' business mix, not from the leadership data it then predicts.  Table~\ref{tab:desc_chain} reports
the full leadership split.

\begin{table}[htbp]
\centering\footnotesize
\setlength{\tabcolsep}{4pt}
\renewcommand{\arraystretch}{0.9}
\caption{The coordination ledger: leadership, regime, and magnitude}
\label{tab:desc_chain}\label{tab:event_window}\label{tab:event_decomp}

% --- Panel A: below-cost exposure + leadership, by chain (full width) ---
\begin{tabularx}{\textwidth}{@{}l*{3}{>{\centering\arraybackslash}X}@{}}
\toprule
\multicolumn{4}{@{}l}{\textbf{Panel A. Below-cost exposure and leadership, by chain}}\\[2pt]
 & CV & FASA & SB \\
\midrule
\multicolumn{4}{@{}l}{\textit{Pre-ban below-cost exposure}} \\
\quad Fraction below cost              & $31.3\%$ & $26.9\%$ & $25.6\%$ \\
\quad Median $(p-c)$ ($10^{3}$ CLP)    & $0.70$   & $0.87$   & $0.98$   \\
\addlinespace
\multicolumn{4}{@{}l}{\textit{Leadership, by period (initiator)}} \\
\quad War (pre-cartel)                 & $0$  & $0$  & $2$   \\
\quad Cartel wave (Dec'07--Mar'08)     & $14$ & $59$ & $211$ \\
\quad Post-investigation rent          & $36$ & $7$  & $34$  \\
\quad All coordinations                & $\mathbf{50}$ & $\mathbf{66}$ & $\mathbf{247}$ \\
\quad\quad of which first-tier ($0\!\to\!1$) & $13$ & $42$ & $165$ \\
\quad\quad of which rent ($1\!\to\!2$)       & $37$ & $24$ & $82$  \\
\quad Share of all ($n=363$)           & $13.8\%$ & $18.2\%$ & $68.0\%$ \\
\bottomrule
\end{tabularx}

\smallskip
% --- Panel B: events by kind and regime window (full width) ---
\begin{tabularx}{\textwidth}{@{}l*{6}{>{\centering\arraybackslash}X}@{}}

  \multicolumn{7}{@{}l}{\textbf{Panel B. Events by kind and regime window}}\\[2pt]
            & Pre-decline & War/dec.\      & Pre-cartel      & Cartel        & Post-invest.   &       \\
Event kind  & (2006)      & (Jan--5Nov'07) & (6Nov--2Dec'07) & (3Dec--31Mar) & (1Apr--Dec'08) & Total \\
\midrule
First coordination (Tier $0\!\to\!1$)   & $0$  & $1$  & $1$  & $216$ & $2$  & $\mathbf{220}$ \\
Rent step (Tier $1\!\to\!2$)            & $0$  & $0$  & $0$  & $77$  & $66$ & $\mathbf{143}$ \\
Failed attempt (reverted)               & $86$ & $99$ & $23$ & $4$   & $67$ & $279$ \\
Unilaterally held                       & $62$ & $50$ & $12$ & $20$  & $42$ & $186$ \\
\textbf{Total events}                   & $\mathbf{148}$ & $\mathbf{150}$ & $\mathbf{36}$ & $\mathbf{317}$ & $\mathbf{177}$ & $\mathbf{828}$ \\
\bottomrule
\end{tabularx}

\smallskip
% --- Panel C: coordination steps by tier transition and magnitude (full width) ---
\begin{tabularx}{\textwidth}{@{}l*{4}{>{\centering\arraybackslash}X}@{}}

\multicolumn{5}{@{}l}{\textbf{Panel C. Coordination steps by tier transition and magnitude}}\\[2pt]
              & Large         & Med.\         & Small         & $<12\%$ \\
              & ($\geq 25\%$) & ($15$--$25$)  & ($12$--$15$)  &         \\
\midrule
$0\!\to\!1$ (first)   & $177$ & $31$ & $3$  & $9$  \\
$1\!\to\!2$ (rent)    & $51$  & $47$ & $21$ & $24$ \\
\textbf{All coord.}   & $\mathbf{228}$ & $\mathbf{78}$ & $\mathbf{24}$ & $\mathbf{33}$ \\
\bottomrule
\end{tabularx}
\par\smallskip
{\scriptsize Magnitude bins sum to $363$ coordination steps ($220$ first $+\,143$ rent).}

\par\smallskip\flushleft
{\footnotesize \textbf{Note}: Chains are Cruz~Verde (CV), FASA, and Salcobrand (SB).
Panel~A decomposes the $363$ coordination events by chain.
Below-cost exposure: drug-weeks with $p_{ijt}<c_{jt}$ in the pre-ban window (weeks $1$--$94$;
$9{,}945$ drug-weeks per chain).  Leadership: the initiating chain of each event (the $220$ first
coordinations $0\!\to\!1$ and the $143$ rent steps $1\!\to\!2$), split by the war (pre-cartel),
cartel ($3$~Dec~$2007$--$31$~Mar~$2008$, to the FNE \emph{Oficio}) and post-investigation periods,
so the rows reconcile to the full ledger ($2+284+77=363$).  Salcobrand leads the war and the cartel
wave; Cruz~Verde overtakes it in the post-investigation rent (the $\mathrm{SB}\!\to\!\mathrm{CV}$
handoff).  Panel~B sub-windows: ``Pre-decline'' $=2006$ (prices flat at $\approx7.7$);
``War/decline'' $=$ Jan~$2007$--$5$~Nov~$2007$ (the decline begins January~$2007$ and Cruz Verde's
August-$2007$ comparative-price campaign intensified it); ``Pre-cartel'' $=6$~Nov--$2$~Dec~$2007$
(civil-court injunction to the day before the first three-chain success); ``Cartel''
$=3$~Dec~$2007$--$31$~Mar~$2008$ (to the day the FNE opened its investigation, notice No.~$419$);
``Post-investigation'' $=1$~Apr--Dec~$2008$.  The $\mathbf{220}$ first coordinations form the cartel
in the Dec--Mar window.  The rent then deepens through $2008$: $77$ of the $\mathbf{143}$ rent steps
fall in the cartel window, the other $66$ post-investigation, as the regime price climbs toward $15$.
Panel~C magnitude is the price increase over the prior tier (the war price for the $0\!\to\!1$
step, the Tier-$1$ price for the $1\!\to\!2$ rent): ``large'' $\geq25\%$, ``medium'' $15$--$25\%$,
``small'' $12$--$15\%$, the rest smaller week-aligned steps.  Most are large ($228$ of $363$ steps are
$\geq25\%$).  Coordinations and rents are from the dynamic ledger (which tracks the rent past the
panel's April cutoff); failed lifts and unilateral holds are the coded event series.}
\end{table}

\begin{figure}[ht]
\centering
\caption{Coordination events over the episode}
\label{fig:fail_rate_time}
\includegraphics[width=\linewidth]{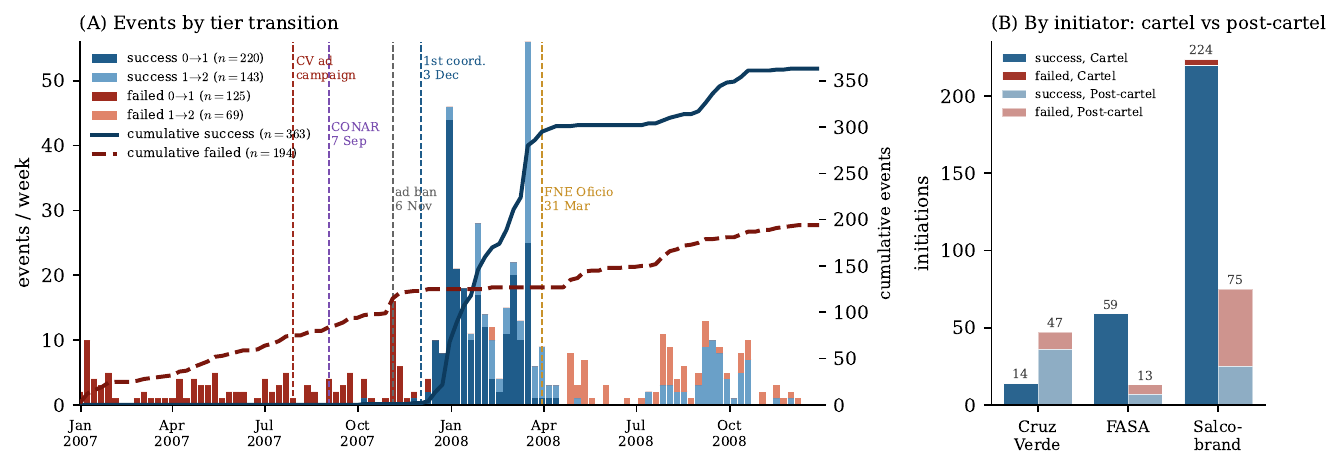}

\par\smallskip\flushleft
{\footnotesize \textbf{Note}: Coordination events over January~$2007$--December~$2008$, decomposed
two ways.  Over the episode the ledger records $363$ successful coordination steps ($220$ first-tier
$0\!\to\!1$, $143$ rent $1\!\to\!2$) against $194$ failed or reverted attempts ($125$ at $0\!\to\!1$,
$69$ at $1\!\to\!2$).  \textbf{Panel A}: weekly events split four ways, successful $0\!\to\!1$ (dark blue) and
$1\!\to\!2$ (light blue) versus failed $0\!\to\!1$ (dark red) and $1\!\to\!2$ (light red); the right
axis plots cumulative successes (solid) and failures (dashed).  (Failed attempts carry no recorded
tier; I assign each to $0\!\to\!1$ or $1\!\to\!2$ by whether the drug had already coordinated by the
attempt date.)  \textbf{Panel B}: initiations by chain, each split successful versus failed and
cartel-window versus post-cartel.  Vertical dashed lines (chronological): Cruz Verde's August
comparative-price campaign, the $7$~September CONAR ruling, the $6$~November ban, the $3$~December
first coordination, and the $31$~March FNE official notice No.~$419$.}
\end{figure}

The wave ran $3$~December~$2007$ to $31$~March~$2008$: $216$ of the $220$ first
coordinations (Tier~$0\!\to\!1$) fall within these $19$ weeks, whereas the rent steps
(Tier~$1\!\to\!2$) ratchet on through year-end.  Over the full episode the chains made
$220$ first coordinations and $143$ rent steps, the $363$ events the structural model
targets, $284$ of them in the cartel-window success moment.  Against these stand $194$
failed or reverted attempts from January~$2007$ on (Panel~B gives the full $2006$--$2008$ ledger by window).
The ordering was organized at the laboratory level: each manufacturer's visitador
delivered a batch across its portfolio, so timing is set by which laboratory engaged
first.  Larger laboratories coordinated first: laboratory membership, not a drug's own
size, elasticity, or below-cost depth, sets the order (Table~\ref{tab:sequence_reg}).
Before the cartel held, the chains were already testing: $35$ attempts in the
$6$~November to $2$~December window, $23$ failed or reverted and $12$ unilaterally
held (Figure~\ref{fig:fail_rate_time}).

\begin{table}[ht]
\centering
\caption{Lab membership, not drug characteristics, sets the coordination order}
\label{tab:sequence_reg}
\small
\begin{tabular}{lc}
\toprule
Predictors of the first-coordination day (OLS, $N=219$) & $R^{2}$ \\
\midrule
Drug covariates (size, elasticity, below-cost gap)            & $0.03$ \\
\quad $+$ drug-type fixed effects                             & $0.09$ \\
\quad $+$ \textbf{manufacturer (laboratory) fixed effects}    & $\mathbf{0.61}$ \\
\bottomrule
\end{tabular}\\[4pt]
\raggedright\footnotesize
\textbf{Note}: Each row regresses the calendar day of a drug's first coordinated increase on the listed
predictors (the $J=220$ coordinated drugs, $N=219$ after dropping one with incomplete pre-ban
characteristics; continuous regressors standardised, larger $=$ later).  The manufacturer fixed effect is
decisive: adding it raises $R^{2}$ from $0.09$ to $\mathbf{0.61}$.  So laboratory membership alone
explains $57\%$ of the cross-drug variance in coordination timing, against only $3\%$ for the drug's own
size, elasticity, and below-cost depth: the order is set by which laboratory engaged, not by the drug.
The laboratories with the highest pre-ban trading volume coordinate first (Figure~\ref{fig:lab_size_timing}).
Standard errors clustered by manufacturer ($G=37$).
\end{table}

\begin{figure}[ht]
\centering
\caption{Larger laboratories coordinate first}
\label{fig:lab_size_timing}
\includegraphics[width=0.70\linewidth]{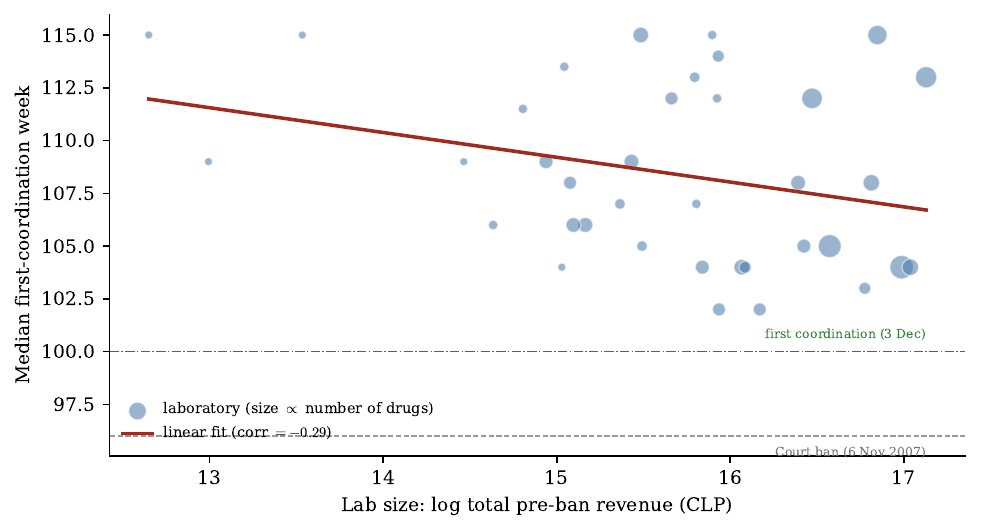}
\par\smallskip\flushleft
{\footnotesize \textbf{Note}: Each point is one laboratory ($37$ labs, $221$ coordinated
drugs), at its size and the median week of its drugs' first three-chain coordination.
Lab size is the laboratory's total pre-ban revenue (log CLP), summed across its
drugs, the stake $W_L$ the structural model assigns it.  Larger labs coordinate earlier
(correlation $-0.29$; the structural model reproduces this at $-0.38$,
Table~\ref{tab:mech_fit}).  Point area is proportional to the laboratory's number of drugs.
Dashed and dash-dotted lines mark the binding $6$~November court ban and the first sustained
coordination, $3$~December~$2007$.}
\end{figure}

Figure~\ref{fig:lab_size_timing} shows the size gradient directly.  A laboratory's median
first-coordination week falls with its size at a correlation of $-0.29$, where size is the log of its
total pre-ban revenue, the same stake $W_L$ the structural model assigns it.  A larger laboratory plausibly
moves first because it carries more SKUs to bring under the coordinated umbrella and so makes a more
focal coordination device: starting with its batch synchronises the most drugs at once.  This negative
correlation does work in the estimation.  The moment $\mathrm{corr}(\log\text{ lab size},\text{ week})=-0.29$
disciplines the lab-engagement parameters $\lambda$ and $\tau_e$ in the structural
model (\S\ref{ssec:mech_estim}, Table~\ref{tab:mech_fit}), so the larger-labs-first ordering is matched by
the model rather than assumed.  Table~\ref{tab:batch_calls}
records the principal batch calls: every batch call in October and November failed, and the first
sustained three-chain coordinations emerged in the week of $3$--$9$~December.

\begin{table}[ht]
\centering\small
\caption{Early lab-batch calls failed; December calls held}
\label{tab:batch_calls}
\begin{tabularx}{0.92\linewidth}{@{}l l r l >{\raggedright\arraybackslash}X@{}}
\toprule
Dates & Laboratory & Drugs & Initiator & Outcome \\
\midrule
Oct~7--13     & Andromaco         & $3$  & Salcobrand           & Failed \\
Nov~8--13     & Laboratorio Chile & $19$ & Salcobrand           & Failed \\
\addlinespace
Nov~30--Dec~8 & Andromaco         & $6$  & Salcobrand, FASA     & Partial ($2/6$; the two holds FASA-led) \\
Dec~7--11     & Bago              & $4$  & Salcobrand, FASA     & Success \\
Dec~12--15    & Tecnofarma        & $4$  & FASA                 & Success \\
Dec~18--20    & Saval             & $5$  & Salcobrand, FASA, Cruz Verde & Success \\
\bottomrule
\end{tabularx}
\par\smallskip\flushleft
{\footnotesize \textbf{Note}: Principal lab-batch calls, October--December $2007$, from the
event panel.  A ``batch call'' is a span of about a week in which a chain raised prices on
several drugs from one laboratory at once; Drugs counts the distinct drugs (DrugIDs) in the
batch and Initiator lists each leading chain with $\geq 1$ drug (ordered by drug count).
Outcome is the three-chain result: ``Failed'' = the initiating chain reverted with no rival
follow, ``Success'' = all drugs held three-chain, ``Partial'' = a fraction held.  The two early
calls (Andromaco, October; Laboratorio Chile, November) failed.  The first sustained three-chain
coordinations came in December: the Andromaco call of late November began holding on $4$~December
(two drugs, FASA-led), and the Bago, Tecnofarma, and Saval calls held in full across the next two
weeks.  The lab-level size gradient (larger laboratories coordinate first) is shown in
Figure~\ref{fig:lab_size_timing}.}
\end{table}

\begin{figure}[ht]
\centering
\caption{Example coordination: Sildenafil price path}
\label{fig:example_coord}
\includegraphics[width=0.92\linewidth]{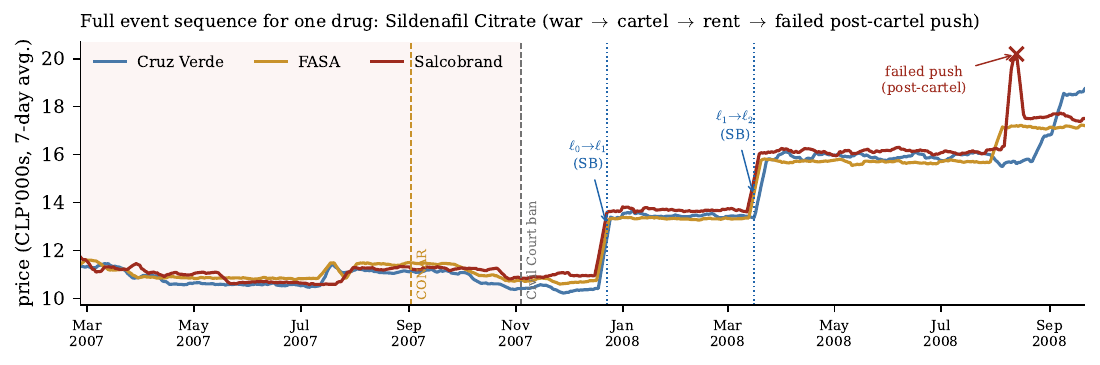}
\par\smallskip\flushleft
{\footnotesize \textbf{Note}: Weekly chain price for Sildenafil Citrate (Drug \#80) with the coded
tier transitions marked.  The three chains move together within the same week at each
tier step; Salcobrand initiates both.}
\end{figure}

Figure~\ref{fig:example_coord} illustrates one episode.  The weekly price for
Sildenafil Citrate (Drug~\#80) drifts down through mid-$2007$, then steps up
within a single week in early December~$2007$ (tier~$0\to1$, led by Salcobrand)
and again in early April~$2008$ (tier~$1\to2$); the three chains co-move tightly
across both transitions.  The official binary indicator $\mathrm{col}_{ijt}$
records only the first step (it is monotone per drug), so it understates the
cartel's reach by about one-third.  It records fewer steps than the event coding in $127$ of
the $137$ multi-round drugs (Online Appendix~\ref{app:event_coding}), which is why I use the
tier ladder throughout.

The failure rate (Figure~\ref{fig:fail_rate_time}) traces the regime. It runs $\approx
97\%$ through the first four weeks of November, below $10\%$ by mid-December, near
zero through March.  Then it snaps back up after the FNE issues Oficio
N$^\circ$~$419$ on $31$~March~$2008$, when new coordination stops while the cartel
holds and the rent deepens.  The reverted lifts in this last window are failed new
attempts on not-yet-coordinated drugs, not defections: the $220$ members largely hold their
coordinated price\footnote{A small number of coordinated drugs defect later: about sixteen show a deep, lasting cut by one chain after the March~2008 FNE Oficio, the chain falling below the held price without rejoining.  These cuts are isolated (the other two chains hold) and fall where undercutting is unprofitable ($G^{01}_j<0$).  They read as enforcement-era retreat rather than competitive undercutting, and their small number matches the model's prediction that post-ban defection is rare.} (Table~\ref{tab:event_window}, Panels~B--C, gives the
full ledger by regime window and magnitude).  Salcobrand leads the failed
re-coordination push ($38$ of the $46$ post-investigation attempts, $83\%$), the same war-exposed chain that led the escape from it.

The dynamic structural model of \S\ref{sec:mechanisms} takes this loss-leader regime
as the period-$0$ state and asks who exits it and when; the next three sections build
that model.

% =====================================================================
% End of descriptive section
% =====================================================================

% =====================================================================
%  Section: Demand (Revision 2, clean draft -- v2 IV-GMM)
%  Standalone .tex; intended to replace pp. 14--17 of
%  manuscript_current_20251102.tex (\section{Demand}...\subsection{Demand
%  Estimation Results}). Numbers and tables track output/r2_demand_*_v2.*
%  and code/03_demand/1_1_demand_r2.py.
%
%  Comments tagged [Rx] in margin show which referee/editor point each
%  paragraph addresses; remove before submission.
% =====================================================================

\section{Demand}\label{sec:demand_estimation}

\subsection{Model}\label{ssec:demand_model}

A consumer in week $t$ decides whether to fill a prescription for drug
$j \in \{1,\dots,J\}$ at one of the three national chains
$i \in \mathcal{I}=\{\mathrm{CV},\mathrm{FA},\mathrm{SB}\}$, or to use the
outside option (an independent pharmacy or no purchase).
I use a two-level nested logit in which the three chains form a single
inside nest for each drug and the outside option is the only competing
nest.\footnote{Drug $j$ is defined at the molecule-dosage level; the three
chains carry the same brand-package portfolio for each $j$ (Section~\ref{sec:background_data}).
Prescription pre-selects $j$, so substitution across drugs is absorbed in
the outside good.}
Consumer $h$'s indirect utility from buying drug $j$ at chain $i$ is
\begin{equation}
  U_{hijt} = u_{ijt} + \nu_{hjt} + (1-\sigma)\,\varepsilon_{hijt},
  \qquad \sigma \in [0,1),
  \label{eq:nl_utility}
\end{equation}
where $\nu_{hjt}$ is the nest-level taste shock common to the three chains,
$\varepsilon_{hijt}$ is an i.i.d.\ Type~I extreme-value shock, and the
nesting parameter $\sigma$ measures within-nest correlation across the three chain alternatives.  Let
$D_t \equiv \mathbf{1}\{t \ge t_{\mathrm{CONAR}}\}$ be the
demand-regime indicator.  The demand split and the structural collapse are dated differently by design.  $D_t$ switches at the September CONAR ruling, which opens the restricted-advertising regime and so marks where the demand data change.  The structural model instead dates the binding collapse in price sensitivity, with the belief shift it brings, to the November court ban that made the restriction enforceable (\S\ref{ssec:background_industry}).  Mean utility is
\begin{equation}
  u_{ijt} \;=\; X_{jt}\beta \;-\;
  \bigl(\alpha_0 + \alpha_1 D_t\bigr) p_{ijt} \;+\;
  \phi_{ij} + \xi_{ijt},
  \qquad \alpha_0 > 0,\ \ \alpha_0 + \alpha_1 > 0,
  \label{eq:meanutility}
\end{equation}
where $p_{ijt}$ is the transaction price (CLP\,000s), $\phi_{ij}$ is a
chain-by-drug fixed effect, $X_{jt}\beta$ collects time-varying drug
characteristics, and $\xi_{ijt}$ is an unobserved demand shock.  The
price coefficient is decomposed into a pre-ban level $\alpha_0$ and a
post-ban shift $\alpha_1$; the null of no ban effect on price sensitivity
is $H_0\!:\!\alpha_1 = 0$.  The sign convention
$\alpha_0,\alpha_0+\alpha_1>0$ makes \eqref{eq:meanutility} consistent
with downward-sloping demand on both sides of the ban.%
\footnote{Throughout, I report estimates of $\alpha_0$ and
$\alpha_0+\alpha_1$ as the relevant price sensitivities;
$\alpha_1$ is the ban-induced shift.}

Aggregating over consumers yields the within-nest share
$s_{i|jt}$, the nest share $s_{jt}=\sum_{i\in\mathcal{I}}s_{ijt}$, and
the outside share $s_{0jt} = 1-s_{jt}$.
Taking logs and differencing relative to the outside good
\citep{berry1994estimating} gives the single pooled estimating equation
\begin{equation}
  \underbrace{\ln s_{ijt} - \ln s_{0jt}}_{Y_{ijt}}
  \;=\;
  X_{jt}\beta \;-\; \alpha_0\,p_{ijt} \;-\; \alpha_1\,(D_t \cdot p_{ijt})
  \;+\; \sigma\, \ln s_{i|jt} \;+\; \phi_{ij}
  + \xi_{ijt}.
  \label{eq:estimating}
\end{equation}
Equation~\eqref{eq:estimating} is estimated in two passes: pre-ban weekly GMM, and a
post-ban re-estimation on the cartel-excluded weeks
(\S\ref{ssec:demand_iv}--\S\ref{ssec:demand_gmm}).  The weekly panel is the natural
prescription-refill unit and the standard frequency in pharmaceutical demand studies
\citep{ellison1997characteristics,bjornerstedt2014does}.  The per-drug coefficients
$\hat\alpha_j$ instead use the daily panel, whose far larger count of within-drug,
cross-chain price observations gives the stronger first stage a drug-specific slope
needs.

The market size for drug $j$ is
\begin{equation}
N_j \;=\; \frac{\max_t\,Q_{jt}^{\mathrm{wk}}}{\hat\rho} + 1,
\label{eq:market_size}
\end{equation}
where $Q_{jt}^{\mathrm{wk}}=q_{\mathrm{CV},jt}+q_{\mathrm{FA},jt}+q_{\mathrm{SB},jt}$
is the total weekly three-chain quantity and $\hat\rho=0.92$ is the
three-chain share of Chilean pharmacy sales (FNE Requerimiento 2008,
p.~7; \citealp{DiazGaletovic2015} estimate $\hat\rho=0.95$).  Dividing
by $\hat\rho$ calibrates the outside good to represent consumers at
independent pharmacies ($\approx 8\%$ of national sales) plus
eligible non-buyers in a given week.  At the peak week, $s_{0j}\approx 8\%$;
at an average week, $s_{0j}\approx 37\%$.  The constant $+1$
prevents a degenerate $s_{0jt}$ in the peak observation; the
chain-by-drug fixed effect $\phi_{ij}$ absorbs any remaining
level mis-scaling.\footnote{Robust to two alternative calibrations
($\hat\rho=0.95$; a constant national population $N^{*}/J$):
$(\hat\alpha,\hat\sigma)$ shift by $<\pm 0.01$ (Online Appendix~\ref{app:ms_robust}).}

The baseline specification imposes a single $\sigma$ and a single pre-ban price
coefficient $\alpha_0$.  With three chains per drug-week, in-sample variation does not
separately identify drug-specific price coefficients from $\phi_{ij}$, so the
cross-product dispersion in implied elasticities reported in
Section~\ref{ssec:demand_fit} reflects variation in prices and shares, not preference
heterogeneity.  I relax the homogeneity restriction in Section~\ref{ssec:demand_est}
below.

\subsection{Identification and Instruments}\label{ssec:demand_iv}

Two regressors in \eqref{eq:estimating} are endogenous: price $p_{ijt}$
(correlated with $\xi_{ijt}$ through unobserved demand shocks) and the
within-nest share $\ln s_{i|jt}$ (correlated with $\xi_{ijt}$
mechanically). The instrument set is four variables:
\begin{enumerate}\itemsep0pt
  \item \textbf{Wholesale cost (drug average)}, $\overline{c}_j$: the
        weekly average of Salcobrand's recorded wholesale cost for drug
        $j$, averaged across weeks. Drug-level cost variation pins down
        the cross-section of $\alpha$.
  \item \textbf{Wholesale cost (weekly)}, $c_{jt}$: Salcobrand's
        wholesale cost in week $t$. Within-drug, week-to-week cost
        variation pins down the time-series identification of $\alpha$.
  \item \textbf{Rival price 1}, $p_{-1,jt}$, the contemporaneous price
        of one rival chain.
  \item \textbf{Rival price 2}, $p_{-2,jt}$, the price of the other
        rival chain.
\end{enumerate}
The two cost instruments shift the marginal cost faced by all three
chains symmetrically (the wholesale market is national), so they enter
$p_{ijt}$ but, conditional on $\phi_{ij}$ and the time controls in
$X_{jt}$, are excluded from $\xi_{ijt}$. The two rival prices follow the
\citet{hausman1996valuation} logic: they shift the within-nest share
$\ln s_{i|jt}$ through demand substitution but are otherwise excluded.

\textbf{Instrument validity across the ban.}  The two cost instruments are
excluded in every regime.  The two rival-price (Hausman) instruments are valid only
when the chains compete: once they coordinate, rival prices co-move with the demand
shock $\xi_{ijt}$ and are no longer excluded.  Pooling all ads-banned weeks, the
over-identifying $J$ rejects ($J=14.5$, $p=.001$).  The rejection is confined to
the collusive cartel window (3~Dec 2007--11~Apr 2008).  Dropping those weeks and
re-estimating on the competitive ads-banned weeks, the same four instruments
pass the over-identifying test ($J=4.4$, $p=.11$;
Table~\ref{tab:demand_hetero}, note~$\ddagger$).  I therefore identify $\alpha_0$ on
the pre-ban weeks and the post-ban price sensitivity on the cartel-excluded
competitive weeks, with $\sigma$ pooled at its pre-ban value $0.393$ so that price is
the only endogenous regressor.

\textbf{The collapse does not hinge on the instruments.}  The two endogenous regressors are not on the same footing.  The within-nest share $\ln s_{i|jt}$ is mechanically tied to $\xi_{ijt}$ through the Berry inversion, so every nested-logit estimate instruments it; an OLS that does not returns an inadmissible $\hat\sigma>1$.  The own price is the contested case.  To isolate it I hold $\sigma$ at its estimated value and re-estimate the price coefficient under OLS, the rival-price instruments, and the wholesale-cost instrument.  Table~\ref{tab:iv_robust} reports the three.  The price coefficient barely moves.  Pre-ban it is $0.093$ under OLS against $0.103$ under the rival IV; post-ban it is $0.032$, $0.029$, and $0.030$ under OLS, the rival IV, and the cost IV.  The collapse is $0.35$ under OLS and $0.29$ under the rival IV.  The own-price endogeneity is therefore small, and the price collapse that drives the structural mechanism does not rest on the rival-price exclusion.  One honest limit: the wholesale-cost series begins in November~2007, so it is flat before the ban and cannot identify $\alpha_0$; pre-ban I use the rival prices, and the OLS agreement shows this does not matter.

\begin{table}[H]
\centering\small
\caption{The price collapse is robust to the identification strategy}
\label{tab:iv_robust}
\begin{tabular}{@{}lccc@{}}
\toprule
                                   & OLS & Rival-price IV & Cost IV \\
\midrule
$\hat\alpha_0$ (pre-ban)           & $0.093$ & $0.103$ & --- \\
$\hat\alpha^{\rm post}$ (post-ban) & $0.032$ & $0.029$ & $0.030$ \\
Collapse ratio                     & $0.35$  & $0.29$  & --- \\
\bottomrule
\end{tabular}
\par\smallskip\flushleft
{\footnotesize \textbf{Note}: The own-price coefficient, with the nesting parameter $\sigma$ held at $0.393$ so that price is the only endogenous regressor, under OLS, the two rival-price (Hausman) instruments, and the Salcobrand wholesale-cost instrument.  The cost instrument is unavailable before the ban (the cost series begins November~2007), so it identifies only the post-ban coefficient.  The collapse ratio is $\hat\alpha^{\rm post}/\hat\alpha_0$.}
\end{table}

\subsection{Estimation and Test Statistics}\label{ssec:demand_gmm}

I estimate \eqref{eq:estimating} by two-step IV-GMM
\citep{berry1994estimating,conlon2020best} on the pooled chain-drug-week
panel, absorbing the chain-by-drug effect $\phi_{ij}$ with a within-group
transform (instruments demeaned likewise).  The moment condition is
$E[Z_{ijt}\,\xi_{ijt}(\theta)]=0$ with $\theta=(\alpha_0,\alpha_1,\sigma,\beta)$;
each instrument enters at both its direct and ban-interacted level
$D_t\cdot Z_{ijt}$, which separates $\alpha_0$ from $\alpha_1$
(\S\ref{ssec:demand_iv}).  Standard errors treat $\xi_{ijt}$ as
heteroskedastic across $(i,j,t)$; drug-level clustering is a robustness check.

On each window I report the first-stage $F$-statistics for $p_{ijt}$ and
$\ln s_{i|jt}$ \citep{stock2005testing}, the \citet{hansen1982large}
overidentification $J$, and the fitted--observed correlation
in $Y_{ijt}$ (the nested-logit analogue of $R^2$).  The $J$ has $q-k=1$ degree of freedom on the pre-ban window, where $\sigma$ is jointly estimated.  It has $q-k=2$ on the post-ban windows, where $\sigma$ is fixed.  I impose no bounds on
$\alpha$ or $\sigma$.

\subsection{Estimates}\label{ssec:demand_est}

Table~\ref{tab:demand_hetero} reports the demand estimates.  The headline row is the
pooled Model~B result ($\hat\alpha_0=0.103$, $\hat\sigma=0.393$; cartel-excluded
post-ban $\hat\alpha^{\rm post}=0.029$) that the structural model consumes; the lower
block reports the per-drug coefficients $\hat\alpha_j$ that give the model its
drug-level heterogeneity.  The headline pooled row (Panel~A) uses the weekly panel and the weekly market-size
calibration \eqref{eq:market_size}; the per-drug coefficients (Panel~B) use the daily
panel ($T=1{,}096$ days, 1~January~2006--31~December~2008, 222 drugs) with the daily analogue
$N_j = 7\cdot\max_t Q_j^{\rm daily}/\hat\rho+1$ (the factor of seven converts
peak-day quantity to a weekly-equivalent market).

\textbf{Per-drug coefficients.}  Per-drug $\hat\alpha_j$ are 2SLS slopes on the
cartel-excluded daily panel ($\hat\sigma$ fixed at $0.393$; rival-price IVs pre-ban,
$+$ Salcobrand cost post-ban); the structural model scales each pre-ban $\hat\alpha_j$
by the pooled ratio $0.29$.  Online Appendix~\ref{app:ij_demand_gof} gives the per-drug
specification.

\begin{table}[H]
\centering
\caption{Heterogeneous price coefficient by drug class}
\label{tab:demand_hetero}
\centering
\small
\begin{threeparttable}
% --- Reference row + Panels A and B side-by-side ---
\flushleft\textbf{Panel A. Pooled headline} (homogeneous $\alpha$, $\sigma$ pooled at the pre-ban value)\\[2pt]
{\centering\footnotesize
\begin{tabularx}{0.9\linewidth}{@{}l c c c c >{\raggedright\arraybackslash}X@{}}
\toprule
 & $\hat\alpha$ & $\sigma$ & $N$ & $J$ ($p$) & \\
\midrule
Pre-ban (cost $+$ rival IV)
 & $0.103$ & $0.393$ & $62{,}985$ & $0.60$ ($.44$) & identifies $\alpha_0,\sigma$ \\
\quad SE
 & $(0.005)$ & $(0.107)$ & & & \\
\addlinespace
Post-ban, all weeks (incl.\ cartel)
 & $0.022$ & $[0.393]$ & $41{,}769$ & $14.5$ ($.001$) & not used$^{\ddagger}$ \\
\quad SE
 & $(0.001)$ & & & & \\
\addlinespace
Post-ban, cartel-excluded
 & $\mathbf{0.029}$ & $[0.393]$ & $29{,}172$ & $\phantom{0}4.4$ ($.11$) & used; $\alpha^{\rm post}\!/\alpha_0\!=\!0.29$ \\
\quad SE
 & $(0.001)$ & & & & \\
\bottomrule
\end{tabularx}\par}

\vspace{0.8em}
% --- Per-drug alpha_j (2SLS, Firm FE, cartel-excluded daily panel; pre-ban rival-price IV, post-ban Salcobrand cost + rival IVs) ---
\flushleft\textbf{Panel B. Per-drug $\hat\alpha_j$ by drug type:} 2SLS, Firm FE, cartel weeks dropped, pooled $\hat\sigma=0.393$; pre-ban rival-price IV, post-ban Salcobrand cost {\it and} rival IVs (daily panel)\\[2pt]
{\centering\footnotesize
\begin{tabularx}{0.9\linewidth}{@{}>{\raggedright\arraybackslash}X r c c c@{}}
\toprule
Drug type & $n_{\rm drugs}$ & Pre-ban $\hat\alpha_j$ & Post-ban $\hat\alpha_j$ & post/pre \\
\midrule
Chronic $\times$ Rx  & $130$ & $0.123$ & $0.023$ & $0.19$ \\
Chronic $\times$ OTC & $\phantom{0}66$ & $0.127$ & $0.061$ & $0.48$ \\
Acute   $\times$ OTC & $\phantom{0}11$ & $0.229$ & $0.026$ & $0.12$ \\
Acute   $\times$ Rx  & $\phantom{0}15$ & $0.214$ & n.i. & --- \\
\midrule
All 222 drugs        & $222$ & $0.130$ & $0.029$ & $0.22$ \\
\bottomrule
\end{tabularx}\par}
\begin{tablenotes}\footnotesize
\item[$\ddagger$] On all ads-banned weeks the $J$ test rejects ($J=14.5$, $p=.001$): cartel weeks are collusive, so rival prices fail exclusion.  Dropping them restores validity ($J=4.4$, $p=.11$) and gives the post-ban $\hat\alpha=0.029$ ($\alpha^{\rm post}/\alpha_0=0.29$).
\item \textit{Per-drug.}  Each $\hat\alpha_j$ is a 2SLS slope (rival-price IV pre-ban, $+$ Salcobrand cost IV post-ban; cartel-excluded), $\hat\sigma$ pooled at $0.393$; the model scales each pre-ban $\hat\alpha_j$ by the pooled collapse $0.29$ (Panel~A): the structural model thus uses the per-drug pre-ban level and the economy-wide collapse.  The all-drug per-drug ratio in Panel~B is $0.22$ rather than $0.29$ because the pooled pre-ban coefficient $\hat\alpha_0=0.103$ lies below the per-drug median $0.130$; the model uses the precisely-estimated pooled collapse, not the noisier per-drug posts.\footnote{A fully drug-specific post-ban coefficient is not used because it is imprecisely identified: post-ban the comparative advertising that identifies cross-chain price response is gone and prices are coordinated, so the cost instrument is thin at the drug level (it cleanly identifies only the Chronic classes, $195$ of the $196$; the Acute classes are wrong-signed).  Estimated per drug, some $48$ drugs receive a spuriously weak collapse, so a single estimated collapse is the right object; dropping all $48$ and re-estimating leaves the structural estimate unchanged (Online Appendix~\ref{app:drop48_robust}).}
\end{tablenotes}
\end{threeparttable}
\end{table}

The per-drug coefficients line up with prior knowledge of pharmaceutical demand on two
points.  First, the acute-over-chronic ranking holds, with a median $\hat\alpha_j$ of $0.23$ for
Acute$\times$OTC against $0.12$ for Chronic$\times$Rx (refill prescriptions create
switching costs that dampen price response).  Second, the per-drug estimates
collapse post-ban to the same all-drug $0.029$, matching the pooled headline.  The cost-only IV ($0.033$) and the post-cartel-only
window ($0.028$) reproduce it without using the
rival prices at all.

\subsection{Goodness of Fit and Implied
Elasticities}\label{ssec:demand_fit}

Three diagnostics describe how well $\hat\theta=(\hat\alpha_0,\hat\alpha_1,\hat\sigma)$ fits the data:
within-nest share fit, the cross-sectional ranking of implied own-price elasticities against
pharmaceutical benchmarks \citep{Goldman2004}, and a tier-$2$ holdout.  The formulas are collected in
Online Appendix~\ref{app:demand_fit_decomp}.

\textbf{(A) Within-nest share fit.}  The structural within-nest share prediction
\eqref{eq:share_pred} uses only the estimated parameters and prices (the observed share never
enters the right-hand side).  It is therefore a genuine prediction, not a mechanical contraction-mapping
$R^2$ \citep[][p.~$32$]{conlon2020best}.  Predicted and observed within-nest shares correlate $0.73$
(pre-ban), $0.86$ (post-ban), and $0.78$ pooled (Table~\ref{tab:demand_fit}).

\textbf{(B) Implied elasticities, benchmarked.}  I evaluate the standard nested-logit formulas
\eqref{eq:own_elas}--\eqref{eq:cross_elas} at each drug's pre-ban prices and shares.  The median own-price
elasticity is $-1.75$, with $99.5\%$ of drugs negatively signed and all cross elasticities positive.
Post-ban the median falls to $-0.6$ at the collapsed price sensitivity and the coordinated prices, with $84\%$ of drugs now inelastic.
The cross-class ranking agrees in sign with the U.S.\ Medicare estimates of \citet{Goldman2004}: chronic
essential categories price-inelastic, discretionary ones price-elastic.  The classes a chain
pushes below cost are the inelastic chronic refills, not the most elastic items (Spearman
$\rho=-0.62$ across $12$ ATC1 classes; Online Appendix Table~\ref{tab:elast_below_cost}).

\textbf{(C) Tier-$2$ holdout.}  Refitting on tier-$\{0,1\}$ weeks and predicting $\hat s_{i|jt}$
on the held-out tier-$2$ weeks via \eqref{eq:share_pred} (with $\hat\phi_{ij}$ recomputed on the
training subset) tests whether demand extrapolates to the fully-coordinated regime in which the
welfare results of \S\ref{sec:welfare} are evaluated; the held-out correlation is $0.85$.

\begin{table}[H]
\centering\small
\caption{Demand goodness-of-fit diagnostics}
\label{tab:demand_fit}
\begin{tabular}{lccc}
\toprule
                       & Pre-ban  & Post-ban  & Pooled \\
\midrule
\multicolumn{4}{l}{(A) Within-nest share fit \eqref{eq:share_pred},
                        non-circular prediction} \\
$\mathrm{Corr}(\hat s_{i|jt},\,s_{i|jt})$
                                      & $0.73$ & $0.86$ & $0.78$ \\
\multicolumn{4}{l}{(C) Tier-$2$ holdout: predict on
                        held-out tier-$2$ weeks} \\
$\mathrm{Corr}$ (tier-$2$ holdout)    & \multicolumn{3}{c}{$0.85$} \\
\bottomrule
\end{tabular}
\par\smallskip\flushleft
{\footnotesize \textbf{Note}: Diagnostic~(A) reports the correlation between predicted and observed
within-nest shares from \eqref{eq:share_pred} on the pooled chain-drug-week panel; diagnostic~(B)
(category ranking) is in Online Appendix Table~\ref{tab:elast_below_cost}; diagnostic~(C) reports the held-out
tier-$2$ correlation.}
\end{table}

A regime decomposition (Online Appendix~\ref{app:demand_fit_decomp}) shows the residual fit gap is
the static logit's week-by-week below-cost selection, immaterial for the tier-level welfare
calculation of \S\ref{sec:welfare}, which uses tier-level prices where shares are stable.

% =====================================================================
%  End of Demand section
% =====================================================================

% =====================================================================
%  Section: Structural Mechanisms (Revision 2)
%  Intended location: after Section 5 (Demand) and before Section 7
%  (Welfare).  Replaces the v1 ``Trust Dynamics + Dynamic Pricing
%  Model'' material with a leaner specification: one free parameter,
%  three calibrated channels (spillover, common belief, threshold),
%  each anchored to an external data source.
%
%  Maps to memo: memo/revision/2026-06-10_belief_threshold_model.pdf
%  Code:        code/06_dynamic_v2/{simple_model,belief_model,ablation}.py
% =====================================================================

\section{Dynamic Structural Model}\label{sec:mechanisms}

This section embeds the store-traffic value and the followers' belief in a
dynamic pricing game and asks three questions.  The game is non-stationary.  The
advertising ban, the February summer holiday, and a finite detection horizon all
fall inside the sample window, so it is solved by backward induction over calendar
weeks rather than as a stationary recursion.  First, is the estimated collapse
in price sensitivity at the advertising ban (by itself, with no change in
costs, patience, or any other fundamental) enough to flip the static
incentive from the loss-leader war to self-enforcing coordination?  Second, does
the same model, rolled forward, replicate the price war in which coordination
attempts are undercut and fail?  Third, does it then reproduce the post-ban
coordination into the cartel and the subsequent post-cartel rent extraction?  The answer
to all three is yes.  Beyond the estimated demand, the model adds a store-traffic value
$\mu$ and a small set of
organizational, belief, and timing parameters governing how the coordination wave
rolls out.

\subsection{The dynamic game}\label{ssec:mech_setup}\label{ssec:mech_estim}

Coordination resolves on two nested timing layers, the structure the case record forces.  A laboratory engages at its own rate, and within an engaged laboratory each drug's lift is then decided.  Larger laboratories engage first, so the documented order in which they coordinate is an outcome of the model, not an assumption.

The dynamic game has five pieces: the state (a); the per-period
payoff and the static-Nash incentive flip that drives the war-to-cartel transition
(b); the leader probability (c); the follower probability (d); and the
belief through which the dated events enter (e).

\textbf{Solution concept.}  The chains play a Markov equilibrium in subjective beliefs, not a
rational-expectations MPE.  Given the market-wide belief $b_t$ that a coordinated lift will be
matched, each chain best-responds: it forms the forward-simulated continuation values
\eqref{eq:mech_V} and chooses among lift, hold, undercut, and deepen.  The belief is not imposed to
equal the equilibrium match probability.  It is a Beta--Binomial posterior updated from the matched
and reverted lifts the chains actually observe \eqref{eq:mech_belief}.  Along the path it therefore
stays consistent with realized play, in the spirit of experience-based equilibrium
\citep{fershtman2012dynamic} and of firms learning toward coordination in a young market
\citep{doraszelski2018just}.   The
war is a low-belief trap in which no chain expects a lift to be followed, and the transition is a
public signal that jumps the belief; a rational-expectations MPE assumes away exactly those dynamics.
Given the belief path, each chain's value recursion \eqref{eq:mech_V} is a contraction (bounded
payoffs, $\delta<1$), so values and choices are well-defined.  The week-by-week roll-forward then
updates the belief from the play it induces.

\textbf{(a) State.}  The model is built drug by drug, with heterogeneous products.  Each of the $J=220$ drugs is sold by the
three chains (each with its own demand quality $\hat\phi_{ij}$, heterogeneous across drugs), belongs to a
laboratory $L(j)$, and carries its own war, restored, and rent prices
$(p^0_j,p^1_j,p^2_j)$, cost $c_j$, and price coefficient $\alpha_j$.  This gives it its
own one-shot undercut gain $G_j(\alpha_t)$.  The three tier prices and the cost are
read from drug $j$'s own entries in the event ledger: $p^0_j$ is the median transacted price across its
below-cost war deviations, $p^1_j$ the median across its first ($0\!\to\!1$)
coordinations, $p^2_j$ the median across its rent ($1\!\to\!2$) coordinations
(imputed as $1.12\,p^1_j$ for drugs that never reach the rent tier), and $c_j$ the
median implied cost (price over markup) across all of $j$'s events; their
cross-sectional distribution, not a representative drug, enters the estimated
model.  Each lab carries its stake $W_L=\sum_{j:L(j)=L}R_j$ (its
drugs' total revenue).  The weekly state is the tier vector
$\boldsymbol\ell_t\in\{0,1,2\}^{J}$ ($0$ war, $1$ restored, $2$ rent), the
advertising regime $\alpha_t$, and the calendar week.

\textbf{(b) Payoff and the static-Nash flip.}  Drug $j$'s within-period demand is the estimated
nested logit of \S\ref{ssec:demand_model}: chain $i$'s share is
\begin{equation}
s_{ij}(\mathbf p_j;\alpha_j)=\frac{e_{ij}}{D_j}\,\frac{D_j^{\,\omega}}{1+D_j^{\,\omega}},
\qquad e_{ij}=\exp\!\Big(\tfrac{\phi_{ij}-\alpha_j\,p_{ij}}{\omega}\Big),\quad
D_j=\textstyle\sum_i e_{ij},\quad \omega=1-\hat\sigma,
\label{eq:mech_shares}
\end{equation}
with $\hat\sigma=0.393$ and the three chains in one nest.  Each chain carries its own estimated
quality $\hat\phi_{ij}$, so the chains split the nest asymmetrically (Cruz~Verde $0.45$, FASA
$0.31$, Salcobrand $0.24$ at the coordinated price, matching the data); the coordinated-window
intercepts predict the war- and rent-period within-shares out of sample to a median $1.6$ percentage
points (Online Appendix~\ref{app:ij_demand_gof}).  The only object the ban changes is the price
coefficient: each drug keeps its own pre-ban elasticity $\hat\alpha_j$, and the ban scales every one
of them down by the same pooled factor $\alpha^{\rm post}/\alpha_0=0.029/0.103$.  The per-drug post-ban interactions are individually too noisy to identify, whereas the
pre-ban $\hat\alpha_j$ that carry the cross-drug heterogeneity are cleanly estimated.  The
comparative-price ads that told shoppers who was cheapest are gone, so shares stop responding to
price.  Leadership, by contrast, is carried by the loss-cushion asymmetry $\psi_i$, so the
least insulated chain, Salcobrand, leads the escape (\S(c), \S\ref{ssec:mech_robust}).

Beyond the drug margin, each customer is worth a store-traffic value $\mu$, the chain's variable
profit on the rest of her basket, so the per-period payoff is
\begin{equation}
\Pi_{ij}(\mathbf p_j;\alpha_j)=M_j\,s_{ij}(\mathbf p_j;\alpha_j)\,\big(p_{ij}-c_j+\mu\big),
\label{eq:mech_payoff}
\end{equation}
with $M_j$ the market size and $c_j$ the wholesale cost.  Store traffic is why pricing below cost
($p_{ij}<c_j$) can be rational, the lost drug margin repaid by basket profit on the extra traffic
\citep{lal1994retail,thomassen2017multicategory}; held fixed across the ban, $\mu$ lets the ban's
entire effect run through $\alpha$.

The transition turns on the traffic a deviation wins.\label{ssec:mech_spillover_id}\label{ssec:mech_id}  A chain undercutting the coordinated price $p^{\mathrm c}$ to the war floor $p^{\mathrm w}$ gains share $\Delta s=0.188$ with ads on ($\alpha=0.103$) but only $0.050$ with ads banned ($\alpha=0.029$), a $3.8$-fold collapse.  The cut pays only if the store-traffic profit on the stolen share clears the drug margin it gives up:
\begin{equation}
G(\alpha)=\underbrace{\mu\,\Delta s}_{\text{traffic profit}}
+\underbrace{\big[s^{\mathrm{dev}}(p^{\mathrm w}\!-\!c)-s^{\mathrm{coord}}(p^{\mathrm c}\!-\!c)\big]}_{\text{drug-margin change}\;<\,0}\;>\;0
\quad\Longleftrightarrow\quad
\mu>\mu^{*}(\alpha)=\frac{-(\text{drug-margin change})}{\Delta s}.
\label{eq:mech_G}
\end{equation}
Because $\Delta s$ collapses $3.8$-fold while the margin barely moves, the threshold jumps about $3.5$-fold, from $\mu^{*}_{\text{ads on}}=8.6$ to $\mu^{*}_{\text{ads banned}}=30.3$.  The estimate $\hat\mu=9.26$ (thousands of CLP, anticipating \S\ref{ssec:mech_estimation}) lies between the two thresholds, $8.6<9.26<30.3$: above the ads-on threshold, so the below-cost cut paid while ads ran, and below the ads-banned threshold, so the same cut stops paying once they are banned.  The follower faces the mirror choice, to match the lift or to harvest by staying cheap while the leader is expensive, and matching always wins.  Operating through demand alone, the ban thus flips the deviation incentive ($G<0$) into a one-shot best response for the inelastic majority ($168$ of $220$ drugs), with no appeal to patience or trigger strategies.  The most-elastic $52$ are held by the common-knowledge belief.  (I test the observed deviations, not a global static best response.)

\textbf{From daily demand to a weekly payoff.}  Demand is estimated on the
daily panel (\S\ref{sec:demand_estimation}), but the coordination game steps
weekly.  Chain $i$'s weekly payoff on drug $j$ at tier $\ell$ therefore scales the daily
flow profit \eqref{eq:mech_payoff} up to a week,
\begin{equation}
\Pi^{\ell}_{ij}=7\,M^{\rm day}_j\,s^{\ell}_{ij}\,\big(p^{\ell}_j-c_j+\mu\big)
=N_j\,s^{\ell}_{ij}\,\big(p^{\ell}_j-c_j+\mu\big),
\label{eq:mech_weekly}
\end{equation}
with $M^{\rm day}_j$ the daily market size and $s^{\ell}_{ij}=s_{ij}(\mathbf
p^{\ell};\alpha_{j,t})$ chain $i$'s estimated tier-$\ell$ share under the prevailing
regime, and $N_j\equiv 7M^{\rm day}_j$ the weekly-equivalent market.

\textbf{(c)--(d) Value, leadership, and the followers' constraint.}  The value function carries the
equilibrium tier transitions that are the substance of the model: a coordinated price is
undercut back into the war, the war is escaped, and a restored price is deepened into the
rent.  I compute the forward-simulated continuation
value, discounting the expected next-period value under the model's own one-step transition
probabilities $P^{\,x\to x'}_j(t)$ (lift, hold, undercut, deepen, revert) at the realized state,
\begin{equation}
V^{x}_j(t)=\Pi^{x}_j(\alpha_t)+\delta\sum_{x'}P^{\,x\to x'}_j(t)\,V^{x'}_j(t+1),
\qquad x\in\{0,1,2\},
\label{eq:mech_V}
\end{equation}
closed by the stationary post-ban perpetuity ($\delta=0.95$).  These transition
probabilities are not free parameters; they are the model's own choices, defined
below: $P^{\,0\to1}$ is the lift $p^{\uparrow}_{ij}$ \eqref{eq:mech_ccp},
$P^{\,1\to2}$ the rent push \eqref{eq:mech_rent_push}, the reversions $1\!\to\!0$
and $2\!\to\!1$ the undercut $p^{\downarrow}_{ij}$~\eqref{eq:mech_cut}, and $P^{\,x\to x}$
the complement.  Value and choice are solved jointly to the fixed point: the value
sets the escape gains $\Delta V$, and $\Delta V$ in turn sets the choices that
generate $P$.  Write $\Delta V^{1}_j=V^{1}_j-V^{0}_j$ for the escape value and
$\Delta V^{2}_j=V^{2}_j-V^{1}_j$ for the rent value; firms escaping the war do not
foresee the rent, so $\Delta V^{1}_j$ excludes it.

% Leadership is not a separate primitive for each tier.  One rule sets who moves
% first, and it switches on the sign of a chain's flow profit.  
Whenever a chain is below cost, in a loss state, the loss cushion $\psi_i$ governs.  The least
insulated chain bears the most loss, so it is the most desperate to act.  I write
$\psi_i\in[0,1]$ for chain $i$'s loss cushion, calibrated outside the model from
its business mix.  Farmacias Ahumada, with half its revenue and two thirds of its
cash flow in Mexico, is the most insulated ($\psi_{\rm FA}\approx0.8$).  Cruz~Verde,
cushioned by its diversified parent, is middling ($\psi_{\rm CV}\approx0.6$).
Salcobrand, purely domestic and drug-focused, is fully exposed
($\psi_{\rm SB}\approx0$).  The outside business cushions a fraction $\psi_i$ of the below-cost war loss,
so $\psi$ enters through the payoff: chain $i$'s war-state flow profit is
\begin{equation}
\Pi^{0}_{ij}(\alpha_t)=\psi_i\,\Pi^{1}_j(\alpha_t)+(1-\psi_i)\,\Pi^{0}_j(\alpha_t),
\label{eq:mech_psi}
\end{equation}
a convex combination that leaves a fully-exposed chain ($\psi_i=0$) bearing the entire loss $\Pi^0_j$ and
moves a fully-insulated chain's war payoff toward the coordinated level $\Pi^1_j$.  Carried through the
value recursion \eqref{eq:mech_V}, the escape value inherits the exposure,
$\Delta V^1_{ij}=V^1_j-V^0_{ij}=(1-\psi_i)\,\Delta V^1_j$.  Under Type-I extreme-value shocks of scale
$\tau$, chain $i$ then leads a move out of a loss state with probability
\begin{equation}
\pi_{ij}(t)=\frac{\exp\!\big(\Delta V^{1}_{ij}(t)/\tau\big)}
{\sum_k\exp\!\big(\Delta V^{1}_{kj}(t)/\tau\big)}.
\label{eq:mech_eta}
\end{equation}
The same $\Delta V^1_{ij}$ enters the follower's incentive constraint below, so $\psi$ is not confined to
the leader.  But the escape value swamps the one-shot undercut gain by an order of magnitude
(Figure~\ref{fig:mech}), so even the most-insulated chain's constraint stays slack: every chain still
holds, and $\psi$ moves who initiates without moving who follows.  The leadership is thus
carried by the loss cushion and the coordination count by the symmetric store traffic.

Two events are loss states, and Salcobrand leads both.  The below-cost war is one,
so Salcobrand leads the escape, the ordering the business-mix $\psi$ predicts.  In the data Salcobrand initiates $74\%$ of cartel-wave increases, which the model reproduces ($69\%$).  A failed
rent push is the other, because a push the followers reject reverts the drug below
cost, and Salcobrand leads those failures too, four-fifths of the post-ban failed
attempts.  Once a price holds above cost the loss is gone and $\psi_i$ no longer
functions.  There the leader is the largest chain, Cruz~Verde, selected by market
share, the $\mathrm{SB}\!\to\!\mathrm{CV}$ handoff that holds the durable rent.

A lift survives only if the followers match and no rival then undercuts.  A
follower's belief $b_t$ is exactly that probability, so it holds rather than
undercuts iff $b_t\,\delta\Delta V^{1}_j>G^{01}_j(\alpha_t)$, where $G^{01}_j$ is
the static gain to undercutting a coordinated price \eqref{eq:mech_G}, positive
while ads run and negative once the ban flips it.  Because each of the two
followers must expect the other to match, the belief enters squared, and a lift
survives with conditional probability
\begin{equation}
p^{\uparrow}_{ij}(t)=\pi_{ij}(t)\;b_t^{2}\;
\operatorname{logit}^{-1}\!\Big(\big[\,b_t\,\delta\Delta V^{1}_j-G^{01}_j(\alpha_t)\,\big]/\tau\Big).
\label{eq:mech_ccp}
\end{equation}
The mirror down-move undercuts a held coordination back into the war.  The slack $s_j=b_t\,\delta\Delta V^{1}_j-G^{01}_j$ is the belief-weighted continuation value the chain forgoes minus the one-shot undercut gain (positive when the chain prefers to hold, negative when undercutting pays), and the most-tempted chain undercuts in week $t$ with probability
\begin{equation}
p^{\downarrow}_{ij}(t)=\lambda_{\rm cut}\,\operatorname{logit}^{-1}\!\big(-s_j/\tau\big),
\label{eq:mech_cut}
\end{equation}
the mirror of the lift~\eqref{eq:mech_ccp}.  Whether a coordinated price is undercut is decided by the slack, and an undercut returns the drug to the war floor ($1\!\to\!0$), reversing the coordination the lift built.  When undercutting pays ($s_j<0$, the gain exceeds the continuation) the probability rises toward $\lambda_{\rm cut}$; when it does not ($s_j>0$) it falls toward zero.  A rent reverts $2\!\to\!1$ by the same rule on its own slack.  The cartel's stability is
therefore an outcome of the ban flipping $G^{01}_j$ negative, not an
assumed no-defection: defection is available every period but unprofitable.  The rate
$\lambda_{\rm cut}$ is itself a free scalar, estimated with the others: the model micro-founds
when a profitable undercut is available (the slack gate), not its baseline intensity.  The deterrence is the slack gate, not the fitted rate: perturbing $\lambda_{\rm cut}$ by $\pm50\%$ leaves the coordination count essentially unchanged and only rescales the war's deviation intensity (Online Appendix~\ref{app:lamcut_robust}).

\textbf{Relation to punishment-supported collusion.}  The slack gate is the no-deviation incentive constraint of repeated-game collusion, written for this setting.  In the textbook account a chain undercuts when the one-shot gain exceeds the discounted punishment it would trigger.  Here it undercuts when the gain $G$ exceeds the belief-weighted continuation value $b_t\,\delta\Delta V$ it forgoes.  The penalty a deviator anticipates is thus endogenous and mild: it drops one tier and re-enters the estimated transition, forgoing that tier's continuation value.  That continuation is read from the forward-simulated equilibrium \citep{igami2022measuring}, not an assumed trigger path.  The two tiers map differently onto the tradition.  For the coordinated price the ban makes $G^{01}_j<0$, so holding it is a one-shot best response and no penalty is needed: this is the static-Nash base, outside the repeated-game frame.  The supra-competitive rent is where the traditional incentive lives, since there $G^{12}_j>0$ and the rent sticks only while $b^{\rm rent}_t\,\delta\Delta V^{2}_j>G^{12}_j$, the forgone rent value clearing the undercut gain.  What the model leaves out is an off-equilibrium punishment phase, in which rivals revert to a war to discipline a deviator.  It leaves this out by design: the coordinated tier needs no punishment, and the rent is disciplined by its own forgone value.  In the data the post-cartel defections are sparse and isolated, an enforcement-era retreat rather than the trigger of a retaliatory war (\S\ref{sec:descriptive}).  This rent discipline is weak by construction, and accordingly the model captures the rent's incentive and Cruz~Verde leadership but not the synchronized late-$2008$ surge, which a smooth-arrival model cannot generate (\S\ref{ssec:mech_robust}).

\textbf{Timing within the week.}  The moves resolve in a fixed order.  A leader is drawn for each not-yet-coordinated drug \eqref{eq:mech_eta}, the two followers simultaneously decide whether to match, and the lift succeeds only if both do \eqref{eq:mech_ccp}.  The most-tempted chain may then undercut a held coordination at rate $\lambda_{\rm cut}$, and the belief $b_t$ updates from the week's matched and reverted lifts \eqref{eq:mech_belief} before the next week begins.

\textbf{(e) Belief and the dated events.}  Coordination is dynamically sustainable in both regimes
(\S\ref{ssec:mech_robust}).  What holds the chains in the war is not that the coordinated price is
unsustainable but that no one expects it to be reached, so the transition selects the
coordinated equilibrium rather than creating it.  The belief is a single, market-wide Beta--Binomial
posterior over a shared prior $(a_0,b_0)$, updated by Bayesian learning from the running tally of matched
($S_t$) and reverted ($F_t$) lifts pooled across all $220$ drugs, so a coordination on one drug raises
the same belief that gates every other (the cross-product channel that carries the wave).  The
\textbf{binding November ban} (week~$t_{\rm ban}=96$) is commonly observed, so every firm performs the same update,
adding $\zeta\approx307$ pseudo-successes at the
ban; the rent step reads the same belief under a fragility discount $e^{-\chi}\approx0.12$.
The three are one object,
\begin{equation}
b_t=\frac{a_0+S_t+\zeta\,\mathbf 1\{t\ge t_{\rm ban}\}}{a_0+b_0+S_t+F_t+\zeta\,\mathbf 1\{t\ge t_{\rm ban}\}},
\qquad b^{\rm rent}_t=e^{-\chi}\,b_t,
\label{eq:mech_belief}
\end{equation}
the war posterior before the ban ($t<t_{\rm ban}$), the coordination posterior after it (a shift, not a
reset, since the war's failures stay in $F_t$), and the discounted rent posterior.  The estimated
prior is low ($a_0/(a_0{+}b_0)\approx0.01$): almost no one expects a lift to be matched, the few
tried are undercut, and the pessimism is self-fulfilling.  The ban lifts the belief in one step to
about one-half, past the squared-belief ignition threshold; it then climbs only to about
seven-tenths over the wave and never near one, so the post-ban dynamics are governed by the
one-time incentive flip rather than by belief continuing to rise.  The belief enters the lift
\eqref{eq:mech_ccp} squared (both followers must expect a follow) and the rent once (the rent
only deepens a coordination the chains already hold).  The \textbf{September CONAR ruling} is a
non-binding precursor; the \textbf{December onset $t_0\approx103$} is endogenous, the lag the
laboratory reach-out generates; the \textbf{March FNE Oficio} raises enforcement on the rent alone
(below).  The rent hazard thus tracks the wave at a lower level and opens slowly.  A rent-eligible drug (one whose
own demand is inelastic enough that the higher price still pays, $\Delta V^2_j>0$) is pushed
$1\!\to\!2$ in week $t$ with probability
\begin{equation}
p^{\,2}_j(t)=b^{\rm rent}_t\,r_t,
\label{eq:mech_rent_push}
\end{equation}
the discounted rent belief (entering once, not squared: the rent only deepens a coordination the
chains already hold jointly) times a learned enforcement belief
$r_t$.\footnote{$r_t$ is the chains' posterior that a rent push is safe from the regulator, not a
calendar trend.  After the Oficio they do not know whether the FNE will sanction the rent, and they
treat each week survived without a sanction as evidence the threat is receding.  This is a
Beta-Binomial survival posterior $r_t=(1+k_t)/(1+k_t+m)$, where $k_t=t-t_{\rm O}$ is the weeks elapsed
since the Oficio and $m$ is the skepticism the notice instils.  It starts low, $r_{t_{\rm O}}=1/(1+m)$,
and recovers as quiet weeks accumulate.  No success term enters, so the belief cannot cascade.}
The Oficio sets off a contested retreat in the rent.  Right after the notice $r_t$ is low, so most rent
pushes are rejected and revert, and these failures persist rather than completing the rent.  The chains
also suspend new pushes for the $\varphi\approx9$-week peak-scrutiny pause, holding the rent without
risking the coordination tally.  As quiet weeks accumulate $r_t$ recovers, the surviving pushes begin to
stick, and the rent climbs to its late peak.\footnote{A one-time negative belief shock at the Oficio,
the natural alternative, does not reproduce this: it depresses the rent hazard uniformly, so the
rent flattens below the observed plateau, never accelerates, and generates no persistent
failures; I verified this directly.  The learned belief instead suppresses early matches, sustains the
contested failures, and recovers only as the cartel survives, producing the late climb.}  The pause
window $\varphi\approx9$ weeks is read from the institutional timeline (the Oficio, $31$~March, plus the
chains' observation lag), not estimated; the skepticism $m$ is the one estimated enforcement parameter
($t_{\rm O}$ a fixed date).

Two things gate the rent, and both are demand- rather than belief-driven: the eligibility
$\Delta V^2_j>0$, which only the post-ban inelastic demand makes true, and the learned enforcement
belief $r_t$.  The rent's timing is therefore orthogonal to the coordination belief that drives
the $0\!\to\!1$ wave.  Forcing that belief to one floods the wave but leaves the rent untouched before
the ban: a further increase never pays while demand is still elastic, so no drug deepens until the ban
makes demand inelastic, regardless of what the chains believe (\S\ref{ssec:mech_robust}).  The rent push
\eqref{eq:mech_rent_push} therefore carries no static-best-response gate of its own; the eligibility
already screens out drugs that would statically defect.

The same primitive \eqref{eq:mech_ccp} runs the pre-ban war: while ads are on $G^{01}_j>0$, so any
coordinated lift is undercut and reverts.  Before the laboratories organize ($t<t_0$) each
not-yet-coordinated drug is tested directly at the follower belief itself (the leader raises only if
it expects a follow),
\begin{equation}
\Pr\{\,\text{drug }j\text{ tested for a lift in week }t\,\}\;=\;b_t\,\big[1+g\,\mathbf 1\{t\ge\text{Aug }2007\}\big],\qquad t<t_0,
\label{eq:mech_war}
\end{equation}
the most war-exposed chain (Salcobrand, lowest $\psi_i$) leading and Cruz~Verde's August Desaf\'io
campaign intensifying by $g$; almost every test is undercut, the reverts concentrating where
$G^{01}_j$ is largest, until the November ban flips $G^{01}_j<0$ and the war collapses.

\textbf{Two-layer organization.}  The cartel was organized lab-by-lab through the manufacturers, the
intermediaries the prosecutor identifies.\footnote{``utilizaron a los laboratorios para coordinar y
monitorear el acuerdo'' \citep{fne2008requerimiento}.}  Each week a chain reaches out to a
not-yet-organized lab $L$ when the coordination it would unlock is worth more than the traffic it
would forgo,
\begin{equation}
\Pr\{\,\text{chain }i\text{ organizes lab }L\text{ in week }t\,\}
=\lambda\,\Big[\tfrac{1}{\tau_e}\big(b_t\,\delta\,\Delta V^{1}_{iL}(t)-G^{01}_L\big)\Big]_{0}^{1}\,\mathrm{hol}(t),
\label{eq:mech_hazard}
\end{equation}
summing the escape value and undercut gain over $L$'s drugs.  Two ordering facts emerge from
this reach-out rather than being imposed: larger labs (a larger summed surplus) cross the threshold
first, so the documented larger-labs-first sequence is an outcome; and the most war-exposed chain,
Salcobrand, with the largest escape value reaches out first and leads.
Engagement is one-time and bounded (active for a spell of $w$ weeks, then disengaged); conditional on
it, each drug is pushed up at its own belief-gated rate \eqref{eq:mech_ccp}, drugs coordinating at
staggered weeks, and a coordinated drug is not profitably undercut once the ban flips $G^{01}_j$ negative, so the cartel once formed holds.

\subsection{Estimation and fit}\label{ssec:mech_estimation}\label{ssec:mech_fit}

\textbf{The parameters.}  The store-traffic value enters as a single common $\mu$,
and the chain asymmetry that selects the leaders is not estimated.  In a loss state
the leader is set by the loss cushion $\psi_i$, calibrated above from each chain's
business mix.  In a profit state it is set by observed market share.  The leadership
ordering is thus predicted, not fitted, and the two dispersion scales that turn each
ordering into a leader share are calibrated to the leader shares rather than
estimated.  The dynamic parameters are the common store-traffic
value $\mu$ and the reach-out scale and temperature $(\lambda,\tau_e)$ governing
laboratory engagement \eqref{eq:mech_hazard} and its onset $t_0$.  The belief block adds the shared
belief prior $(a_0,b_0)$ together with the ad-ban pseudo-successes $\zeta$ and the rent-tier
signal penalty $\chi$ that together micro-found the two belief shifters.  The remaining parameters are the
post-Oficio enforcement skepticism $m$ that the learned rent-match belief works
off, the symmetric down-cut rate $\lambda_{\rm cut}$, and the campaign strength $g$.  Fixed are the weekly discount $\delta=0.95$, the
Type-I EV choice scale $\tau$, the holiday factor $\mathrm{hol}$, and the engagement window
$w=26$ weeks, calibrated to the cartel's roughly six-month documented active span rather than
estimated.  The
per-drug objects that drive the dynamics (each drug's undercut and rent gains
$G^{01}_j,G^{12}_j$ and shares, computed once from its own $\alpha_j$,
prices, and cost) are inputs, not estimated here.

\textbf{Forward simulation.}  For a candidate $\theta$ I first build the per-drug
value functions from the rolling-horizon recursion \eqref{eq:mech_V}, then roll the weekly state
$\boldsymbol\ell_t$ forward from January~$2007$ (all drugs in the war,
$\boldsymbol\ell_0\equiv\mathbf 0$) over the $105$-week window through December~$2008$.  Week numbers are inherited from the full panel beginning January~$2006$, so January~$2007$ is week~$52$, the window ends at week~$156$, the binding
November ban falls at week~$96$, and the coordination onset $t_0\approx103$ in late
December~$2007$, about seven weeks later.  Each week resolves in order: while $t<96$ the
below-cost war cheating \eqref{eq:mech_war}, its failed attempts the only events; the ban at
week~$96$ both collapses price sensitivity and, as a public event, jumps the coordination belief;
and from the endogenous onset $t_0\approx103$ on (once the laboratories have organized,
the lag the reach-out generates) laboratory
engagement \eqref{eq:mech_hazard}, the belief-gated $0\!\to\!1$ coordinate step (which
succeeds only if both followers match and every chain holds its incentive constraint, else a
failed attempt that reverts),
and the $1\!\to\!2$ rent step for the drugs whose own demand makes it pay (post-Oficio, an
unmatched rent push reverts, a coordinated drug otherwise rarely defecting).  $B$ replications under common random numbers are
averaged into the moment vector $\widehat m(\theta)$,\footnote{The criterion stacks $38$ moments against the $11$ estimated parameters, so the model is over-identified by $27$ restrictions; the moments are grouped in Online Appendix Table~\ref{tab:mech_moments}.} and I minimize the weighted
distance to the data moments,
\begin{equation}
  \hat\theta \;=\; \arg\min_{\theta}\ \sum_{k}
  \Big(\frac{\widehat m_k(\theta)-m^{\rm data}_k}{w_k}\Big)^{2},
  \label{eq:smm}
\end{equation}
\citep{mcfadden1989smm,pakespollard1989simulation,bajari2007estimating} by a
multi-start derivative-free search (Powell from several random starts), chosen
because the objective is a step function of the discrete tier transitions; search
and simulation use fixed seeds, so $\hat\theta$ is reproducible.  The hyperparameters I set
rather than estimate are the replication count $B$, the moment weights $w_k$, the
number of starts and the per-start iteration budget, the random seeds, and the
three fixed values $(\delta,\tau,\mathrm{hol})$.  The weights are not all neutral: the $0\!\to\!1$
wave-path checkpoints are deliberately down-weighted so that, being largely redundant with the
onset/peak/spread moments, they do not crowd out the leadership fit.

\textbf{Standard errors.}  The step-function criterion has no well-defined moment Jacobian, so the standard sandwich variance does not apply.  I instead report observed-information standard errors.  I profile the weighted criterion in each parameter at the estimate with the others fixed, and read each standard error off the local curvature as $\mathrm{SE}=\sqrt{2/H}$, with $H$ the fitted second difference (Table~\ref{tab:mech_theta}).

\textbf{Identification.}  Each parameter maps to a distinct moment (Online Appendix
Table~\ref{tab:mech_moments}).  Two cross-sectional moments discipline the micro-foundations.
The who-cheats-most moment, the rank correlation between a drug's pre-ban cheat frequency and its undercut
gain $G^{01}_j$, identifies the benefit-ordered war \eqref{eq:mech_war}.  The who-takes-rent moment, the
$\alpha_j$ gap between drugs that reach the rent tier and those that stop, identifies the endogenous
$1\!\to\!2$ selectivity.  The store-traffic mean $\mu$ scales every payoff and rests on no single
moment, though the below-cost war floor is the most informative and bounds it from below\footnote{The complement basket whose profit $\mu$ captures is not in the case record, so $\mu$ is identified from the chains' revealed willingness to price below cost (the war floor), not measured from non-drug sales; its magnitude is that revealed-preference inference, an identifying assumption on the spillover rather than a direct measurement.}; the belief prior $(a_0,b_0)$ and the ad-ban signal $\zeta$ are read from the regime split of
the failed attempts, $\chi$ from the smaller and later $1\!\to\!2$ count, $m$ from the
post-cartel failure rebound, $(\lambda,\tau_e,t_0)$ from the onset, peak and lab-size$\times$timing
gradient, and $g$ from the August intensification.

% [Table ``Targeted moments and what they identify'' (tab:mech_moments) moved to
%  Appendix~\ref{app:structural_sample}; referenced from the main text here.]

\textbf{Estimates.}  The eleven estimated parameters are in Table~\ref{tab:mech_theta} (Panel~A); the
table carries the cell-by-cell fit, and I read off only the interpretive content.  The store-traffic
value is now a single common $\hat\mu=9.26$, sitting just inside the ads-on threshold above which
undercutting pays; the chain asymmetry has moved out of $\mu$ and into the calibrated loss cushion
$\psi_i$ and market share.  Cruz~Verde's category financials support the loss-leader economics:
it sold chronic drugs below cost (chronic-drug profit $-28$~M~CLP/month) while harvesting acute
($+2{,}299$) and non-pharma ($+1{,}088$~M~CLP/month) sales off the resulting traffic, a basket many
times the drug loss; I do not use these aggregates to estimate $\mu$, because they mix drug-driven and
autonomous purchases.  The data contain no basket-level complement sales, so $\mu$ is identified only
from the chains' own below-cost pricing; its level is bounded by that revealed behavior rather than
measured against observed cross-category margins.

The model coordinates $216$ of the $220$ drugs and reaches $141$ of the $143$ rent steps: the $168$ inelastic drugs hold on their own because undercutting is unprofitable, and the focal belief shift carries another $48$ of the most-elastic $52$.  Coordination
on a drug holds only if it is incentive-compatible for every chain, so the binding constraint is
whichever chain is most tempted to undercut (the smallest per-chain slack).  A drug fails to coordinate
when even one chain's undercut gain $G^{01}_{ij}$ stays positive.  That is why the binding object is the
product-firm incentive compatibility of \eqref{eq:mech_shares}, not a single product-level constraint.
All three leadership moments
land in place.  In the loss states $\psi_i$ governs: Salcobrand leads $69\%$ of the cartel wave (data
$74\%$) and $91\%$ of the post-early failed attempts (data $83\%$).  In the profit state market share
governs: Cruz~Verde leads $68\%$ of the post-investigation rent (data $65\%$), the SB$\to$CV handoff
(Table~\ref{tab:mech_fit}).  These three leader shares are the targeted moments; the calibrated
primitives pin each phase's leader but not the full ordering, so the runner-up positions are
noisier (in the cartel wave, for instance, the model places Cruz~Verde second where the data place
FASA).  The whole fit comes at an SMM loss of $16.59$.

\begin{table}[H]
\centering\footnotesize\renewcommand{\arraystretch}{0.92}
\caption{Estimated parameters and model fit}
\label{tab:mech_theta}\label{tab:mech_fit}

\begin{tabularx}{0.9\linewidth}{@{}>{\raggedright\arraybackslash}X r r@{}}
\toprule
\multicolumn{3}{@{}l}{\textbf{Panel A. Estimated parameters (SMM, eleven parameters)}}\\[2pt]
Parameter & Est. & (SE) \\
\midrule
$\mu$\quad store-traffic value (common)             & $9.26$      & $(0.15)$ \\
$\lambda$\quad reach-out scale                      & $0.63$      & $(0.18)$ \\
$\tau_e$\quad reach-out temperature                 & $124$       & $(37)$ \\
$t_0$\quad organization onset (week)                & $103$       & $(1.2)$ \\
$a_0,b_0$\quad shared belief prior                  & $1.8,\,129$ & $(0.32,\,61)$ \\
$\zeta$\quad ad-ban pseudo-successes                & $307$       & $(41)$ \\
$\chi$\quad rent-tier belief penalty ($e^{-\chi}\!\approx\!0.12$)   & $2.08$      & $(0.10)$ \\
$g$\quad Aug.\ campaign strength                    & $0.57$      & $(0.19)$ \\
$\lambda_{\rm cut}$\quad symmetric down-cut rate    & $0.088$     & $(0.006)$ \\
$m$\quad post-Oficio enforcement skepticism         & $14.8$      & $(7.1)$ \\
\midrule[\heavyrulewidth]
\multicolumn{3}{@{}l}{\textbf{Panel B. Estimated model versus record}}\\[2pt]
Moment & Model & Data \\
\midrule
First coordinations (Tier $0\!	o\!1$)             & $214$ & $220$ \\
Rent steps (Tier $1\!\to\!2$)                       & $141$ & $143$ \\
Onset / peak week                                  & $103/107$ & $102/107$ \\
Wave spread (SD, weeks)                            & $4.5$ & $4.6$ \\
$1\!\to\!2$ lag behind $0\!\to\!1$ (weeks)         & $8.1$ & $8.0$ \\
\midrule
\textit{Coordination successes, by period} & & \\
\quad war\,/\,cartel                               & $0/289$ & $2/284$ \\
\quad post-early\,/\,post-late                     & $36/34$ & $28/49$ \\
\textit{Failed lifts, by period} & & \\
\quad war\,/\,cartel                               & $110/0$ & $123/4$ \\
\quad post-early\,/\,post-late                     & $46/17$ & $46/21$ \\
\midrule
Below-cost war markup                              & $-0.09$ & $-0.10$ \\
August war intensification ($1\!+\!g$)             & $1.57$ & $1.68$ \\
\textbf{corr(log lab size, coord.\ week)}          & $\mathbf{-0.38}$ & $\mathbf{-0.29}$ \\
Lab-wave SD (weeks)                                & $5.8$ & $4.9$ \\
\midrule
\textit{Leadership (predicted from calibrated primitives), CV$/$FA$/$SB \%} & & \\
\quad Cartel wave (loss state, $\psi$) & $24/7/\mathbf{69}$ & $5/21/\mathbf{74}$ \\
\quad Post-early failures (loss state, $\psi$) & $6/3/\mathbf{91}$ & $9/9/\mathbf{83}$ \\
\quad Post-late rent (profit state, share) & $\mathbf{68}/20/12$ & $\mathbf{65}/12/22$ \\
\bottomrule
\end{tabularx}

\par\smallskip\flushleft
{\footnotesize \textbf{Note}: Panel~A: eleven parameters estimated by SMM (loss $16.6$ at $B=240$; the objective adds weekly rent-path checkpoints disciplining the post-Oficio dynamics).
The store-traffic value $\mu$ and the tier prices $p^{x}$ are in thousands of CLP, so $\hat\mu=9.26$ is about $9{,}300$~CLP, the basket profit a drug customer brings.
The chain-asymmetry primitives are calibrated, not estimated: the loss cushion $\psi=(0.60,0.80,0)$ for (CV,\,FA,\,SB) and the within-market shares $(0.45,0.31,0.24)$, with the two leadership dispersions $\tau_{\rm lead}=3.0,\ \tau_{\rm rent}=0.12$ set to the leader shares; the engagement window $w=26$ weeks, $\delta=0.95$, $\tau=0.13$ and the holiday factor are fixed.  Standard errors are the profile SMM-criterion curvature (observed information), profiling the criterion in each parameter at the estimate with the others fixed.
Panel~B: each drug uses its own prices, cost, $\alpha_j$, lab and lab size.}
\end{table}

\textbf{The belief transient.}  Traced along the simulated path
\eqref{eq:mech_belief}, the follower belief is a fast transient: low
through the war, lifted by the ad-ban signal to about one-half at the jump and climbing to about
seven-tenths over the wave, never near one (the war's failures stay in $F_t$).  Squared, this is
already past ignition ($b_t^2\gtrsim0.36$) once $G_j<0$, so over December--March belief is no
longer the binding margin: the wave is the mechanical consequence of the static flip, not a
gradually-built belief.  Belief is thus decisive for the transition: it sustains the pre-ban war and
times the onset, and neither a no-belief nor a gradual alternative fits
(\S\ref{ssec:mech_robust}).

\begin{figure}[H]
\centering
\caption{Structural specifications versus the data}
\label{fig:price_path}

\includegraphics[width=0.95\textwidth]{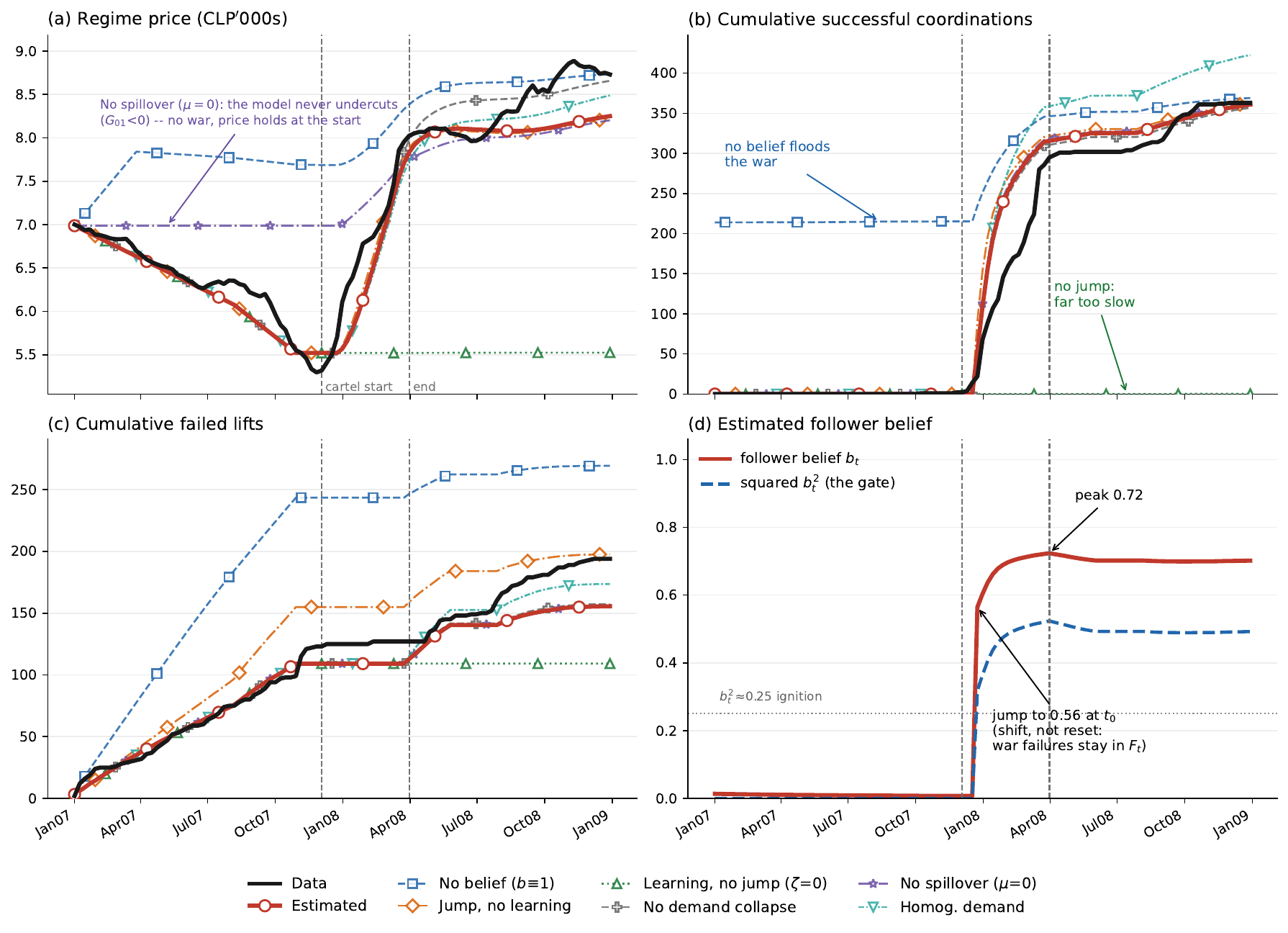}
\par\smallskip\flushleft
{\footnotesize \textbf{Note}: four panels, January~$2007$--December~$2008$.  The data (solid black) are
set against the estimated model and six knockouts of it, each switching off one ingredient:
no belief ($b\!\equiv\!1$); jump, no learning; learning, no jump ($\zeta\!=\!0$);
no demand collapse ($\alpha\!\equiv\!\alpha_0$); no spillover ($\mu\!=\!0$); and a
misspecified homogeneous demand.
(a) regime price, quantity-weighted.
(b) cumulative successful coordinations ($0\!\to\!1$ wave and $1\!\to\!2$ rent).
(c) cumulative failed lifts.
(d) the estimated belief $b_t$ and its square $b_t^2$ (the coordination gate).
Each belief or spillover knockout breaks one panel; the homogeneous spec tracks the counts but
mis-shapes the wave (\S\ref{ssec:mech_robust}, Table~\ref{tab:spec_fit}).}
\end{figure}

\textbf{The post-investigation rent.}  The cartel does not break: the $220$ coordinated drugs hold their
price through the FNE's $31$~March Oficio~$419$ (Figure~\ref{fig:price_path}a), and the learned
enforcement belief of \S\ref{ssec:mech_setup} reproduces the contested recovery: failed lifts rebound
($48$ vs.\ near zero in the cartel) and the rent climbs late to $141$ of $143$ (Table~\ref{tab:mech_fit},
Panel~B, skepticism $m=14.8$).  The fit here is honest but imperfect: a smoothly learned belief
reproduces the direction of the late surge yet undershoots its sharpness (post-late successes
$33$ against $49$), because a smooth posterior cannot fully match a sharp freeze-then-surge.  Because the held rent is a profit state, market share orders it: the
dominant chain, Cruz~Verde, leads the post-investigation rent, the SB$\to$CV handoff.

% [Model-fit table merged into Table~\ref{tab:mech_theta}, Panel B.]

\begin{figure}[H]
\centering
\caption{Coordination selected by the deviation gain}
\label{fig:mech}
\includegraphics[width=0.72\textwidth]{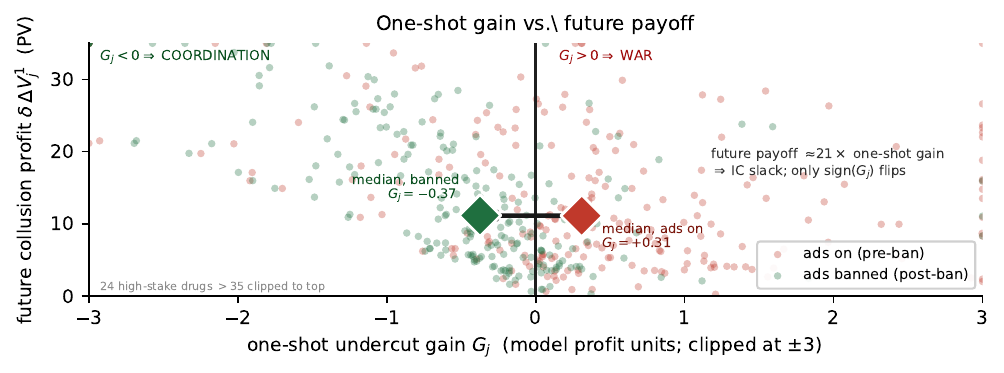}

\par\smallskip\flushleft
{\footnotesize \textbf{Note}: Each drug's one-shot gain to undercutting a coordinated price,
$G_j$ \eqref{eq:mech_G} evaluated on the estimated demand (horizontal), against the forgone
discounted escape value $\delta\,\Delta V^1_j$ \eqref{eq:mech_V} (vertical); red $=$ pre-ban
(ads on), green $=$ post-ban (ads banned), diamonds $=$ regime medians, axes clipped to
$G_j\in[-3,3]$ and $\delta\Delta V^1\le35$ ($22$ high-stake drugs lie above).}
\end{figure}

\begin{figure}[H]
\centering
\caption{Model reproduces the coordination sequence}
\label{fig:mech_seq}
\includegraphics[width=\textwidth]{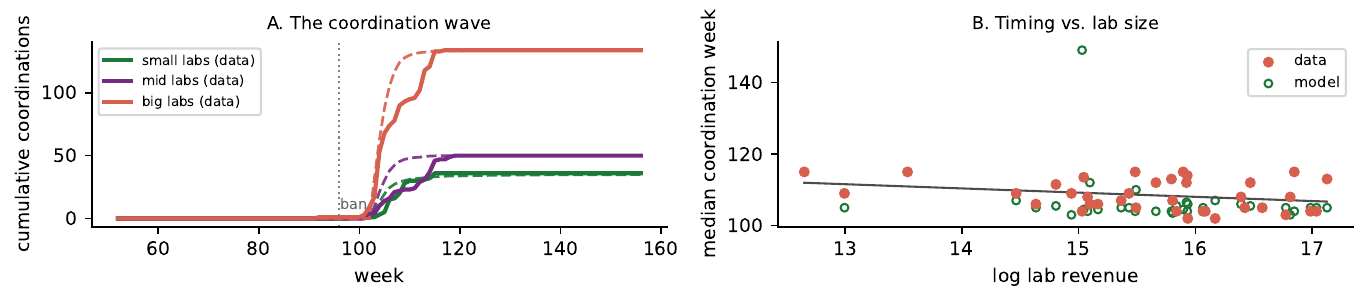}

\par\smallskip\flushleft
{\footnotesize \textbf{Note}: \textbf{Panel A:} cumulative coordinations, model (dashed) vs.\
the coded data (solid), by lab-size tercile. The big- and mid-lab cohorts lead
the small-lab cohort; flat through the pre-ban war, then the post-ban wave.
\textbf{Panel B:} lab-level coordination week vs.\ log lab revenue; larger labs
coordinate first (rank correlation $-0.39$ model vs.\ $-0.29$ data).}
\end{figure}

\textbf{The price path.}  Figure~\ref{fig:price_path}a checks the model against the price series itself,
with one caveat: the regime price levels are calibrated, not predicted.  The coordinated tiers
are the markup-implied prices of \S\ref{ssec:mech_setup}, and the war floor is the pre-ban symmetric
Bertrand--Nash price with store traffic, each chain maximizing $(p-c+\mu)\,s(p;\alpha_{\mathrm{pre}})$.
Because the below-cost war markup is itself a matched moment, the fitted floor reproduces the observed
war price near-mechanically.  Panel~(a)
traces the symmetric regime path against the quantity-weighted data price: the January-$2007$ decline to
the war floor while ads are permitted, then the recovery into the coordinated tiers once the November
ban flips the incentive (the regime price holds at the floor through the war, recovers from week
$\sim\!102$, and climbs to about $9$ by late 2008).  This is a consistency check on the whole trajectory,
joined by the estimated timing, not an independent test of the level.  Panels~(b) and~(c) make it
discriminating.  The cumulative wave and the failed lifts separate the estimated mechanism from the
re-estimated alternatives: the belief and spillover knockouts each break a panel (the war floods, the
wave is far too slow, or coordination collapses), whereas a misspecified homogeneous-demand spec tracks the
aggregate counts but mis-shapes the wave (\S\ref{ssec:mech_robust}).

\begin{table}[htbp]\centering\scriptsize
\caption{Structural specifications versus the data: per-moment fit}
\label{tab:spec_fit}
\centering
\scriptsize\setlength{\tabcolsep}{4pt}%
\begin{tabularx}{\linewidth}{@{}>{\raggedright\arraybackslash}X r r r r r r r@{}}
\toprule
Moment & Data & Estimated & No belief & \makecell{Jump,\\no learning} & \makecell{Learning,\\no jump} & \makecell{No demand\\collapse} & \makecell{No spillover\\($\mu\!=\!0$)} \\
\midrule
First coordinations ($0\!\to\!1$) & 220 & 216 & 218 & 216 & 0 & 191 & 220 \\
Rent steps ($1\!\to\!2$) & 143 & 141 & 150 & 140 & 0 & 132 & 136 \\
Peak week & 107 & 107 & 52 & 107 & --- & 107 & 106 \\
Wave spread (SD, weeks) & 4.6 & 4.5 & 9.9 & 4.8 & --- & 5.8 & 3.7 \\
\textit{Coordination successes} & & & & & & & \\
\quad war & 2 & 0 & \textbf{220} & 0 & 0 & 0 & 0 \\
\quad cartel & 284 & 289 & 137 & 282 & 0 & 274 & 294 \\
Failed lifts, war & 123 & 110 & 385 & 130 & 162 & 132 & 102 \\
War deviations (below-cost) & 516 & 512 & \textbf{22} & 536 & 386 & 509 & \textbf{155} \\
corr(log lab size, week) & $-0.29$ & $-0.38$ & $-0.14$ & $-0.36$ & --- & $-0.36$ & $-0.49$ \\
SB leads the cartel wave & 0.74 & 0.69 & 0.12 & 0.68 & --- & 0.64 & 0.70 \\
\midrule
\textbf{SMM criterion (loss)} & --- & \textbf{16.6} & \textbf{4154} & \textbf{21} & \textbf{2323} & \textbf{28} & \textbf{612} \\
\bottomrule
\end{tabularx}
\par\smallskip\flushleft{\footnotesize \textbf{Note}: per-moment companion to Figure~\ref{fig:price_path}.  Each
column is re-estimated under the identical SMM criterion, so every restricted spec gets its own
best-fit parameters and the loss row is a fair horse race.  ``No demand collapse'' shifts beliefs at the ban but
holds price sensitivity at its pre-ban level (so $G^{01}$ never flips); ``No spillover'' ($\mu\!=\!0$) removes the
store-traffic floor.  Figure~\ref{fig:price_path} additionally plots a misspecified homogeneous-demand spec (Online Appendix Table~\ref{tab:spec_robustness}).}
\end{table}

From one set of primitives the estimated model reproduces the whole episode: the below-cost
war, the post-ban cartel wave, the post-Oficio pause, and the late rent coordination.  The
war's two signatures, the intensification of below-cost cheating and the sporadic,
mostly-failing coordination that preceded the cartel, come out of the same simulation that
produces the wave.  While ads are permitted
$G^{01}_j>0$, so undercutting for store traffic pays: chains cut below cost, and a
coordinated lift, even when matched, is undercut, so it reverts.  The below-cost
war is broad from the January-$2007$ decline onset and intensifies as Cruz
Verde's August-$2007$ ``Desaf\'io'' campaign \eqref{eq:mech_war} escalates the
advertising arms race, peaking in the weeks before the ban; once the ban flips
$G^{01}_j<0$ the cheating unwinds and coordination becomes self-enforcing.  The
weekly below-cost deviation series (its pre-ban level, its August intensification,
and its post-ban collapse) is matched through the war-series and war-floor moments
of Online Appendix Table~\ref{tab:mech_moments}.  Coordination meanwhile is
sporadic and reverts throughout the war (the cumulative coordinated stock stays
near zero until the ban, then rises along the wave) which is just the high
war-window failure rate of Table~\ref{tab:event_window}.  The estimated campaign strength $g=0.57$ reproduces the August
intensification ($1.57\times$ vs.\ the data's $1.68\times$), and the cheating collapses once the ban flips
$G^{01}_j<0$.

The structure of the value functions \eqref{eq:mech_V} sharpens the
interpretation.  In the estimate the forward-looking cooperation premium, the discounted escape
value $\delta\Delta V^1_j$, exceeds the one-shot undercut gain by more than an order of magnitude
at the median drug (Figure~\ref{fig:mech}), so the dynamic incentive constraint is slack for all
$220$.  Coordination is therefore dynamically sustainable in both regimes, even before the ban.
Pre-ban the war is the stage-game Nash ($G>0$, undercutting pays), and pre-ban the coordinated price
is not a static Nash at all (every chain gains by undercutting it), so it could be sustained only as
repeated-game collusion with punishment strategies, which the chains, as the failed
war-period attempts show, did not in fact coordinate on.  The ban
flips the undercut incentive to $G<0$, making the coordinated price self-enforcing for these drugs, and the common-knowledge belief jump of the previous paragraph
shifts expectations onto it; the ban supplies both, and so selects the cooperative equilibrium.  The static
flip holds for $168$ of the $220$ drugs ($76\%$).  The other $52$ (the most elastic, whose undercut gain
$G_{ij}$ stays positive for at least one chain even post-ban) are not a static Nash, and the ban selects them through the belief
shift alone, as dynamically sustainable coordination.  Two consequences follow.  First, for the $168$ static-Nash drugs, because
post-ban coordination is a one-shot best response for every chain it holds at every date including the
terminal one; backward induction yields no finite-horizon unraveling.  Second, their observed
failures to match are not incentive-constraint violations (undercutting is myopically unprofitable
post-ban) but the idiosyncratic Type-I EV shocks of \eqref{eq:mech_ccp}, infrequent and
dispersed across all three chains; the $52$ non-static-Nash drugs, where undercutting still
pays for some chain, supply the rest.  That the model's static-Nash selection flags $52$ non-static-Nash
drugs whereas only a score fail to coordinate, and pure dynamic enforcement would flag none,
places the truth between the two (partial enforcement), and is why I read that
count as supporting, not identifying.

\subsection{Identification and robustness}\label{ssec:mech_robust}

The structural exercise makes one point: the advertising ban turns the loss-leader war into self-enforcing coordination through two estimated changes together, the collapse in price sensitivity that makes undercutting unprofitable for the inelastic majority and the focal belief shift that releases the wave, with no appeal to patience or trigger strategies.  A set of re-estimations shows it does not rest on a tuned parameter.  Each check
re-estimates the model with one ingredient switched off, plotted against the
data in Figure~\ref{fig:price_path}.  Table~\ref{tab:spec_fit} gives the per-moment fit.  Each column is
re-estimated under the identical SMM criterion, so the loss row (headline $16.6$ against $21$--$4154$ for
the broken specifications) is a direct horse race.
Online Appendix Table~\ref{tab:struct_robust} adds the value-function checks.

\textbf{Robustness checks.}  Belief is essential and dated to the binding November ban: a no-belief
model floods the war ($220$ pre-ban coordinations against the data's $2$, loss $4154$).  Without the
focal jump the wave never ignites inside the sample: a learning-only belief ($\zeta=0$)
coordinates essentially no drugs and fits orders of magnitude worse ($2323$ against the headline $16.6$).
A jump-only belief that drops the event-tally, re-estimated, fits essentially as well ($21$ against
$16.6$).  The static flip needs the spillover: with $\mu=0$ the undercut gain collapses to its
drug-margin term, so no chain prices below cost and the war barely forms (war deviations $155$ against the
data's $516$; loss $612$).  The flip is not a knife edge: the post-ban threshold
$\mu^{*}$ jumps about $3.5\times$ at the estimated $\alpha$-drop and the ratio
$\mu^{*}_{\text{post}}/\mu^{*}_{\text{pre}}$ stays in $[3.3,3.8]$ under $\pm10$--$30\%$ perturbations of
wholesale cost and $\sigma$.  The model is a forward-simulated Markov dynamic; with the learned enforcement
belief it reproduces the late rent ($141$ of $143$ rent steps) and the post-cartel success/failure
crossover.

\textbf{What the post-cartel rent does not capture.}  The model matches the rent's total, its
Cruz~Verde leadership, and the incentive that drives it, but not the timing of one feature: $45$ of
the $143$ rent steps ($31\%$) arrive in a synchronized cluster over September--October~$2008$, after the
cartel window closes and during the FNE investigation.  Because the model coordinates each drug
independently at a common arrival rate, it fills the same drugs gradually rather than in a burst; it
cannot generate a synchronized spike, and the source of that late synchronization lies outside the
model.  I report the rent's magnitude and leader, which the model does reproduce, and flag this timing
gap rather than let the cumulative-fit panel of Figure~\ref{fig:price_path} obscure it.

\textbf{The dynamic value function.}  Three checks probe the forward-looking value \eqref{eq:mech_V}, each
re-simulated at the estimate (Online Appendix~\ref{app:struct_robust}).  The model is genuinely dynamic.  A
myopic firm that ignores the continuation value ($\delta\!\to\!0$) coordinates only the static-flip
drugs and never builds the rent, a loss of $555$, because escaping the war and deepening the rent each
trade a current cost for a future payoff.  The regulatory change was also unanticipated.  A
perfect-foresight firm that reads the September CONAR precursor as the coordination trigger starts the wave
two months early, in September (onset week $88$ against the data's $102$; loss $2550$).  The data thus place
the operative shift at the binding, unforeseen November injunction, not the foreseeable precursor.  Finally
the continuation value must be transition-aware.  Valuing each tier as an absorbing perpetuity
($V^x=\Pi^x+\delta V^x$) treats a coordinated drug as if it never reverts and a rent as if it never fails.
That overstates the tier values and mistimes the rent, a loss of $23$, so the forward simulation over the
model's own transition probabilities is what reproduces the post-cartel dynamics.

\textbf{Leadership rests on the calibrated loss cushion, not on $\mu$.}  The chain asymmetry is the
loss cushion $\psi_i$, calibrated from the business mix and held fixed, so the leadership ordering is a
prediction, not a fitted split.  Setting $\psi$ symmetric collapses Salcobrand's lead to chance
while leaving the coordination count untouched (Online Appendix~\ref{app:struct_robust}), the signature of
a leadership channel that is decoupled from the incentive constraint.  Putting the asymmetry back into a
chain-specific store-traffic spread $\mu_i$ instead faces a structural bind.
Store traffic enters both the escape value that selects the leader and the incentive constraint that holds
the followers.  A spread wide enough to make Salcobrand lead therefore also makes its coordination incentive
differ from the rivals', distorting the count, so the spread cannot separate who leads from who follows.
The loss cushion avoids this because it shifts only who initiates and leaves the incentive constraint
symmetric.  Matching the leadership with $\mu_i$ also requires tuning the spread to the leadership it then
``explains,'' the circularity a calibrated loss cushion avoids.  The share magnitudes are then calibrated to the data (Salcobrand leads $74\%$ of cartel-wave increases, the model $69\%$).  The ordering does not depend on that calibration: fixing the two dispersions at a common value still leaves Salcobrand leading the wave ($69$--$80\%$ across the range, Online Appendix~\ref{app:lead_robust}).
Freeing $\psi_{\rm SB}$ and re-estimating returns it to zero, so the most-exposed corner is the estimate,
not an assumption; the August~2007 acquisition of Salcobrand by the private-equity fund Southern~Cross,
which inherited the intensifying below-cost war and led the December escape, is its institutional reading.

\textbf{If prices are deflated for inflation.}  Deflating prices to the war base for the $2007$--$2008$ inflation leaves the war below cost and the cartel, the rent, and the leadership intact (Online Appendix~\ref{app:macro_own}).

\textbf{If the demand estimate is wrong.}  Because the whole flip is read off the estimated demand, I
re-estimate the structural model under alternative demand specifications (Online Appendix
Table~\ref{tab:spec_robustness}): the per-drug heterogeneous $\hat\alpha_j$ specifications reproduce the
leadership and the count, while a homogeneous demand at the median $\hat\alpha$ fits far worse because,
without the per-drug heterogeneity, the wave is too concentrated and the rent too slow.

\subsection{Welfare of the equilibrium transition}\label{sec:welfare}

What did the transition cost consumers?  Because the war was below-cost and the cartel a coordinated
increase, the answer depends on the counterfactual.  I evaluate the settled coordinated prices of
January--April~$2008$ (fourteen weeks, the back of the nineteen-week wave) against two benchmarks: the
below-cost war prices of mid-$2007$ (what consumers faced when the cartel formed) and the pre-war
$2006$ baseline, computing for each drug the consumer transfer and, from the nested logit, the
deadweight loss from the induced quantity distortion.\footnote{Per drug, with
the estimated nested logit ($\sigma=0.393$; post-ban $\alpha=0.029$, the same demand
the structural model uses).  The consumer-surplus change between the coordinated price
$p^{\mathrm c}$ and benchmark $p^{\mathrm b}$ is $M\,[\mathrm{CS}(p^{\mathrm
b})-\mathrm{CS}(p^{\mathrm c})]$, with $\mathrm{CS}(p)=\alpha^{-1}\log(1+[N
e^{(\phi-\alpha p)/(1-\sigma)}]^{1-\sigma})$, decomposed into the transfer
$(p^{\mathrm c}-p^{\mathrm b})\times$ quantity and the residual deadweight loss; $M$ is
backed out from the cartel-window quantity and the observed $0.92$ nest share; here $N=3$ counts the inside chains, not the market size $N_j$ of \eqref{eq:market_size}, which enters only through $M$.}

The measure is within the pharmaceutical market: the drug overcharge and the
quantity-distortion deadweight loss, both identified by the estimated drug demand.  The coordinated and
benchmark price levels are read directly from the data (the same median tier prices the structural
model is calibrated to), so the transfer is the realized overcharge, not a price path the model predicts.
I do not extend it
to the non-pharmaceutical basket, which the case record cannot price; the coordination raised
drug prices, so this counts the pharmaceutical overcharge alone.

The coordinated path extracted supra-competitive rents on many products
(Table~\ref{tab:welfare}).  Within the cartel window $37\%$ of the coordinating drugs reached
Tier~2 rent extraction, deepening to $143$ drugs ($65\%$ of all $220$) by year-end as it ratcheted
post-Oficio.  These rents were priced at a median $1.48\times$ wholesale cost and
$1.11\times$ the $2006$ baseline ($11\%$ above it), the margins the conduct was fined
for \citep{fne2008requerimiento,tdlc2012sentencia}.

\begin{table}[H]
\centering\small
\caption{Welfare of the coordinated price path}
\label{tab:welfare}
\begin{tabularx}{0.9\linewidth}{@{}>{\raggedright\arraybackslash}X r@{}}
\toprule
\multicolumn{2}{l}{\textbf{Panel A. Coordinated prices above margin}}\\
Drugs reaching Tier~2 within the window             & $37\%$ \\
Median Tier-2 price $/$ wholesale cost              & $1.48\times$ \\
Coordination events priced $>1.4\times$ cost        & $43\%$ \\
Median Tier-2 price $/$ 2006 baseline ($\%$ above)  & $1.11\times$ ($11\%$) \\
\midrule[\heavyrulewidth]
\multicolumn{2}{l}{\textbf{Panel B. Consumer harm over the window}}\\
Transfer vs.\ below-cost war (\% of sales)          & $+$CLP\,$3.4$bn \ ($+11\%$) \\
\quad of which from the Tier-2 rent drugs           & $\sim$CLP\,$2.7$bn \ ($80\%$) \\
Gross rent vs.\ 2006 baseline (above-historical drugs) & $+$CLP\,$1.1$bn \\
Deadweight loss (quantity distortion)               & $<$CLP\,$0.03$bn \ ($<0.1\%$) \\
\bottomrule
\end{tabularx}
\par\smallskip\flushleft
{\footnotesize \textbf{Note}: Over the settled fourteen-week window (Jan--Apr~2008, three
chains; window sales $\approx$CLP\,$30.7$bn).  Panel~A: how far
coordinated prices stood above margin.  Panel~B: consumer harm, the overcharge on
the transacted quantity (transfer) plus the deadweight loss from the quantity
response in the estimated nested logit.  The transfer and the deadweight loss are
deliberately distinct objects (the first the overcharge incidence on the
transacted quantity, the second the efficiency loss from the induced quantity
distortion), reported separately, not summed.}
\end{table}

Relative to the below-cost war prices consumers had been paying, the coordinated path cost them about
$11\%$ (CLP\,$3.4$bn over the fourteen-week window), four-fifths of it from the Tier-2 rents.  The closest like-for-like enforcement figure is the $\sim$CLP\,$2.1$bn the chains later paid in consumer compensation, of the same order as this transfer and the same kind of object.  The other two figures are on a different time frame or basis, so a fourteen-week transfer is not directly comparable to them.  The FNE's $\sim$CLP\,$27$bn is a gross-revenue figure over the full conduct; annualized, the window transfer is about CLP\,$13$bn a year, the same order.  The $\sim$US\$$38$M fine ($\approx$CLP\,$19$bn at the $2012$ exchange rate) upheld by the Supreme Court is a deterrence penalty, not a measure of consumer harm \citep{fne2008requerimiento,tdlc2012sentencia}.  The deadweight loss is small ($<$CLP\,$0.03$bn,
under $0.1\%$ of sales): post-ban demand is inelastic, so the increase moved little quantity and the
harm is overwhelmingly a transfer.  The extraction was selective: the cartel held the high-volume,
salient staples near competitive levels (volume--overcharge correlation $-0.14$) while taking rents on
the less-salient products.

\section{Conclusion}\label{sec:conclusion}

I have analysed the equilibrium transition from loss-leader
competition to coordinated pricing among the three large Chilean
retail pharmacy chains in 2007--2008.  The analysis yields three
results.

First, the transition needs two changes at once, and neither alone is sufficient.  The
war was a low-belief trap: each chain expected a price rise to be
undercut, so none held one.  The advertising ban broke the trap in two
ways.  As a public, dated event it made the chains expect a coordinated
rise to hold this time.  And by removing the comparative-price advertising
that told shoppers who was cheapest, it cut the estimated price
coefficient from $0.103$ to $0.029$, so undercutting no longer won
customers.  Once undercutting stopped paying, the coordinated price held on its own for those drugs, with no patience and no
punishment threat, and the finite horizon did not unravel it.  The belief
jump and the demand change are each necessary; only together do they
reproduce the path (\S\ref{ssec:mech_robust}).

Second, the cartel's leadership and its slow rollout are predictions of
the model, not facts I put into it.  The case file records who led, and
that the laboratories organized the coordination, but it does not say why.
The model does.  Salcobrand leads because it has the least non-drug business to cushion
the below-cost war, so it has the most to gain from ending it.  The larger
laboratories move first because they carry the biggest prize, so the
model's reach-out engages them earliest.  Both come from primitives I
calibrate outside the model, the chains' loss cushions and the labs'
stakes, and the orderings they imply match the record out of sample: the model picks the
right leader and the larger-labs-first sequence it was never told, along
with the war, the post-ban wave and its February pause, and the split
between margin restoration and rent.  The post-ban failures are mostly stray
unmatched lifts and reverted rent pushes, rather than defections from coordinated
prices.

Third, the harm is a transfer, not an efficiency loss.  The coordinated
path raised prices well above cost on many products; $37\%$ reached rent
prices near $1.5\times$ wholesale cost, costing consumers about
CLP\,$3.4$bn over the cartel window, four-fifths of it on the rent
products.  Because post-ban demand is inelastic, the deadweight loss is
tiny.  The cartel mainly moved money from consumers to firms, and it did
so where consumers paid least attention.

The model's economic content rests on objects estimated or documented
outside it: the demand system, the wholesale costs and tier prices,
and the laboratory structure and leadership from the legal record.
The store-traffic value is bounded by the chains' own below-cost
behaviour, and a handful of organizational-timing parameters are
calibrated by simulated method of moments.  The gradual, four-month
rollout, which the prosecutor's complaint describes as expanding as the
agreement's success was verified,\footnote{Original: ``a medida que se
verificaba el éxito del acuerdo'' \citep{fne2008requerimiento}.} is
reproduced by the endogenous laboratory reach-out rather than by a
separate contagion term.  More chains find it worth organizing a
laboratory as undercutting stops paying and beliefs shift, drug by drug.  What gradual belief-learning adds beyond
this is a secondary refinement (\S\ref{ssec:mech_robust}).  Disentangling
it from antitrust-detection risk and operational complexity is a useful
direction for future work.

% =====================================================================
% End of conclusion
% =====================================================================

% =====================================================================
%  appendix_r2_stubs.tex -- expanded appendix sections.
% =====================================================================

\appendix

\section{Demand-fit decomposition}\label{app:demand_fit_decomp}

\textbf{Goodness-of-fit formulas.}  The within-nest share prediction used for the diagnostics of
\S\ref{ssec:demand_fit} is
\begin{equation}
  \hat s_{i|jt}
  \;=\;
  \frac{\exp\!\bigl(\hat u_{ij(t)}/(1-\hat\sigma)\bigr)}
       {\sum_{i'\in\mathcal{I}}\exp\!\bigl(\hat u_{i'j(t)}/(1-\hat\sigma)\bigr)},
  \qquad
  \hat u_{ij(t)} = \hat\phi_{ij} - \hat\alpha_{\rm eff}(t)\,p_{ijt},
  \label{eq:share_pred}
\end{equation}
with $\hat\phi_{ij}$ the within-(chain, drug) fixed effect and
$\hat\alpha_{\rm eff}(t) = \hat\alpha_0 + \hat\alpha_1 D_t$.  Only the structural parameters and
prices enter the right-hand side, so \eqref{eq:share_pred} is a non-circular prediction, not the
mechanical contraction-mapping $R^2$ of \citet[][p.~$32$]{conlon2020best}.  Own- and cross-price
elasticities use the standard nested-logit formulas
\begin{align}
  \varepsilon_{ijj,t}^{\rm own}
  &= -\alpha\, p_{ijt}\Big[
       \tfrac{1}{1-\sigma}
       - \tfrac{\sigma}{1-\sigma}\, s_{i|jt}
       - s_{ijt}\Big],
       \label{eq:own_elas}\\
  \varepsilon_{ijk,t}^{\rm cross}
  &= \alpha\, p_{kjt}\, s_{kjt}\Big[
       1 + \tfrac{\sigma}{(1-\sigma)\, s_{jt}}\Big],
       \quad i \ne k. \label{eq:cross_elas}
\end{align}

\begin{table}[htbp]
\centering\small
\caption{Below-cost share and implied elasticity}
\label{tab:elast_below_cost}
\begin{tabular}{lrrr}
\toprule
ATC1 class                       & $N$ drugs & Below-cost share & $|\varepsilon^{\rm own}|$ \\
\midrule
H (systemic hormones)            & $6$  & $0.45$ & $1.16$ \\
L (antineoplastics)              & $2$  & $0.40$ & $0.31$ \\
S (sensory organs)               & $3$  & $0.30$ & $0.19$ \\
C (cardiovascular)               & $38$ & $0.26$ & $1.35$ \\
G (genito-urinary)               & $42$ & $0.23$ & $1.28$ \\
A (alimentary tract / metab.)    & $20$ & $0.13$ & $0.78$ \\
D (dermatologicals)              & $1$  & $0.10$ & $3.47$ \\
B (blood / blood-forming)        & $10$ & $0.10$ & $1.24$ \\
N (nervous system)               & $59$ & $0.07$ & $1.08$ \\
R (respiratory)                  & $22$ & $0.06$ & $2.15$ \\
M (musculo-skeletal)             & $14$ & $0.06$ & $1.65$ \\
J (anti-infectives, systemic)    & $5$  & $0.01$ & $3.26$ \\
\bottomrule
\end{tabular}
\par\smallskip\flushleft
{\footnotesize \textbf{Note}: Below-cost share is the fraction of chain-drug-week observations in the
pre-ban window with $p_{ijt}<c_{jt}$; $|\varepsilon^{\rm own}|$ is the class-mean of
$|\varepsilon_{ijj,t}^{\rm own}|$ from \eqref{eq:own_elas} at median pre-ban prices.  Spearman rank
correlation across the $12$ classes: $\rho = -0.62$.}
\end{table}

The pre-ban share-fit correlation $0.73$ reported in
Table~\ref{tab:demand_fit} is below the post-ban value of $0.86$.
This appendix decomposes the gap.  The three relevant sub-regimes
are: pre-ban competitive weeks (panel-wide below-cost
prevalence $\leq 25\%$), pre-ban war weeks ($> 25\%$
below-cost prevalence), and post-ban weeks.  At the headline
parameters $(\hat\alpha_0, \hat\alpha_1, \hat\sigma) = (0.103,
-0.074, 0.393)$:

\begin{table}[H]
\centering\small
\caption{Share-fit decomposition by regime}
\label{tab:demand_fit_breakdown}
\begin{tabular}{lrc}
\toprule
Regime                 & $N$       & $\mathrm{Corr}(\hat s_{i|jt},\,s_{i|jt})$ \\
\midrule
pre-ban competitive    & $44{,}421$ & $0.67$ \\
pre-ban war            & $18{,}564$ & $\mathbf{0.88}$ \\
pre-ban (all)          & $62{,}985$ & $0.73$ \\
post-ban               & $41{,}769$ & $0.86$ \\
\bottomrule
\end{tabular}
\par\smallskip\flushleft
{\footnotesize \textbf{Note}: Pre-ban "war
weeks" are pre-ban weeks with panel-wide below-cost prevalence
above $25\%$ ($28$ weeks).  Pre-ban "competitive" are the
remaining pre-ban weeks.}
\end{table}

The pre-ban war weeks fit better than the post-ban regime
($\rho = 0.88$ vs.\ $0.86$).  The fit gap is therefore concentrated
in the pre-ban competitive weeks, where the static
nested-logit struggles to match the strategic dimension of
loss-leader behaviour: in those weeks each chain was independently
selecting which drugs to push below cost and by how
much, drug-by-drug and week-by-week.  These are coordinated
intertemporal pricing decisions that do not enter the right-hand
side of \eqref{eq:meanutility}; the nested-logit sees them as
residuals $\xi_{ijt}$.  Post-ban, the three chains converge to a
single coordinated tier and the strategic dimension flattens.

I tested allowing the price coefficient to differ in war
weeks, $\alpha_t = \alpha_0 + \alpha_1 D_t + \alpha_2 W_t$ where
$W_t$ is the war-week indicator, by a grid search that maximises
pre-ban share correlation.  The optimum is $\hat\alpha_2 = +0.05$
but the gain in pre-ban $\rho$ is negligible.  Removing the regime
split entirely and fitting a whole-period $(\hat\alpha, \hat\sigma)$
by share-fit grid search yields $(\hat\alpha, \hat\sigma) =
(0.05, 0.15)$ with whole-period $\rho = 0.80$, marginally above
the pooled spec's $\rho = 0.78$.  The
modest gain confirms that the binding constraint is the
within-competitive-week strategic dimension, not the
between-regime price coefficient.  A fuller resolution would
require a dynamic model of consumer search or chain pricing
choices; I leave that to future work and proceed with the
single $(\hat\alpha_0, \hat\alpha_1, \hat\sigma)$ for the welfare
calculation in \S\ref{sec:welfare}, which uses tier-level prices
where shares are stable.

\subsection{The product-firm demand the structural model
inherits}\label{app:ij_demand_gof}

The dynamic game inherits demand at the chain-by-drug level.  Each
chain $i$'s quality intercept $\hat\phi_{ij}$ is recovered from the
within-nest shares $\hat s^{w}_{i|j}$ observed in the coordinated
window via \eqref{eq:share_pred} at the common coordinated price; the
instrumented coefficient $\hat\alpha_j$ governs the price response and
the market size $N_j$ is anchored on a normal-period median quantity.
Two out-of-sample checks confirm the mapping reproduces both the
market shares and the quantities the game needs.

\textbf{Per-drug specification.}  Holding the nesting parameter at the pooled
$\hat\sigma=0.393$, I estimate the per-drug $\hat\alpha_j$ for each of the $222$ drugs by
two-stage least squares on the competitive (cartel-excluded) daily panel
($\approx 3{,}288$ obs per drug across three chains), instrumenting each drug's price (and its
post-cartel interaction) with the two rival-chain prices.\footnote{The nesting parameter $\sigma$
governs the single chain nest (the three chains compete for the same drug), so it is a
market-level, not a drug-level, object.  I therefore identify it once on the pooled pre-ban data
(Panel~A of Table~\ref{tab:demand_hetero}) and condition the per-drug slopes on it; a separate
per-drug $\sigma_j$ is not identified, as it would need within-drug substitution variation that
three chains over a short window cannot supply.  Estimating a common $\sigma$ with heterogeneous
price coefficients $\alpha_j$ is the standard sequential nested-logit approach
\citep{berry1994estimating}.  Table~\ref{tab:mech_robust} (Panel~B) re-solves the structural model
at $\sigma\in\{0.30,0.50\}$: the $\mu$ band and the war-to-coordination flip are unchanged, so the
fixed-$\sigma$ choice does not drive the results.}
The within-transform removes chain (Firm) effects only: a two-way Firm$\times$Day demeaning would
force the three chains' prices to sum to zero within each day, making own price a mechanical function
of the rivals and the rival instrument degenerate.  The within-drug estimating equation is
\begin{equation}
  \tilde{Y}_{ijt} \;=\; -\,\alpha_j\,P_{ijt} \;-\;
  \gamma_j\,(P_{ijt}\times\mathrm{PostCartel}_t)
  \;+\; \phi_i \;+\; \varepsilon_{ijt},
  \label{eq:per_drug}
\end{equation}
with $\tilde{Y}_{ijt}=Y_{ijt}-\hat\sigma\ln s_{i|jt}$, $\mathrm{PostCartel}_t=1$ for days
$829$--$1{,}096$, $\phi_i$ a chain fixed effect, and $P_{ijt}$ (and its interaction) instrumented
by the rival prices.  Identification of $\alpha_j$ relies on within-drug, cross-chain price
dispersion in the competitive windows.  The ban interaction
$\gamma_j$ enters only as a control, so $\hat\alpha_j$ is the clean pre-ban within-window
price slope; the post-ban $\hat\alpha_j$ is re-estimated with the
Salcobrand-cost instrument, since post-ban rival prices are coordination-contaminated, and the
structural model scales each $\hat\alpha_j$ to its post-ban value by the common pooled ratio
$\alpha^{\rm post}/\alpha_0=0.29$ (\S\ref{sec:mechanisms}).  The pooled analog of the ban shift is
$\hat\alpha_1=-0.074$.

\textbf{Market share.}  Intercepts recovered at the coordinated price
predict the within-nest shares in two regimes they were not fit
to (the price war and the supra-competitive rent) to a median
absolute error of $1.6$ percentage points (mean $2.5$; $90$th
percentile $5.6$) across the $216$ drugs.  The coordinated window is
in-sample and reproduces exactly.

\textbf{Quantity.}  With $N_j$ anchored on the normal-period quantity,
the implied three-chain quantity $N_j\sum_i \hat s_{ij}$ matches the
observed quantity to a median ratio of $0.97$, $0.99$, and $0.98$ at
the war, coordinated, and rent tiers (interquartile range
$\approx[0.85,1.13]$; log-correlation $0.75$--$0.85$).  Anchoring
instead on the absolute-peak quantity (the price-war stocking-up
spike) over-states quantity by a third and is rejected.

The chain-specific undercut gains reduce to the symmetric gain when
the three chains' within-shares are equal and correlate $0.94$ with it
across drugs, so the product-firm layer nests the symmetric mapping as
a special case.

\begin{figure}[htbp]
\centering
\caption{Out-of-sample market-share and quantity fit}
\label{fig:demand_ij_gof}
\includegraphics[width=\textwidth]{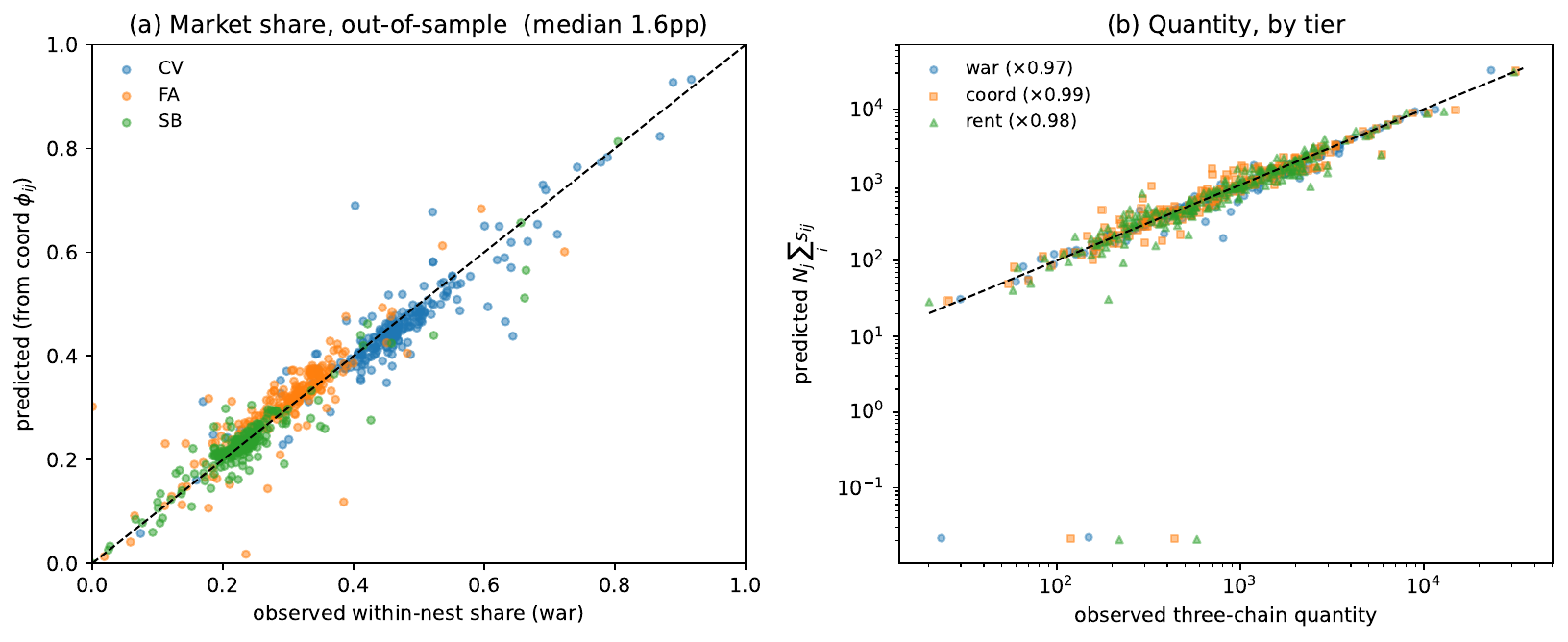}
\par\smallskip\flushleft
{\footnotesize \textbf{Note}: Panel (a): predicted versus observed
within-nest shares in the price war, using intercepts
$\hat\phi_{ij}$ recovered from the coordinated window
(out-of-sample); the $45^{\circ}$ line is exact fit.  Panel (b):
predicted three-chain quantity $N_j\sum_i\hat s_{ij}$ versus observed,
by tier (log scale).  $216$ drugs with a complete war / coordinated /
rent price triple.}
\end{figure}

\section{The event ladder vs.\ the binary cartel
indicator}\label{app:event_coding}

% =====================================================================
%  APPENDIX fragment: event coding vs. the binary cartel indicator
%  Place Fig_event_coding_vs_col.png in your figures path before compiling.
% =====================================================================
\subsection{Event coding versus the binary cartel indicator}

The case data include a coordination indicator, $\mathrm{col}_{fjt}\in\{0,1\}$, equal to one
when chain $f$'s price for drug $j$ on day $t$ is in the coordinated (post-increase) regime.
This indicator records whether a firm has joined the coordinated price but not how
far prices have risen: a $0\!\to\!1$ flip marks the first coordinated increase, yet the variable
cannot distinguish a first increase from a second or third. The event ladder instead records the
full sequence of coordinated increases for each drug: every transition from price tier $\ell$ to
$\ell+1$, its date, and the firm that moved first.

\begin{figure}[t]\centering
\caption{Event taxonomy versus the binary indicator}
\label{fig:event_coding_vs_col}
\includegraphics[width=\textwidth]{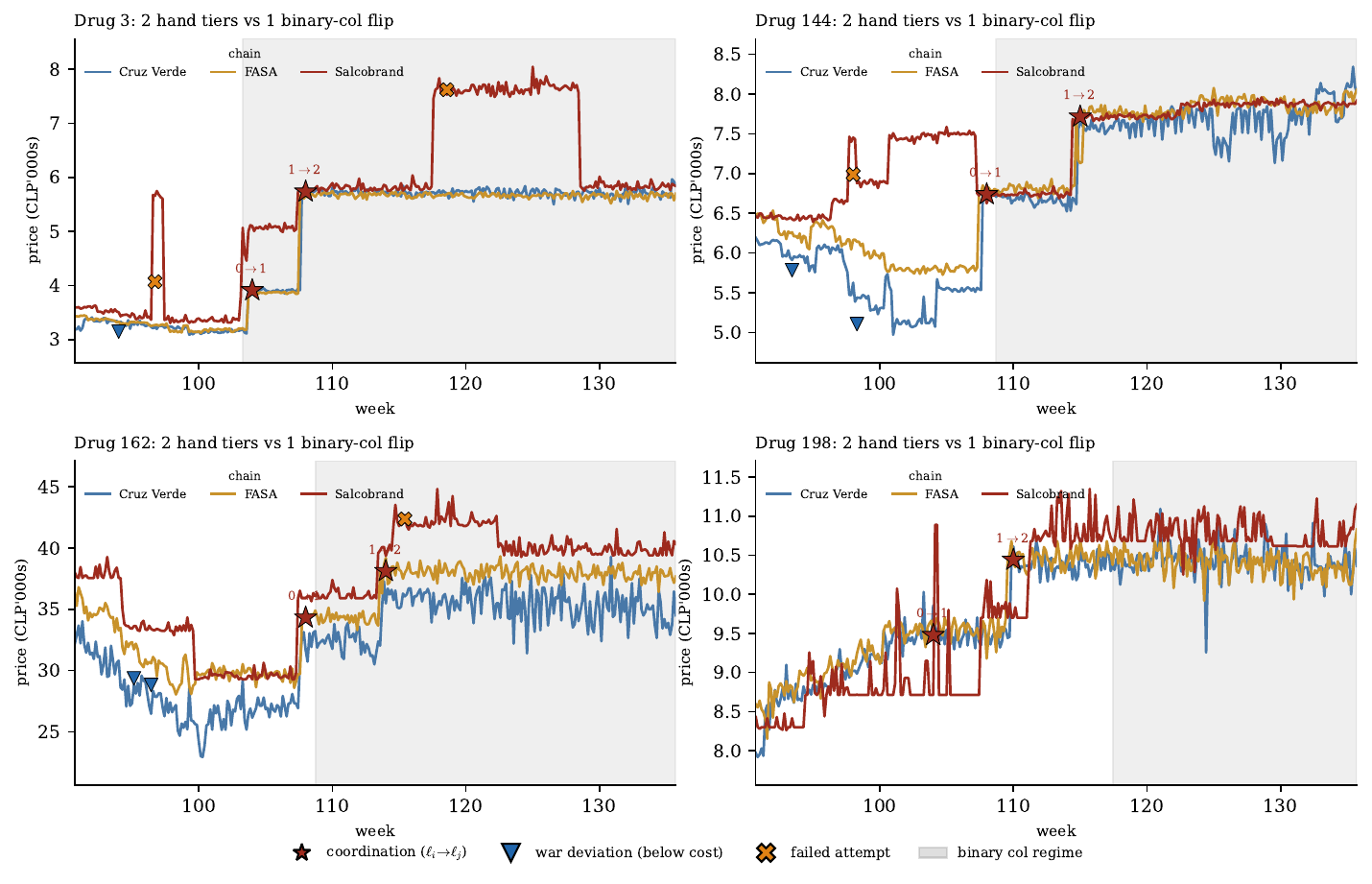}

\par\smallskip\flushleft
{\footnotesize \textbf{Note}: Each event is marked at its own date and price by nature: coordination
tier transitions (stars, labelled $\ell_i\!\to\!\ell_j$), below-cost war deviations (triangles),
and failed attempts (crosses); post-coordination defections use diamonds where present. Prices move
in discrete tiers that the event coding records individually, with their nature distinguished,
whereas the binary indicator marks only the onset of the coordinated regime.}
\end{figure}

Table~\ref{tab:event_coding_compare} compares the two. The binary indicator coincides with most
first-round onsets but with only $22\%$ of the second- and third-round increases, and in
$127$ of the $137$ multi-round drugs it records fewer steps than the event ladder. Because the mechanism
I study is the sequential, multi-round escalation itself, the tier ladder, not a single on/off
flag, is the economically relevant unit of observation. Figure~\ref{fig:event_coding_vs_col}
illustrates with four representative drugs: prices climb in discrete plateaus that the event ladder tracks
step by step, while $\mathrm{col}$ collapses the entire episode into one block.

\begin{table}[t]\centering
\caption{Event ladder versus binary indicator}
\label{tab:event_coding_compare}
\begin{tabular}{lr}
\hline
 & Value \\
\hline
Drugs in the coordination sample & 222 \\
Coded coordinated price increases & 442 \\
\quad first-round onsets (tier $0\!\to\!1$) & 293 \\
\quad later-round steps (tier $1\!\to\!2$, $2\!\to\!3$) & 149 \\
Drugs with $\ge 2$ coordinated increases & 137 \\
$\mathrm{col}$ $0\!\to\!1$ flips (drug level) & 249 \\
First-round onsets matched by a $\mathrm{col}$ flip & 188 (64\%) \\
Later-round steps matched by a $\mathrm{col}$ flip & 33 (22\%) \\
Multi-round drugs where $\mathrm{col}$ records fewer steps & 127 of 137 \\
Leader agreement ($\mathrm{col}$ first-mover vs.\ coded leader) & 57\% \\
\hline
\end{tabular}
\par\smallskip\flushleft
{\footnotesize\raggedright Notes: a $\mathrm{col}$ flip is matched to a coded event if the two
occur within 15 days. Later-round steps have no counterpart in a binary indicator by construction.\par}
\par\smallskip\flushleft
{\footnotesize \textbf{Note}: The event ladder vs.\ the binary $\mathrm{col}$ indicator}
\end{table}

\textbf{Event-time comparison.} I further compare the
event ladder with the FNE official cartel indicator
$\mathrm{col}_{ijt}\in\{0,1\}$ at the drug level.  For each drug $j$
I define the FNE official event time as the first day on which
$\mathrm{col}_{\mathrm{CV},jt} = \mathrm{col}_{\mathrm{FA},jt} =
\mathrm{col}_{\mathrm{SB},jt} = 1$.  Of the $205$ drugs that have an
event in both panels, the median difference (FNE day minus
ladder day) is $-2$ days, with inter-quartile range $[-3,+5]$ days
and Pearson correlation $r=0.768$.  Seventy-nine percent of drugs
are within seven days of each other across the two panels, and
$95\%$ are within sixty days.

\section{Market-size robustness}\label{app:ms_robust}

The demand estimator of Section~\ref{sec:demand_estimation} uses
$N_j = \max_t Q_{jt}^{\rm wk}/\hat\rho + 1$ with $\hat\rho=0.92$
(three-chain market share, FNE 2008).  I re-estimated the model under
two alternative calibrations.

\begin{enumerate}\itemsep0pt
  \item \textbf{Higher chain-share estimate.} Set $\hat\rho=0.95$
        (\citealp{DiazGaletovic2015}).  This lowers $N_j$ by $3.3\%$
        and raises the outside share at the peak week from $8\%$ to
        $5\%$.  Point estimates of $(\hat\alpha,\hat\sigma)$ shift by
        less than $\pm 0.005$.
  \item \textbf{Constant national population.} Set $N_j = N^{*}/J$
        for all $j$, with $N^{*}=4\!\times\!10^{6}$ (Chile's adult
        prescription-buying population).  $(\hat\alpha,\hat\sigma)$
        shift by less than $\pm 0.01$.
\end{enumerate}

The own-price elasticity dispersion across drugs is essentially
unchanged across the three specifications, confirming that the
chain-by-drug fixed effect $\phi_{ij}$ absorbs level mis-scaling
of $N_j$.

\section{Data construction}\label{app:data}

The transaction data, obtained from the FNE
under Chile's transparency law (Ley No.\ 20.285), cover all purchases of the
$222$ case products from the three chains over $2006$--$2008$; each record carries
the drug, chain, date and time, list price, transacted price, and units.  I
aggregate to weekly revenue-weighted averages: daily data are noisy at low
volumes, while monthly aggregation would blur the $2$--$4$-day coordination
windows.  Product attributes (active ingredient and ATC class; package size,
pill count, brand-vs-generic status, prescription status) were hand-compiled and
cross-checked from three public sources (an internal catalog, DrugBank, and
the Chile-specific listing Farmazon) with the small share of conflicts
($<3\%$) resolved in favour of the Farmazon listing.  Wholesale costs are
Salcobrand's wholesale prices for November $2007$--May $2008$, submitted to the
Competition Tribunal as expert evidence \citep{tdlc2012sentencia}.  For
chain-product-periods with no recorded sales I use the chain-product median
price; periods with zero quantity are treated as missing, not as a zero price;
and within-chain coordination statistics use only chains that stock the product.

\textbf{Mapping to the $222$-drug TDLC list.}  Of the $222$ drugs
named in the TDLC \emph{Sentencia}, $220$ ($99.1\%$) show at least one
successful three-chain coordination in the event panel; $211$ clear the
$15\%$ coding threshold, and $9$ are recovered only by the more
permissive daily-panel detector or the later-window pass (small
tier-$0\!\to\!1$ events of $12$--$15\%$).  The remaining two, Losopil
(Mediapharm) and Progyluton (Bayer), show only single-firm unilateral
lifts of $12$--$14\%$, with no event in which all three chains follow.
Both count as coordinated in the official \texttt{col\_ijt} indicator (a
binary flip covering the full $222$-drug list by construction), but my
finer panel, which separates single-firm attempts from three-chain
agreements, does not classify them as coordinated.  The implied
three-chain coordination rate on the TDLC list is $220/222 \approx
99.1\%$.

\section{The structural estimation sample}\label{app:structural_sample}

The dynamic model of \S\ref{sec:mechanisms} is estimated on the $J=220$ drugs that
complete the war-to-coordination price transition the model is built around.  A drug
enters the structural panel only if the event ledger records all three of (i)~a
war/deviation price (a pronounced dip during the price-war window), (ii)~a Tier-1
coordinated price, and (iii)~an imputed cost; the Tier-2 rent price is imputed when absent, and the
per-drug price coefficient $\alpha_j$ falls back to its type median when the demand
regression is uninformative.

\begin{figure}[htbp]
\centering
\caption{Six borderline drugs: four recovered, two excluded}
\label{fig:dropped}
\includegraphics[width=\linewidth]{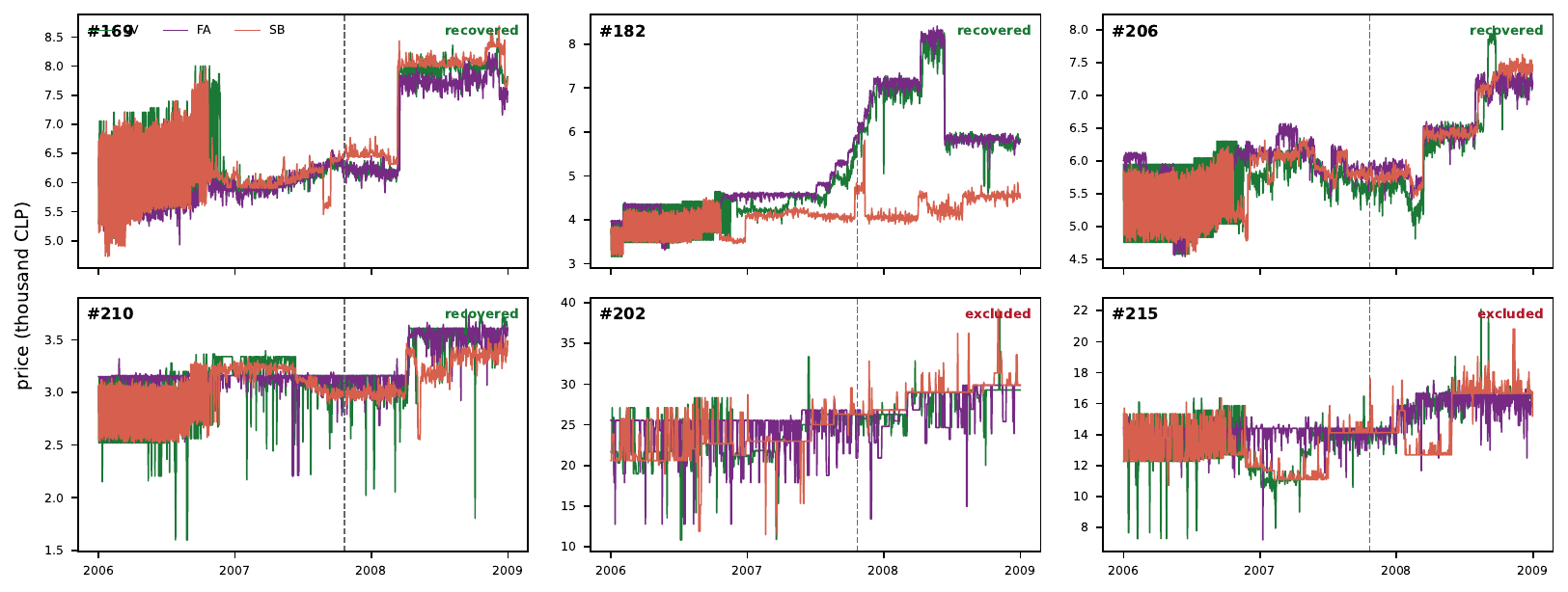}
\par\smallskip\flushleft
{\footnotesize \textbf{Note}: All six borderline drugs I adjudicated are plotted: the four
recovered (green) I admit to the sample and the two excluded (red) I drop, so only
the two red drugs leave the sample (the figure shows both decisions, not only the drops).  Each panel
plots the three chains' daily price (thousand CLP, cleaned of abnormal jumps as in the
demand panel) for one drug; the dashed line marks the advertising ban.  The four
recovered (green) each step up to a sustained higher level in $2008$ (a
coded three-chain coordination), but their war-period dip is shallow or unrecorded,
so I impute the war-deviation price from the demand panel and admit them, taking the
structural sample to $J=220$.  The two labelled excluded (red) stay out: \#$202$
rises monotonically and then defects without a preceding loss-leader war, and \#$215$
never reaches a sustained Tier-1 level.}
\end{figure}

An initial pass over the event ledger excluded fifteen drugs.  Cross-checking
each against its raw price path showed that nine of them do complete the
transition, a marked dip during the war window followed by a sustained higher level,
that the ledger had failed to record; I reclassified these (restoring the
missing war and/or Tier-1 events), raising the sample from $207$ to $216$.  A second
pass then recovers four more (Table~\ref{tab:dropped}): Carvedilol,
Chlorpheniramine Maleate, Ethinyl Estradiol, and Oxolamine Citrate are all coded
three-chain coordinations (Figure~\ref{fig:dropped} shows each stepping up to a
sustained higher level in $2008$) whose only missing ledger leg is a flagged
war-deviation price.  I impute that leg from the drug's demand-panel war-window median
(at the $0.82$ ratio that aligns the demand-window median with the recorded
war-deviation price across the in-sample drugs; \texttt{full\_structural.build\_panel}),
which brings the structural sample to $J=220$, exactly the descriptive count of
coordinated drugs in Table~\ref{tab:event_window}.  This price-path audit doubles as a
completeness check on the event coding.  Two case drugs remain excluded: the Neuractin
presentation of Valproic Acid (\#$202$) coordinates in early $2008$ but then defects
after a monotonic price rise, with no preceding loss-leader war to identify the
transition; and Estradiol Valerate/Norgestrel (\#$215$) never reaches a sustained
Tier-1 level.  Excluding these two is conservative for the coordination-timing moments.

\medskip
Table~\ref{tab:mech_moments} lists the moments the SMM criterion targets and the parameter each
primarily identifies, the mapping summarised in \S\ref{ssec:mech_estim}.

\begin{table}[htbp]
\centering\small
\caption{Targeted moments and what they identify}
\label{tab:mech_moments}
\begin{tabular}{l >{\raggedright\arraybackslash}p{6.5cm} >{\raggedright\arraybackslash}p{4.6cm}}
\toprule
Group & Moment & Primarily identifies \\
\midrule
Counts      & first ($0\!\to\!1$) and second ($1\!\to\!2$) coordination totals & $\lambda$; rent penalty $\chi$ \\
Timing      & onset week, peak week, wave SD                       & $t_0,\ \lambda$ \\
            & $1\!\to\!2$ lag behind $0\!\to\!1$                    & rent penalty $\chi$ \\
Laboratory  & lab-wave SD; labs coordinated by wk.\ $104$; within-lab SD & reach-out $\lambda,\tau_e$ \\
            & corr(log lab size, coordination week)                & reach-out $\lambda,\tau_e$ \\
War floor   & below-cost war markup                                & $\mu$ (joint) \\
War series  & weekly below-cost deviations: pre / mid / late / post & $g$ \\
Micro       & corr(pre-ban cheat freq., $G^{01}_j$)                & benefit-ordered war \\
            & $\alpha_j$/margin gap: rent-takers vs.\ not          & rent selectivity \\
Belief      & failed attempts by period (war high, cartel $\approx\!0$, post rebound) & war prior $(a_0,b_0)$, signal $\zeta$, skepticism $m$ \\
Rent        & rent steps by period (in-cartel + post-cartel deepening) & rent penalty $\chi$, skepticism $m$ \\
Leadership  & leader share by period (SB wave, CV late rent)       & outside option $\psi$ (calibrated) \\
\bottomrule
\end{tabular}
\end{table}

\begin{table}[htbp]
\centering
\footnotesize
\caption{Borderline case drugs: recovered or excluded}
\label{tab:dropped}
\begin{tabular}{rllll}
\toprule
Drug ID & Active ingredient & Type & Status & Ledger note \\
\midrule
169 & Carvedilol & Chronic OTC & Recovered & war price imputed \\
182 & Chlorpheniramine Maleate & Acute OTC & Recovered & all legs imputed \\
206 & Ethinyl Estradiol & Acute Rx & Recovered & war price imputed \\
210 & Oxolamine Citrate & Acute OTC & Recovered & war price imputed \\
\midrule
202 & Valproic Acid (Neuractin) & Chronic Rx & Excluded & coordinates, then defects; no war \\
215 & Estradiol Valerate/Norgestrel & Chronic Rx & Excluded & never reaches Tier-1 \\
\bottomrule
\end{tabular}
\par\smallskip\flushleft
{\footnotesize \textbf{Note}: Four of the six are recovered into the structural sample, each a
coded three-chain coordination whose only missing ledger leg is a war-deviation price
imputed from the demand panel.  This raises $J$ from $216$ to $220$, the full count of
coordinated drugs; two remain excluded.  ``Type'' is the chronic/acute $\times$
prescription/OTC classification.}
\end{table}

\section{Robustness to the demand specification}\label{app:spec_robust}

The structural estimate is robust to how the demand side is specified.
Table~\ref{tab:spec_robustness} re-estimates the model under three demand specifications on the
identical headline criterion: the heterogeneous per-drug $\hat\alpha_j$ from the rival-price IV
(Het.\ IV, the headline; Table~\ref{tab:demand_hetero}); the same heterogeneous $\hat\alpha_j$
from the OLS (Het.\ OLS); and a homogeneous demand that gives every drug the median $\hat\alpha$.
The store-traffic level ($\bar\mu\approx9.0$--$9.3$) and the leadership ordering are essentially unchanged
across all three: Salcobrand leads the cartel wave (SB-share $0.69$--$0.70$) and Cruz~Verde takes over the
post-investigation rent (CV-share $0.67$--$0.68$).  The two heterogeneous specifications give the same
structural estimate and fit comparably (loss $16.6$ and $17.3$; the two $\hat\alpha_j$ series are nearly
collinear).  Homogeneous demand fits worse (loss $27.4$): without the per-drug heterogeneity the rent
over-deepens and arrives too late ($170$ steps against $143$, a $1\!\to\!2$ lag of $13.7$ against the
data's $8.0$ weeks), so the per-drug heterogeneity is what sizes and times the rent.

\begin{table}[H]
\centering\small
\caption{Structural estimates across three demand specifications}
\label{tab:spec_robustness}
\begin{tabularx}{0.9\linewidth}{@{}>{\raggedright\arraybackslash}X c c c c@{}}
\toprule
                                & Data & Het.\ IV & Het.\ OLS & Homog. \\
\midrule
\multicolumn{5}{@{}l}{\textbf{Panel A. Estimated parameters (SMM)}}\\[2pt]
$\bar\mu$\quad mean store-traffic value &        & $9.26$ & $9.26$ & $9.02$ \\
$\lambda$\quad reach-out scale          &        & $0.63$ & $0.63$ & $0.34$ \\
$\tau_e$\quad reach-out temperature     &        & $124$ & $124$ & $54$ \\
$t_0$\quad organization onset (wk)      &        & $103$  & $103$  & $102$  \\
$a_0,b_0$\quad belief prior             &        & $1.80,129$ & $1.80,129$ & $2.11,129$ \\
$\zeta$\quad ad-ban pseudo-successes    &        & $307$ & $307$  & $325$  \\
$\chi$\quad rent-tier discount          &        & $2.08$ & $2.08$ & $2.59$ \\
$m$\quad enforcement skepticism         &        & $14.8$ & $14.8$ & $17.1$ \\
$g$\quad Aug.\ campaign strength        &        & $0.57$ & $0.57$ & $0.64$ \\
\textbf{SMM loss}                       &        & $\mathbf{16.6}$ & $\mathbf{17.3}$ & $\mathbf{27.4}$ \\
\midrule[\heavyrulewidth]
\multicolumn{5}{@{}l}{\textbf{Panel B. Fit on selected moments}}\\[2pt]
First coordinations          & $220$  & $214$ & $220$ & $220$ \\
Rent steps                   & $143$  & $141$ & $140$ & $\mathbf{170}$ \\
Failed war attempts          & $123$  & $110$ & $110$ & $126$ \\
Cartel successes             & $284$  & $290$ & $293$ & $303$ \\
Post-late successes          & $49$   & $35$  & $34$  & $55$  \\
Peak week                    & $107$  & $107$ & $107$ & $106$ \\
Wave SD (weeks)              & $4.6$  & $4.5$ & $4.6$ & $4.1$ \\
$1\!\to\!2$ lag (weeks)      & $8.0$  & $8.0$ & $8.0$& $\mathbf{13.7}$\\
Between-lab var.\ share      & $0.59$ & $0.54$& $0.61$& $0.49$\\
Spearman(size, week)         & $-0.30$& $-0.24$& $-0.28$& $-0.25$\\
SB leads cartel wave         & $0.74$ & $0.69$& $0.69$& $0.69$\\
CV leads post-invest.\ rent  & $0.65$ & $0.68$& $0.67$& $0.68$\\
\bottomrule
\end{tabularx}
\par\smallskip\flushleft
{\footnotesize \textbf{Note}: each column re-estimates the eleven structural parameters (symmetric
$\bar\mu$ plus the belief, timing and rent parameters) by SMM on the headline criterion; the engagement
window $w=26$, $\delta$, $\tau$ and the holiday factor are fixed.  Het.\ IV and Het.\ OLS
differ only in the per-drug $\hat\alpha_j$ (Table~\ref{tab:demand_hetero}, Panel~B: rival-price 2SLS
vs.\ OLS); the two series are nearly collinear, so they yield the same structural estimate.  \textbf{Homog.}\
sets every $\hat\alpha_j$ to the median and selects rent-eligible drugs by the flow condition alone.  The
main text reports the rival-IV estimate; the mechanism (belief transition, lab-organized wave, and the
Salcobrand$\to$Cruz~Verde leadership handoff) survives all three demands, only the homogeneous spec
mistiming and over-deepening the rent.}
\end{table}

\medskip\noindent\textbf{Estimating the collapse.}  The headline applies the pooled post-ban collapse $\alpha^{\rm post}/\alpha_0=0.29$ (Table~\ref{tab:demand_hetero}) to each drug's pre-ban $\hat\alpha_j$.  As a further check I estimate the collapse as a single parameter jointly with the per-drug pre slopes, in two forms: a proportional reduction $\alpha^{\rm post}_j=\kappa\,\hat\alpha_j$ ($\hat\kappa=0.347$, SE $0.006$) and an additive reduction $\alpha^{\rm post}_j=\hat\alpha_j-\delta$ ($\hat\delta=0.053$, SE $0.001$).  Re-estimating the structural model under each gives SMM loss $19.2$ (proportional) and $33.5$ (additive), against the headline $16.6$.  The degradation is modest and the cartel-formation result and the Salcobrand-first leadership survive both, so the mechanism does not hinge on the exact collapse.

\section{Structural robustness checks}\label{app:struct_robust}

Each check re-simulates the headline at the estimate with one feature switched off, reporting the fit
moments of Table~\ref{tab:mech_fit} (Panel~B).  Table~\ref{tab:struct_robust} collects the three
value-function checks that justify the dynamic structure; the belief and payoff knockouts are in
Table~\ref{tab:spec_fit}.

\textbf{Myopic decisions ($\delta\to0$).}  A myopic firm weighs only the current period, so the
continuation value $\Delta V^1$ drops out of the incentive constraint.  Coordination then survives only
where holding is already a one-shot best response: the count falls to $164$ of $220$ and the rent to $97$
of $143$, and because the escape value no longer differs across chains the leadership collapses toward
chance (Salcobrand $0.30$).  Escaping the war and deepening the rent each trade a current cost for a
future payoff, so the forward-looking value is necessary; the fit degrades to a loss of $555$.

\textbf{Perfect foresight of the ban.}  The headline treats the November injunction as an unanticipated
public event.  The alternative lets firms anticipate it from the first public signal, the
$7$~September CONAR ruling, so the coordination belief jumps there.  Coordination then starts in
September, two months before the binding event (onset week $88$ against the data's $102$), and because
the pre-ban demand still rewards undercutting the premature attempts flood as failures (war-period failed
lifts $720$ against $110$).  The data place the operative shift at the binding November event, not the
foreseeable precursor; the fit degrades to a loss of $2550$.

\textbf{Absorbing value functions.}  The headline values each state by forward-simulating the model's own
transitions, so a firm in the war anticipates it may coordinate and a coordinated firm anticipates the
rent may revert.  Valuing each tier instead as a permanent perpetuity ($V^x=\Pi^x+\delta V^x$), as if a
coordinated drug never reverts and a rent never fails, overstates the tier values and mistimes the rent
(rent steps $127$ of $143$, loss $24$).  This is the mildest of the three, as it should be: it refines
the continuation value rather than removing a channel, but the transition-aware forward simulation is
what reproduces the post-cartel dynamics.

\textbf{Leadership.}  Two further checks confirm that leadership is the calibrated outside option, not the
spillover.  Setting $\psi$ symmetric collapses Salcobrand's lead toward chance ($\sim\!1/3$) while leaving
the coordination count untouched, the signature of a leadership channel decoupled from the incentive
constraint.  Restoring the asymmetry through a chain-specific store-traffic spread $\mu_i$ instead
reproduces the Salcobrand-first split only by fitting $\mu_i$ to the leadership it then explains, the
circularity a calibrated outside option avoids (\S\ref{ssec:mech_estim}).

\begin{table}[H]
\centering\small
\caption{Value-function robustness checks: fit moments}
\label{tab:struct_robust}
\begin{tabular}{l r r r r r}
\toprule
Moment & Data & Estimated & Myopic ($\delta\!\to\!0$) & Perfect foresight & Absorbing \\
\midrule
First coordinations ($0\!\to\!1$)   & $220$ & $214$ & $151$ & $188$ & $214$ \\
Rent steps ($1\!\to\!2$)            & $143$ & $141$ & $107$ & $108$ & $138$ \\
Onset / peak week                   & $102/107$ & $103/107$ & $105/118$ & $\mathbf{88}/90$ & $103/105$ \\
$1\!\to\!2$ lag (weeks)             & $8.0$ & $8.1$ & $23.7$ & $28.8$ & $7.5$ \\
Cartel successes                    & $284$ & $289$ & $95$ & $81$ & $298$ \\
Failed lifts: war                   & $123$ & $110$ & $110$ & $\mathbf{717}$ & $111$ \\
SB leads the cartel wave            & $0.74$ & $0.69$ & $\mathbf{0.26}$ & $\mathbf{0.34}$ & $0.68$ \\
\midrule
\textbf{SMM loss}                   & --- & $\mathbf{16.6}$ & $\mathbf{555}$ & $\mathbf{2550}$ & $\mathbf{23}$ \\
\bottomrule
\end{tabular}
\par\smallskip\flushleft
{\footnotesize \textbf{Note}: each column re-simulates the headline at the estimate ($B=240$) with one
feature of the value function switched off.  Myopic: $\delta\to0$ (no continuation value).
Perfect foresight: the coordination belief jumps at the $7$~September CONAR precursor rather than
the binding November ban.  Absorbing: each tier valued as a permanent perpetuity.  Compare
``Estimated'' to Table~\ref{tab:mech_fit}, Panel~B.}
\end{table}

\begin{figure}[H]\centering
\caption{The cost of leading}
\label{fig:leader_cost}
\includegraphics[width=0.78\linewidth]{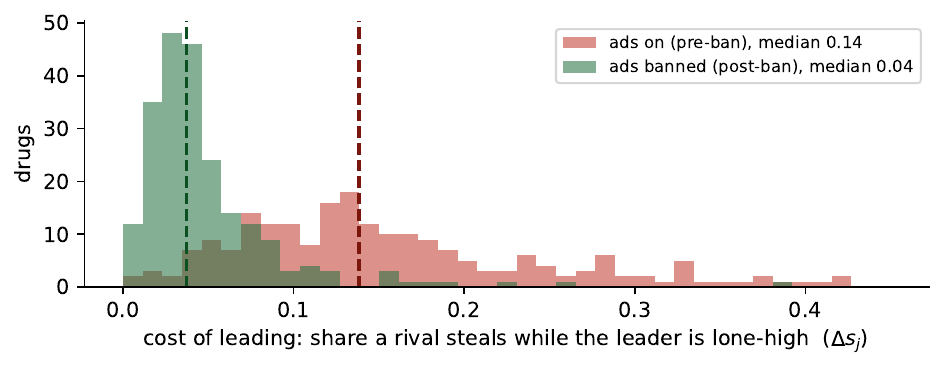}
\par\smallskip\flushleft
{\footnotesize \textbf{Note}: distribution across the $220$ drugs of the cost of leading, the
within-period share $\Delta s_j$ a rival steals while the leader is the lone high-price seller, computed
from the estimated demand.  Dashed lines mark the medians.  The ban collapses the cost from a median
$0.14$ (ads on) to $0.04$ (ads banned).}
\end{figure}

\textbf{The cost of leading.}  A chain that raises a drug first is the lone high-price seller, and a rival
that stays low steals a share $\Delta s_j$ of the drug's volume; that share loss is the cost of leading.
Figure~\ref{fig:leader_cost} plots its distribution across the $220$ drugs, before and after the ban.
While comparative-price advertising is on the cost is large (median $\Delta s\approx0.14$), so the
lone mover is undercut and the war persists; the ban collapses it (median $\approx0.04$), so leading
becomes cheap.  The cost never gates coordination, however.  The discounted escape value exceeds even the
pre-ban cost by an order of magnitude (\S\ref{ssec:mech_setup}, the slack incentive constraint), so the
most-exposed chain leads despite it, and the coordination count is the same whether the store-traffic
spillover is large, small, or shut off entirely (Table~\ref{tab:spec_fit}).

\textbf{Robustness of the coordination flip.}  Table~\ref{tab:mech_robust}
reports the threshold cushion behind the parameter-free flip discussed in
\S\ref{ssec:mech_robust}.  Panel~A varies the size of the post-ban drop in
$\alpha$; Panel~B perturbs the wholesale-cost and nesting calibration.  The band
$[\mu^{*}_{\text{pre}},\mu^{*}_{\text{post}}]$ never closes.

\begin{table}[H]
\centering\small
\caption{Robustness of the coordination flip}
\label{tab:mech_robust}
\begin{tabularx}{0.9\linewidth}{@{}>{\raggedright\arraybackslash}X r r r@{}}
\toprule
\multicolumn{4}{l}{\textbf{Panel A. Cushion vs.\ the size of the $\alpha$ drop}}\\
post-ban $\alpha$ & $\mu^{*}_{\text{post}}$ & cushion $\mu^{*}_{\text{post}}/\mu^{*}_{\text{pre}}$ & band \\
\midrule
$0.029$ (cartel-excl.\ est.) & $30.3$ & $3.5\times$ & open \\
$0.060$                      & $14.6$ & $1.7\times$ & open \\
$0.100$ (rival-IV upper bound)& $\phantom{0}8.8$ & $1.0\times$ & open (sliver) \\
\midrule[\heavyrulewidth]
\multicolumn{4}{l}{\textbf{Panel B. Sensitivity of $[\mu^{*}_{\text{pre}},\mu^{*}_{\text{post}}]$ to the calibration}}\\
perturbation & $\mu^{*}_{\text{pre}}$ & $\mu^{*}_{\text{post}}$ & ratio \\
\midrule
baseline                 & $8.6$ & $30.3$ & $3.5\times$ \\
wholesale cost $-10\%$    & $7.9$ & $29.6$ & $3.8\times$ \\
wholesale cost $+10\%$    & $9.3$ & $31.0$ & $3.3\times$ \\
nesting $\sigma=0.30$     & $9.8$ & $34.8$ & $3.6\times$ \\
nesting $\sigma=0.50$     & $7.2$ & $25.0$ & $3.5\times$ \\
\bottomrule
\end{tabularx}
\par\smallskip\flushleft
{\footnotesize \textbf{Note}: $\mu^{*}(\alpha)$ is
the store-traffic value at which undercutting the cartel ties coordinating; the flip
requires $\mu^{*}_{\text{pre}}<\mu<\mu^{*}_{\text{post}}$.  Panel~A
varies the size of the post-ban drop in $\alpha$ ($\mu^{*}_{\text{pre}}=8.6$
fixed at $\alpha=0.103$); Panel~B perturbs the calibration.}
\end{table}

\textbf{Belief and payoff specifications.}  The main-text Table~\ref{tab:spec_fit}
re-estimates the model with each belief channel and each payoff ingredient switched
off in turn, the evidence behind \S\ref{ssec:mech_robust} and the per-moment companion
to Figure~\ref{fig:price_path}.  The focal jump dominates: dropping the Bayesian
learning (jump-only) still reproduces the cartel wave and fits essentially as well as
the headline (loss $21$ against $16.6$), so the event-tally learning is a minor refinement.
The focal jump itself is what ignites the wave: with it switched off ($\zeta=0$,
learning-only) the coordination never fires in-sample, a degenerate corner ($0$
coordinations, loss $\sim\!2300$).  The commonly observed ban, not the slow tally, is
what selects the coordinated equilibrium.  The
outside option's separate role (selecting who leads) is the re-simulated check in
Table~\ref{tab:struct_robust} (RC3: without it leadership collapses to uniform).

\textbf{The rent's belief gate, and its orthogonality to the coordination belief.}  The rent push
\eqref{eq:mech_rent_push} deepens an eligible coordinated drug ($\Delta V^2_j>0$) at a discounted belief
gate $b^{\rm rent}_t=e^{-\chi}b^{\rm coord}_t$ times the learned enforcement belief $r_t$.
Table~\ref{tab:rent_gates} probes both.  The belief gate is essential and times the rent: forcing
$b^{\rm rent}_t\!\equiv\!1$, the deepening fires eight weeks early and floods ($184$ steps against $141$),
and the loss rises more than fivefold ($88$ against $16.6$).  The rent's pre/post-ban timing, by contrast, is
orthogonal to the coordination belief: forcing $b^{\rm coord}\!\equiv\!1$ floods the $0\!\to\!1$ wave
before the ban ($215$ coordinations, against zero in both the data and the estimate) yet leaves the rent at
zero before the ban, exactly as the estimate has it.  The rent only deepens once the demand collapse makes
the further increase pay, regardless of what the chains believe; the $\Delta V^2_j>0$ eligibility, not a
static-best-response gate, screens which drugs deepen.

\begin{table}[H]\centering\small
\caption{The rent belief gate, and its orthogonality to the coordination belief}
\label{tab:rent_gates}
\begin{tabular}{@{}l r r r r@{}}
\toprule
Moment & Data & Estimated & $b^{\rm rent}\!\equiv\!1$ & $b^{\rm coord}\!\equiv\!1$ \\
\midrule
$0\!\to\!1$ before the ban    & $0$   & $0$   & $0$            & $\mathbf{215}$ \\
Rent steps before the ban     & $0$   & $0$   & $0$            & $0$ \\
Rent steps ($1\!\to\!2$)      & $143$ & $141$ & $\mathbf{184}$ & $145$ \\
Rent peak week                & $117$ & $115$ & $\mathbf{107}$ & --- \\
\midrule
\textbf{SMM criterion (loss)} & ---   & $\mathbf{16.6}$ & $\mathbf{88}$ & $\mathbf{4154}$ \\
\bottomrule
\end{tabular}
\par\smallskip\flushleft{\footnotesize \textbf{Note}: $b^{\rm rent}\!\equiv\!1$ drops the rent's
discounted belief gate (re-estimated under the identical criterion); the rent then floods and fires early,
so the gate is essential and times the rent.  $b^{\rm coord}\!\equiv\!1$ forces the coordination
belief to one (the no-belief column of Table~\ref{tab:spec_fit}): the $0\!\to\!1$ wave floods before the
ban, but the rent stays at zero before the ban because the $\Delta V^2_j>0$ eligibility (the demand
collapse) gates it, not belief.  The rent push therefore carries no static-best-response gate; the
eligibility already screens drugs that would statically defect.}
\end{table}

\subsection{Macro and ownership controls}\label{app:macro_own}

Two contemporaneous shocks could in principle confound the structural reading: the August~2007 acquisition
of Salcobrand by the private-equity fund Southern~Cross, and the Chilean inflation of $2007$--$2008$, which
neared $10\%$ at its mid-$2008$ peak.  Table~\ref{tab:macro_own} re-estimates the model against each.

\textbf{Ownership.}  The headline makes Salcobrand lead the cartel formation through a fixed outside option
$\psi_{\rm SB}=0$: purely domestic, it bears the full below-cost war loss, so it has the largest escape
value and the strongest incentive to coordinate out of the war.  The August~2007 acquisition is the
institutional counterpart of that calibration: a leveraged buyer inheriting an intensifying below-cost war
had every reason to end it, and Salcobrand does, four months later, when it leads the December wave.  The
timing identifies the channel: the cartel forms at the ban, not at the acquisition, so the buyout does not
create the cartel; it rationalizes why Salcobrand, not Cruz~Verde, is the eager leader.  Freeing
$\psi_{\rm SB}$ and re-estimating returns it to $0.00$, with the fit and the leadership ordering unchanged
(loss $16.6$): the most-exposed corner is the estimate, not an assumption.

\textbf{Inflation.}  The deviation gain that drives the model, $G^{01}_j=\mu\,\Delta s_j+\Delta(\text{margin})$,
runs on price--cost margins and within-market relative prices (the rival-price instrument
differences out common shocks), so inflation that lifts prices and costs together is, to first order,
neutral.  The magnitudes confirm it: the war-trough-to-coordinated jump is about $45\%$ over the four-month
formation window, against roughly $4\%$ inflation over those months.  Re-estimating on prices deflated to
the war base by the Chilean CPI (the coordinated tier by $4\%$, the rent tier by $10\%$) leaves the war
below cost (markup $-0.083$ against $-0.091$) and the coordination count, the rent, and the leadership
intact, at a slightly higher loss ($17.7$): the cartel is a real markup, not an artifact of the
$2007$--$2008$ inflation.

\begin{table}[H]\centering\small
\caption{Macro and ownership controls}
\label{tab:macro_own}
\begin{tabular}{@{}l r r r r@{}}
\toprule
Moment & Data & Headline & Real prices & Free $\psi_{\rm SB}$ \\
\midrule
$\bar\mu$\quad store-traffic value     &         & $9.26$ & $9.22$ & $9.26$ \\
$\psi_{\rm SB}$\quad SB outside option &         & $0$\,(fix) & $0$\,(fix) & $\mathbf{0.00}$ \\
War markup ($p-c$)                     & $-0.10$ & $-0.091$ & $-0.083$ & $-0.091$ \\
First coordinations                    & $220$   & $214$ & $216$ & $214$ \\
Rent steps ($1\!\to\!2$)               & $143$   & $141$ & $141$ & $141$ \\
SB leads cartel wave                   & $0.74$  & $0.69$ & $0.66$ & $0.69$ \\
CV leads late rent                     & $0.65$  & $0.68$ & $0.68$ & $0.68$ \\
\midrule
\textbf{SMM criterion (loss)}          & ---     & $\mathbf{16.6}$ & $\mathbf{17.7}$ & $\mathbf{16.6}$ \\
\bottomrule
\end{tabular}
\par\smallskip\flushleft{\footnotesize \textbf{Note}: Real prices deflates the coordinated and rent
tiers to the war-price base by the $2007$--$2008$ Chilean CPI ($\sim\!0.7\%$/month: $-4\%$ on the
coordinated tier, $-10\%$ on the rent) and re-estimates.  Free $\psi_{\rm SB}$ frees Salcobrand's
outside option (fixed at $0$ in the headline) as a twelfth parameter.  Both leave the war below cost, the
coordination count and timing, and the Salcobrand$\to$Cruz~Verde leadership handoff essentially unchanged.}
\end{table}

\subsection{The war forms endogenously, and experimentation cannot escape it}\label{app:experimentation}

\begin{figure}[H]\centering
\caption{The war forms endogenously from the belief dynamics}
\label{fig:full_path}
\includegraphics[width=\linewidth]{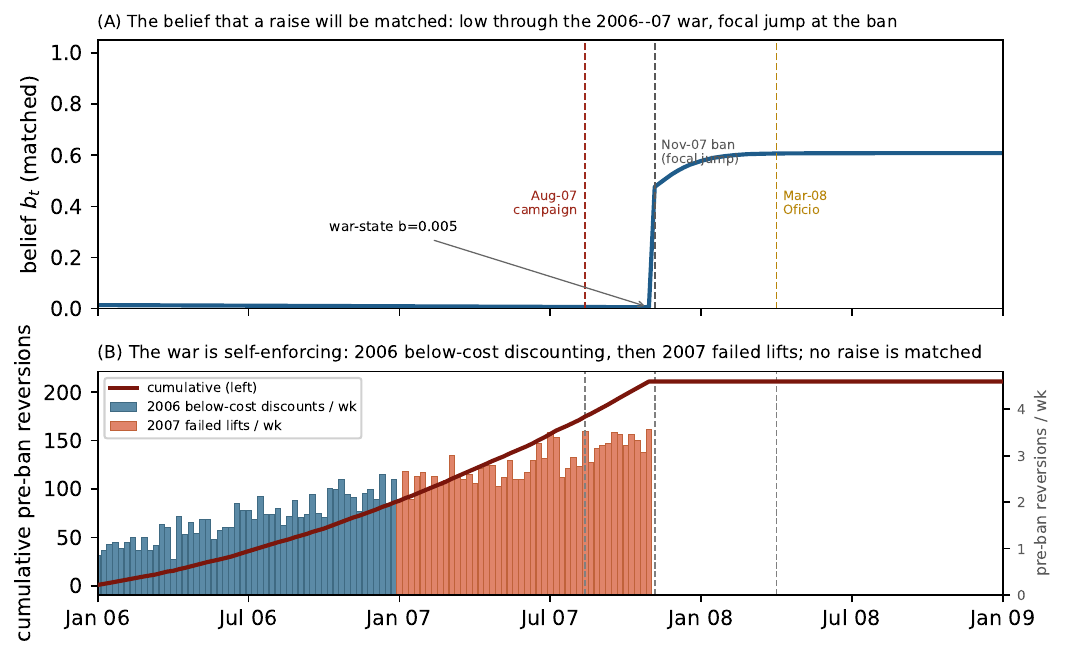}
\par\smallskip\flushleft
{\footnotesize \textbf{Note}: belief and price dynamics rolled from January~$2006$ (week~$0$).
\textbf{(A)}~the belief that a raise will be matched: low and self-fulfilling through the $2006$--$07$
war, then the focal jump at the $6$~November ban.  \textbf{(B)}~the pre-ban price reversions that keep it
low: $2006$ is below-cost discounting (the war itself, Fact~1), $2007$ the clean failed coordination
lifts; the simulation matches the record ($86$/$125$ versus $87$/$123$) with zero pre-ban successes.  The
reversion rate is calibrated to the two annual totals; the belief, hold and jump are the estimated model.}
\end{figure}

The main-text Table~\ref{tab:spec_fit} shows the focal jump is essential and the Bayesian learning a contributing refinement.  To see why,
I roll the belief and price dynamics from January~$2006$ (a year before the structural
window) through the war and into the cartel (Figure~\ref{fig:full_path}).  The war is the low-belief one of two coexisting equilibria, and it forms on its own.
High-belief coordination is dynamically
sustainable throughout (\S\ref{ssec:mech_robust}), so the question is never whether the chains
can coordinate but which equilibrium is selected.
Through $2006$ the war is fought by below-cost discounting: a chain cuts to steal store traffic, the
rivals match within days, and the cut springs back (Fact~1).  This is the war, and no chain yet
holds a coordinated increase.  As the below-cost pain deepens into $2007$, clean attempts to lift out
of it appear; but pre-ban every lift is undercut ($G^{01}_j>0$) and reverts, so none sticks.  The
simulation reproduces the record on both counts: about $86$ below-cost reversions in $2006$ and $125$
failed lifts in $2007$, against $87$ and $123$, with zero successes.  Discount or failed lift, every
pre-ban reversion teaches the same lesson, that a raise will not be matched, so the belief never climbs
above its low prior: it drifts from $0.013$ to $0.004$ by the ban, and the pessimism is self-fulfilling.
The $6$~November ban then delivers the focal jump and flips $G^{01}_j<0$; the belief leaps to $0.81$ and
the cartel forms.

\textbf{What the focal jump adds: ignition on the observed timescale.}  Suppose the ban's payoff effect
occurs but there is no focal jump ($\zeta=0$), so the chains raise the belief through the tally alone.
Post-ban the static flip ($G^{01}_j<0$) makes a coordinated raise a best response for the static-Nash
majority, and a co-led raise faces only the $b_t^{3-k}$ gate for $k$ simultaneous leaders, far weaker
than the single-leader $b_t^{2}$, so coordination is not logically impossible.  But it is far
too slow to explain the record.  Re-estimated without the jump, the wave never ignites within the
sample: no coordinated wave forms by the window's end, fitting an order of magnitude worse ($\sim\!2300$
against the headline $16.6$).  By pure experimentation the static-Nash majority would coordinate only
over a horizon far longer than the observed window; its exact length turns on the learning rate, which
the data identify only loosely, but on any reading it is far beyond the four-month wave, and the
most-elastic drugs, whose $G^{01}_j$ stays positive even post-ban, never coordinate without the belief
at all.  The focal jump's role is therefore to make the observed wave possible on the observed
timescale: the commonly observed event resolves the higher-order belief in one step, compressing what
bare experimentation would reach only much later into the sharp December onset the record shows.

\textbf{The low belief is disciplined out of sample.}  The pessimism is not fit to the estimation window.
Through $2006$ the chains fought a deepening below-cost war and never sustained a coordinated
increase, an observed coordination frequency of $0.00$ against the estimated prior
$a_0/(a_0+b_0)=0.013$; folding the $2006$ record into the tally only lowers it.  By the time the
structural window opens, no raise had ever been matched.

% [tab:horizon removed: its single-leader ~56-yr horizon is superseded by the co-leadership bootstrap (~1 yr); see the reframed passage above.]

\subsection{Leadership-dispersion robustness}\label{app:lead_robust}
The two leader-softmax dispersions $\tau_{\rm lead}$ and $\tau_{\rm rent}$ are calibrated to the leadership shares, so the share magnitudes are calibration targets, not predictions; the escape \emph{ordering} is the prediction of the loss cushion $\psi$.  Table~\ref{tab:lead_robust} separates the two.  Fixing $\tau_{\rm lead}=\tau_{\rm rent}$ at a common value, Salcobrand still leads the cartel wave throughout ($69$--$80\%$), so the escape ordering does not depend on the calibration.  The rent leader is more sensitive: Cruz~Verde leads the held rent at the calibrated and low common scales, but the rent leadership flattens toward uniform at a high common dispersion.  The robust object is the escape ordering; the share magnitudes and the rent leader are the calibration-dependent parts.

\begin{table}[H]
\centering\small
\caption{Leadership robustness to a common softmax dispersion}
\label{tab:lead_robust}
\begin{tabular}{@{}lcc@{}}
\toprule
Dispersion & SB leads cartel wave & CV leads the rent \\
\midrule
$\tau_{\rm lead}{=}3.0,\ \tau_{\rm rent}{=}0.12$ (headline) & $69\%$ & $68\%$ \\
common $\tau=0.12$ & $73\%$ & $68\%$ \\
common $\tau=1.0$  & $79\%$ & $37\%$ \\
common $\tau=3.0$  & $76\%$ & uniform ($35/32/33$) \\
\bottomrule
\end{tabular}
\par\smallskip\flushleft
{\footnotesize \textbf{Note}: Leadership shares at the headline estimate, re-simulated with the two leader-softmax dispersions set equal.  Salcobrand leads the cartel wave under every dispersion (the $\psi$ ordering); the rent leader and the magnitudes depend on the calibrated scales.}
\end{table}

\subsection{Defection-rate robustness}\label{app:lamcut_robust}
The per-week down-cut rate $\lambda_{\rm cut}$ is a free scalar, but the post-ban cartel stability is an outcome of the slack gate, not of the rate.  Table~\ref{tab:lamcut_robust} perturbs $\lambda_{\rm cut}$ from half to double its estimate.  The coordination count is essentially invariant (first coordinations $214$--$215$ of $220$; cartel successes $288$--$291$), while only the pre-ban war-deviation intensity scales with the rate.  Post-ban the slack is positive, so undercutting does not pay at any plausible rate and the cartel holds; the rate governs how violent the \emph{war} is, not whether the \emph{cartel} survives.

\begin{table}[H]
\centering\small
\caption{Cartel stability is invariant to the down-cut rate}
\label{tab:lamcut_robust}
\begin{tabular}{@{}lccc@{}}
\toprule
$\lambda_{\rm cut}$ & First coord.\ (data $220$) & Cartel succ.\ (data $284$) & War deviations \\
\midrule
$0.044$ ($-50\%$)  & $215$ & $286$ & $253$ \\
$0.088$ (headline) & $215$ & $291$ & $513$ \\
$0.133$ ($+50\%$)  & $214$ & $291$ & $770$ \\
$0.177$ ($+100\%$) & $214$ & $292$ & $1019$ \\
\bottomrule
\end{tabular}
\par\smallskip\flushleft
{\footnotesize \textbf{Note}: Headline estimate re-simulated with $\lambda_{\rm cut}$ perturbed.  The coordination count barely moves; only the war's below-cost deviation count scales with the rate.}
\end{table}

\subsection{Dropping the noisiest drugs}\label{app:drop48_robust}
The pooled demand collapse is used because the standalone per-drug post-ban coefficient is too thinly identified for some drugs: estimated drug by drug, $48$ of the $220$ receive a spuriously weak collapse.  As a direct check that these noisy drugs do not drive the structural estimate, I drop all $48$ and re-estimate on the remaining $172$.  Table~\ref{tab:drop48_robust} shows the estimate is unchanged: the store-traffic value ($\bar\mu=9.39$ against $9.26$), the down-cut rate, and the leadership shares all hold, and the model reproduces the same path (first coordinations $168$ of $172$ against $214$ of $220$; cartel successes $226$ of $229$).  The criterion level is not comparable across the two samples because the count moments scale with the number of drugs, but the estimated primitives and the fit ratios are stable, so the weak-collapse drugs neither drive nor distort the estimate.

\begin{table}[H]
\centering\small
\caption{Estimate is stable to dropping the 48 weak-collapse drugs}
\label{tab:drop48_robust}
\begin{tabular}{@{}lcc@{}}
\toprule
                                       & Headline ($J=220$) & Drop weak-collapse ($J=172$) \\
\midrule
$\bar\mu$\quad store-traffic value     & $9.26$    & $9.39$    \\
$\lambda_{\rm cut}$\quad down-cut rate & $0.088$   & $0.084$   \\
SB leads cartel wave                   & $69\%$    & $66\%$    \\
CV leads post-rent                     & $67\%$    & $68\%$    \\
First coordinations (model/total)      & $214/220$ & $168/172$ \\
Cartel successes (model/data)          & $289/284$ & $228/229$ \\
\bottomrule
\end{tabular}
\par\smallskip\flushleft
{\footnotesize \textbf{Note}: The $48$ drugs whose standalone per-drug post-ban collapse is weakest are dropped and the model re-estimated on the remaining $172$.  The store-traffic value and the leadership shares are unchanged; the coordination counts scale with the smaller sample.}
\end{table}

\section{Example war attempts}\label{app:war_attempts}

\begin{figure}[htbp]
\centering
\caption{Six example war attempts}
\label{fig:war_attempts}
\includegraphics[width=\textwidth]{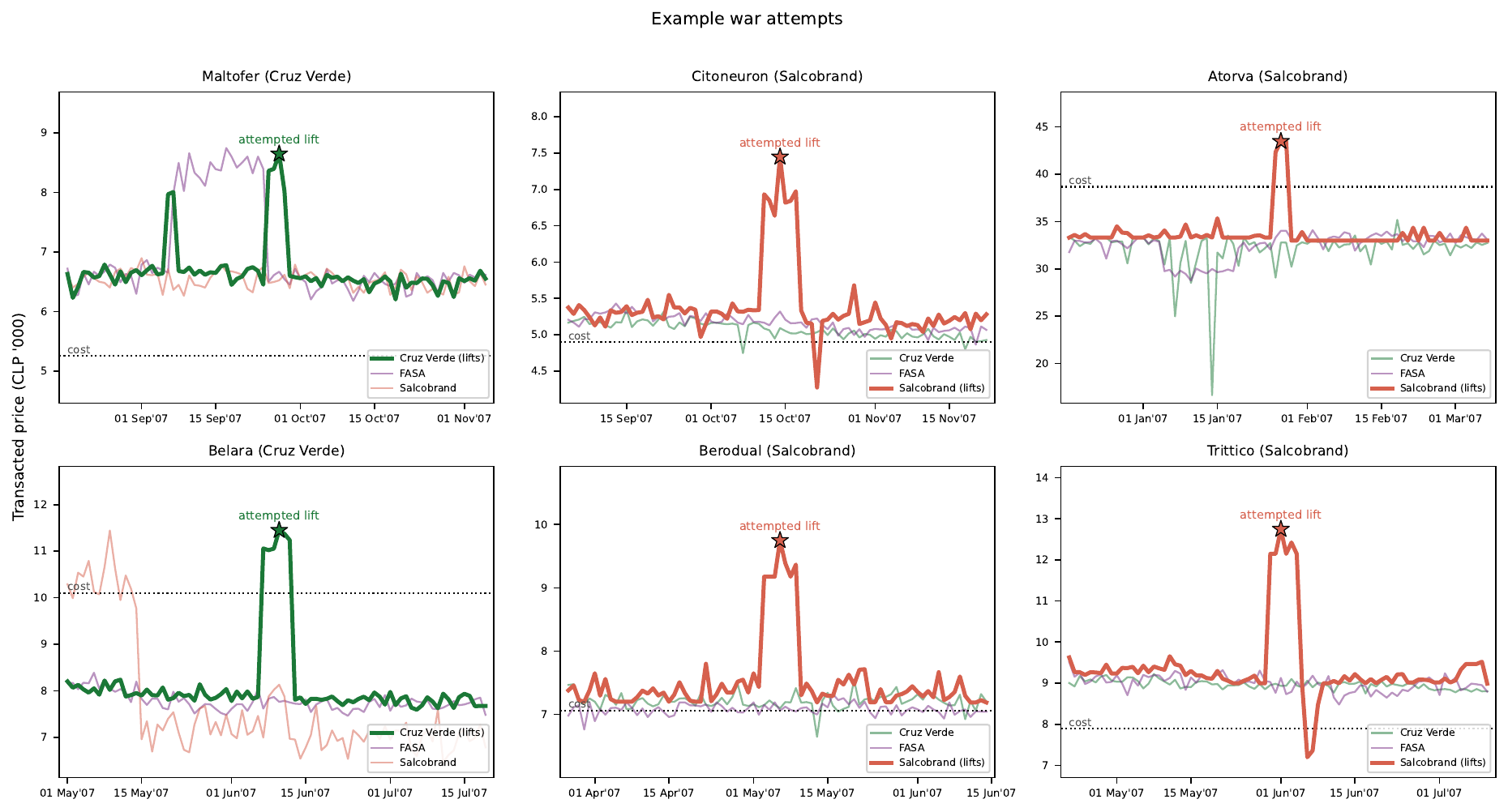}

\par\smallskip\flushleft
{\footnotesize \textbf{Note}: In each
panel the heavy line is the chain that raised price (marked $\star$ at the lift); the two
light lines are the rivals, which stay near or below cost (dotted line) and undercut the
lift, so it reverts within weeks.  Transacted prices, three chains, $\pm 6$-week window
around the event.}
\end{figure}

Figure~\ref{fig:war_attempts} shows six representative war attempts from the
pre-ban period: episodes in which one chain raised a drug's price toward a coordinated
level, the two rivals did not follow but undercut it, and the lift reverted to the
below-cost war floor within weeks.  These are the ledger's clean \texttt{failed\_attempt}
events ($101$ lift-then-revert episodes in the January--November $2007$ war window, after
setting aside $34$ lower-amplitude price wobbles that do not clear a rise-and-revert
threshold relative to the drug's own volatility), which the structural model reproduces
through its belief-driven, sporadic war lifts (\S\ref{sec:mechanisms};
$100$ simulated).  Salcobrand is the lifter in four of the six (Cruz Verde in the other two),
consistent with its role as the coordination leader.

\section{Regulatory timeline and case outcome}\label{app:timeline}

The episode's documented chronology: a comparative-price advertising campaign by
Cruz Verde (August $2007$); the CONAR self-regulatory ruling (CONAR $704$/$07$,
$7$ September $2007$) and the binding Civil Court \emph{medida precautoria}
prohibiting comparative price advertising ($6$ November $2007$); the first
three-chain coordinated increase ($3$ December $2007$) and the wave running
through April $2008$; the FNE's complaint (Requerimiento, Rol C No.\ $184$-$08$,
$2008$; \citealp{fne2008requerimiento}); and the Competition Tribunal's
conviction on $206$ drugs with fines of about US\$$38$M (Sentencia
No.\ $119$/$2012$; \citealp{tdlc2012sentencia}), upheld by the Supreme Court
(Rol No.\ $2578$-$2012$).  Civil compensation followed: Cruz Verde and Salcobrand
settled in November $2020$ (CLP $1.1$bn to $\sim$$53{,}000$ consumers) and FASA in
August $2025$ (CLP $980$M to $\sim$$34{,}000$), for total consumer compensation
exceeding CLP $2.08$bn ($\sim$US\$$3.9$M) across more than $87{,}000$ consumers
(SERNAC settlement records).  These realized magnitudes are an external check on
the welfare accounting of \S\ref{sec:welfare}.

\bibliographystyle{plainnat}
\bibliography{reference}

\end{document}